\documentclass[twocolumn]{aastex62}

\usepackage{natbib}
\usepackage{amssymb}
\newcommand{\ciii}{C{\small III}]}
  \newcommand{\oiii}{[O{\small III}]}
  \newcommand{\oii}{[O{\small II}]}
  \newcommand{\hii}{H{\small II}}
  \newcommand{\ha}{H$\alpha$}
  \newcommand{\ew}{EW(C{\small III}])}
  
  \newcommand{\haro}{{\small Haro\hspace{.5mm}11}}
  
\received{\today}
\revised{\today}
\accepted{\today}
\shorttitle{CIII] in Haro 11}
\shortauthors{Micheva et al.}
\begin{document}

\title{Spatially resolved CIII]$\lambda1909$ emission in Haro 11}

\correspondingauthor{Genoveva Micheva}
\email{gmicheva@aip.de}

\author[0000-0002-0786-7307]{Genoveva Micheva}
\affiliation{Leibniz-Institute for Astrophysics Potsdam, An der Sternwarte 16, 14482 Potsdam, Germany}

\author{G\"oran \"Ostlin}
\affiliation{Stockholm University, Department of Astronomy and Oskar Klein Centre for Cosmoparticle Physics, AlbaNova University Centre, SE-10691, Stockholm, Sweden}

\author{Jens Melinder}
\affiliation{Stockholm University, Department of Astronomy and Oskar Klein Centre for Cosmoparticle Physics, AlbaNova University Centre, SE-10691, Stockholm, Sweden}

\author{Matthew Hayes}
\affiliation{Stockholm University, Department of Astronomy and Oskar Klein Centre for Cosmoparticle Physics, AlbaNova University Centre, SE-10691, Stockholm, Sweden}

\author{M. S. Oey}
\affiliation{University of Michigan, 311 West Hall, 1085 South University Ave, Ann Arbor, MI 48109-1107, USA}

\author{Akio K. Inoue}
\affiliation{Department of Physics, School of Advanced Science and Engineering, Waseda University, 3-4-1 Okubo, Shinjuku, Tokyo 169-8555, Japan}
\affiliation{Waseda Research Institute for Science and Engineering, Faculty of Science and Engineering, Waseda University, 3-4-1 Okubo, Shinjuku, Tokyo 169-8555, Japan}

\author{Ikuru Iwata}
\affiliation{National Astronomical Observatory of Japan, 2-21-1, Osawa, Mitaka, Tokyo 181-8588, Japan}
\affiliation{Department of Astronomical Science, The Graduate University for Advanced Studies (Sokendai), 2-21-1, Osawa, Mitaka, Tokyo}

\author{Angela Adamo}
\affiliation{Stockholm University, Department of Astronomy and Oskar Klein Centre for Cosmoparticle Physics, AlbaNova University Centre, SE-10691, Stockholm, Sweden}

\author{Lutz Wisotzki}
\affiliation{Leibniz-Institute for Astrophysics Potsdam, An der Sternwarte 16, 14482 Potsdam, Germany}

\author{Kimihiko Nakajima}
\affiliation{National Astronomical Observatory of Japan, 2-21-1 Osawa, Mitaka, Tokyo 181-8588, Japan}



\begin{abstract}
  The CIII]1909 (hereafter, \ciii) line is the strongest ultraviolet emission line after Ly$\alpha$ and is therefore of interest to high redshift studies of star-forming (SF) galaxies near the epoch of reionization. It is thought that \ciii\ emission is strongest in galaxies with subsolar metallicity and low mass, however, spectral observations of numerous such galaxies at high and low redshift produce inconclusive or even contradictory results. We present the first ever \ciii\ imaging, obtained with HST/STIS for the low-redshift SF galaxy Haro 11. Cluster parameters like stellar mass, dust fraction and dust attenuation, and ionization parameter, obtained through spectral energy distribution fitting, show no correlation with the CIII] equivalent width (EW), which may be due to a combination of the limitation of the models and the age-homogeneity of the cluster population. Comparing the ratio of \ciii\ emission line flux from individual clusters to that of H$\alpha$, \oiii, and \oii\ we find that the clusters with the highest \ew\ can be reconciled only with Cloudy models with an extremely high C/O ratio of $\ge1.4$(C/O)$_\odot$ for an ionizing population of single stars, binary stars, or a mixture of binary stars and active galactic nuclei. Given the point-like nature of strong \ciii, the integrated total strength of \ew\ becomes dependent on the morphology of the galaxy, which would explain the large scatter in \ew\ strengths, observed in galaxies with otherwise similar SF properties, and of similarly low metallicity and stellar mass.
                      
\end{abstract}

\keywords{Haro 11 --- starburst galaxies --- interstellar medium --- photoionization --- ultraviolet astronomy}

\section{Introduction} \label{sec:intro}

The [CIII]$\lambda1906$, CIII]$\lambda1909$ emission line doublet (hereafter \ciii) is the strongest UV line after Ly$\alpha$ in star-forming (SF) galaxies \citep[e.g.,][]{Stark2014,Jaskot2016}. It is not a resonant line, and hence can provide systemic redshifts for galaxies near the epoch of reionization, which will make it a frequent target for James Webb Space Telescope observations. \ciii\ is also the primary emission line of C$^{++}$ in the UV and at low metallicities and is therefore necessary for carbon abundance measurements \citep[e.g.][]{Garnett1995, Kobulnicky1998,Shapley2003,Erb2010,Berg2016}. The strength of the equivalent width of \ciii\ (\ew) has been proposed as an indirect indicator of Lyman continuum (LyC) espace, since it decreases with decreasing LyC optical depth \citep{Jaskot2016}.

  The behavior of the \ciii\ line is currently not well understood. Theory predicts that production of \ciii\ should be enhanced in galaxies with young stellar ages and high excitation, traced by the ionization parameter $\log{U}$ \citep[e.g.,][]{Gutkin2016,Feltre2016,Jaskot2016}. This is supported by observations of young, low-mass, low-metallicity galaxies with low dust extinction at redshifts $1.5\le z \le 7.7$, which show very strong \ciii\ emission with \ew$\sim22$\AA\ \citep[e.g.,][]{Erb2010,Stark2014, Stark2015, Berg2016,Stark2017, Senchyna2017,Maseda2017,Berg2019}. Galaxies with strong \ciii\ detections seem to fall in metallicity range of approximately $12+\log{O/H}\sim7.7$ \citep{Berg2016} to $\sim8.5$ \citep{Senchyna2017}. The strength of \ciii\ decreases significantly in more metal-poor or metal-rich galaxies, because in the former case the carbon abundance is low, and in the latter the cooling becomes very efficient, decreasing the strength of collisionally excited lines. For example, for galaxies with $12+\log{O/H}=8.6$, \citet{Steidel2016} measure a \ciii\ of only $1$\AA. The metallicity does not directly correlate with the strength of \ciii, however. \citet{PenaGuerrero2017} find extremely low \ew\ of $\lesssim2$\AA\ for all of their local starburst galaxies ($z\lesssim0.06)$, spanning the metallicity range $7.2\mbox{-}8.14$. \citet{Senchyna2017} demonstrate that for the same metallicity ($12+\log{O/H}=7.81$), a galaxy can be a strong emitter, with \ew$=14.86$\AA, or a non-detection, with an upper limit of $<3.9$\AA (their figure 4). Large samples of hundreds of galaxies at redshifts $0.8 \lesssim z \lesssim3.0$ show low \ew\ of $\lesssim 2$\AA\ \citep[e.g.,][]{Shapley2003,Rigby2015, Du2016,Maseda2017}. These samples have comparable metallicities to the range in the \citet{Senchyna2017} sample, e.g., $12+\log{O/H}$ of $8.3$ \citep{Shapley2003}, $8.2$ \citep{Rigby2015}, and $<8.4$ \citep{Du2016}. They also have SF properties similar to the strong \ciii-emitters at high redshifts, and hence their low \ciii-emission levels are puzzling. The discrepancy cannot be attributed to a redshift evolution of the \ciii\ strength, since strong (and weak) \ciii\ emitters can be found at any redshift. For example, the local star-forming objects Mrk 71 (distance $3.5$ Mpc) and Tol 1214-277 ($z=0.026$), are among the strongest \ciii\ emitters known in the universe, with \ew$\gtrsim14$ \AA\ \citep{Rigby2015}. A LyC-leaking Green Pea galaxy at $z=0.37$ \citep{Schaerer2018} and among the samples of \citet{Berg2016,Berg2019} and \citet{Senchyna2017}, there are galaxies with similarly high \ew$\sim14$\AA. 
  
  The literature studies of \ciii\ are mostly spectroscopic, and hence use integrated \ciii\ emission, resulting in an effective \ew. Although \citet{Patricio2016} and \citet{James2018} present MUSE observations covering the \ciii\ line, the target galaxies are at high redshifts ($z\ge2.38$), and hence the spacial resolution is low. Integrated \ciii\ emission may be difficult to interpret due to the dilution of the total \ew\ resulting from integrating over regions with different SF properties. Spatially resolved \ciii\ emission on the scales of star clusters has to date been unavailable. Therefore, in the Hubble Space Telescope (HST) cycle 25 we proposed imaging of the three local starbursts Haro 11, ESO 338-04, and Mrk 71 in the \ciii\ emission line with STIS/NUV-MAMA. In this paper we present the results for Haro 11, a.k.a. ESO-350-IG-038, which is at a distance of $\sim87$ Mpc, and is well-studied.

  Haro 11 is a strongly star-forming (SF) blue compact galaxy (BCG) and a Lyman Break Analog (LBA) at redshift $z=0.02$ ($\sim87$ Mpc), with a metallicity of $12+\log{O/H}=7.9$ \citep{Bergvall2002} for the galaxy as a whole and $8.09\pm0.2$, $8.25\pm0.15$, and $7.8\pm0.13$ for knots A, B, and C, respectively \citep{James2013}. In other words, both the galaxy as a whole and the individual knots populate a metallicity range seemingly hospitable to strong \ciii\ emission \citep[e.g.,][]{Berg2016,Senchyna2017}. Haro 11 is undergoing a merger, as indicated by its complex kinematics \citep{Ostlin2001, Ostlin2015, James2013,Menacho2019}. Its three main starburst regions, called knots A, B, and C \citep{Kunth2003}, have very different properties. Knot A consists of individual young clusters, and ionization parameter mapping indicates that the entire knot region is likely optically thin \citep{Keenan2017}. Knot B is unresolved, is dissected by a visible dust lane \citep{Ostlin2015} and has the highest extinction of all three \citep{Adamo2010}. Knot C is a strong Ly$\alpha$ emitter \citep{Hayes2007}, and its Ly$\alpha$ line profile is consistent with a density-bounded low-column density region \citep{Rivera2017}. Haro 11 has a Ly$\alpha$ escape fraction of $\sim9\%$ \citep{Hayes2007} and is the first spectroscopically detected Lyman continuum (LyC) leaker in the local universe \citep{Bergvall2006,Leitet2011}, with $f_{esc}=3.3\%$ \citep{Leitet2011}. By all accounts, Haro 11 has all the SF properties and the right metallicity to be a strong \ciii\ emitter. Yet from Internation Ultraviolet Explorer (IUE) archival spectra, we measure a total effective \ew$\sim3$ \AA. In this paper we study the \ciii\ emission with the high spatial resolution offered by the HST, in an attempt to understand the behavior and properties of the \ciii\ line. 
  
This paper is structured as follows: in Section \ref{sec:data} we present the HST observations. The four schemes used to extract signal from the data are described in Section \ref{sec:signalextraction}. Section \ref{sec:powlaw} details the method for estimating and removing the continuum from the on-line narrowband filter. The resulting \ew\ is presented in Section \ref{sec:ew}, and discussed in Section \ref{sec:discuss}. In the discussion, we compare the strength of \ciii\ to other emission lines in Haro 11, look for correlations between the the \ciii\ line strength and star cluster properties derived from fitting their spectral energy distribution (SED), as well as compare various line ratios to Cloudy model predictions. The details of the powerlaw, SED and radiative transfer modeling are given in the Appendix.

Throughout this paper we adopt a flat cold dark matter cosmology with $H_0= 70$ km s$^{-1}$ Mpc$^{-1}$, $\Omega_M=0.3$, $\Omega_\Lambda=0.7$.

\section{HST observations} \label{sec:data}
\begin{figure}[ht!]
\plotone{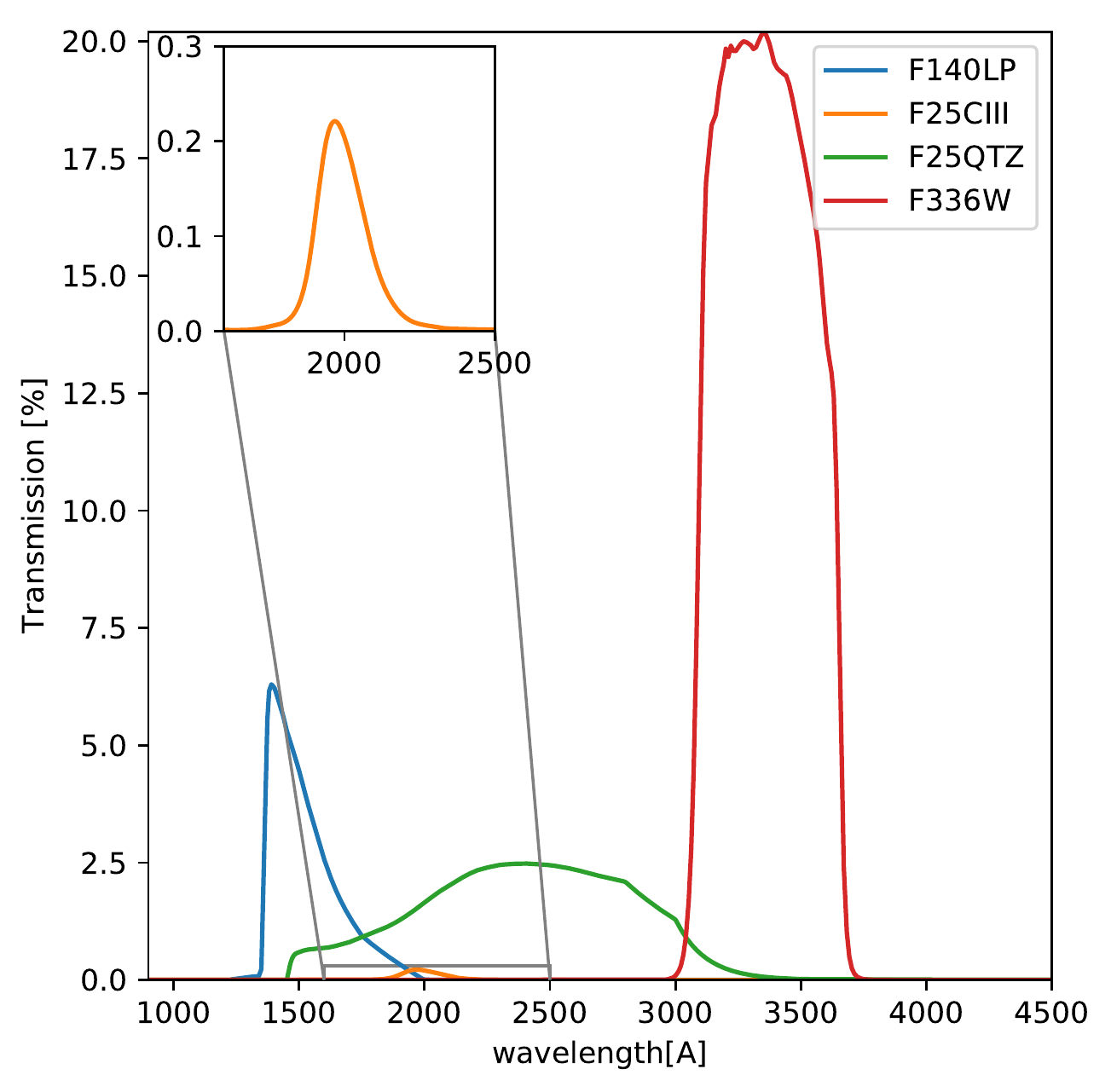}
\caption{Filter transmissions used in the powerlaw fit. The zoomed inset is of the F25CIII filter, which has a $\sim0.2\%$ maximum transmission. \label{fig:filters}}
\end{figure}

We obtained 4 orbits of STIS/NUV-MAMA imaging of Haro 11 with the F25CIII filter, which covers the \ciii\ line, and the broad continuum filter F25QTZ. A summary of the observations is shown in Table \ref{tab:data}, and filter transmission curves in Figure \ref{fig:filters}. In addition, our analysis uses ``auxiliary'' imaging from the archive, spanning the UV and optical spectral ranges. The auxiliary UV filters are ACS/SBC F140LP, WFC3/UVIS1 F336W, and the narrowband WFC3/UVIS1 FQ378N (\oii$\lambda3727$). The auxiliary optical filters are the narrowband ACS/WFC1 FR505N (\oiii$\lambda\lambda4959,5007$) and ACS/WFC1 FR656N (\ha), and the broadband ACS/WFC1 F435W, ACS/WFC1 F550M, and WFC3/UVIS1 F763M. 

%% The "ht!" tells LaTeX to put the figure "here" first, at the "top" next
%% and to override the normal way of calculating a float position

%\startlongtable
\begin{deluxetable*}{cccccccc}
\tablecaption{Summary of the HST data used in the powerlaw fits.\label{tab:data}}
\tablehead{
\colhead{Instrument} & \colhead{Filter} & \colhead{PID} & Exp. time & $\lambda_{Pivot}$ & Rectangular Width & PHOTFLAM & Comment\\
\colhead{} & \colhead{} & \colhead{} & \colhead{[sec]} & \colhead{[\AA]} & \colhead{[\AA]}&\colhead{erg/s/cm$^2$/\AA} &\colhead{[\AA]}
}
%\colnumbers
%\decimals
\startdata
%ACS          & F122M   & 9470    &	9095  & 1272.5  & 129.8 \\	
ACS/SBC       & F140LP  & 9470    &	2700  & 1519.4  & 236.2 & $1.941046\times10^{-17}$ & archive\\	
%WFPC2        & F218W   & 6708    &	1800  &   & \\	
%WFPC2        & F439W   & 6708    &	1400  &   & \\	
%ACS          & F550M   & 10575   &	800   & 5581.3  & 533.1 \\	
%WFPC2        & F555W   & 6708    &	260   &   & \\	
%WFPC2        & F814W   & 6639    &	840   &   & \\	
%ACS          & FR656N  & 10575   &	1214  & 6624.6  & 133.0 \\	
STIS/NUV-MAMA & F25CIII & 15088   &     10000 & 2010.7  & 200.7 & $4.5746211\times10^{-16}$ & this work\\
STIS/NUV-MAMA & F25QTZ  & 15088   &     1544  & 2360.0  & 1139.2& $6.4743538\times10^{-18}$ & this work\\
WFC3/UVIS1    & F336W   & 13702   &	2664  & 3354.8  & 511.5 & $1.2661227\times10^{-18}$ & archive\\	
\enddata
\end{deluxetable*}

After downloading the calibrated {\tt\string FLT} frames from the archive, we used {\tt\string AstroDrizzle} \citep{Fruchter2011,Fruchter2012} to drizzle the images and resample them from their original pixel scale of $0.025$ arcsec to a pixel scale of $0.04$ arcsec, which is the common scale of the auxiliary HST imaging available for \haro. To obtain error images, we run {\tt\string AstroDrizzle} a second time to obtain inverse variance maps ({\tt\string WHT}), which then give the error as $1/\sqrt{{\tt\string WHT}}$. Upon comparing the {\tt\string PHOTFLAM} values in the headers of the F25CIII images to output from {\tt\string pysynphot} with {\tt\string oref} file {\tt\string n181340bo\_pht.fits}, we found discrepancies of up to $0.1$ mag. In this work, we use the {\tt\string PHOTFLAM} values from the FITS headers, listed in Table \ref{tab:data}\footnote{The STSci help desk usually recommends using {\tt\string pysynphot} since it should have the most up-to-date photometry and time-dependent sensitivity changes. However, we have performed the entire analysis in this paper with the {\tt\string pysynphot} {\tt\string PHOTFLAM} and compared it to that of the header {\tt\string PHOTFLAM}. The pysynphot version gives $\left<EW(CIII])\right>=34.5$\AA, with a maximum of $44.9$\AA. These numbers are unphysically large and also do not fit with any Cloudy simulation models for stellar or AGN ionizing sources, for any assumed C/O ratio, including $1.4\times$, $3\times$, and even $10\times$ solar. We therefore must conclude that the {\tt\string pysynphot} {\tt\string PHOTFLAM} value for F25CIII is incorrect.}.

After drizzling, the data were aligned with {\tt\string IRAF GEOMAP/GEOTRAN} to counter any residual misalignement. To account for different point spread functions (PSF) in the different filters, we performed PSF equalization. The PSF for the Solar Blind Channel (SBC) filters has a non-standard shape, with a narrow core and extended wings \citep[ISR ACS 2016-05,][]{Hayes2016}. The extended wings in particular present a challenging problem for detecting low surface brightness emission in continuum subtracted images. In order to take this into account we employ a PSF matching technique where the optimum matching kernel (based on multiple delta functions, rather than a single Gaussian function as is standard) can be found and used to match the PSFs of the
FUV filters to each other and to the optical filters \citep[][Melinder et al. in prep.]{Becker2012}.

The PSF of the F25QTZ was constructed from $\sim100$ stars in the globular cluster NGC 6681 from available archival data in the same filter. The same stars were available in filter F25CN182 but not in F25CIII. In fact, other than our \ciii\ data, there are no other observations in existence with the STIS F25CIII filter, which makes it problematic to obtain a reliable PSF in this filter. We attempted to replace F25CIII with F25CN182 because at first glance the STIS handbook suggests that the PSF of F25CIII is nearly identical to that of F25CN182, to within $200$ milliarcseconds, or $\sim70\%$ encircled energy. However, upon comparing the F25CN182 empirical PSF, obtained from $\sim50$ NGC 6681 stars, with the F25CIII PSF obtained from the observation of a single star in the year 1998, we found significant differences in both the core and the wings of the PSF, with F25CN182 having a narrower core and a wider wing profile than F25CIII. The age of the F25CIII star observations makes it risky to use this PSF, since the shape of the profile may have changed over the years due to thermal breathing of the instrument, which would affect the core more than the wings.  Since there are also no suitably isolated point sources in our F25CIII data of Haro 11, we are left with little choice but to use the F25CIII PSF based on a single star observations from 1998 throughout this work. To alleviate any concerns about incorrect PSFs or spurious \ciii\ detections due to lingering residuals after PSF equalization, we perform the signal extraction on several spatial scales, as described in Section \ref{sec:signalextraction}.

During the writing of this work, the Space Telescope Science Institute (STSci) released a notice\footnote{ISR 2019-05, \url{http://www.stsci.edu/files/live/sites/www/files/home/hst/instrumentation/acs/documentation/instrument-science-reports-isrs/_documents/isr1905.pdf}} of a $\sim30\%$ reduction of the HST ACS/SBC zeropoints for all data obtained after $2002$. This affects the auxiliary F140LP imaging, for which STSci provided us with a new zeropoint of PHOTFLAM $=1.941046\times10^{-17}$ erg/s/cm$^2$/\AA. Throughout this work we use the new zeropoint in both powerlaw and spectral energy distribution fits.

\section{Signal Extraction} \label{sec:signalextraction}
In an attempt to learn as much as possible from the data, we extract the \ciii\ signal in four complementary ways - one using unbinned and three using binned images. First, if the signal is high enough, working with unbinned pixels is ideal and offers the greatest spatial resolution. However, the pixels are quite small ($0.04^{''}\times0.04^{''}$) and the transmission of the F25CIII on-line filter is at most $\sim0.2\%$ (Figure \ref{fig:filters}). This makes the resulting CIII] image quite noisy and its interpretation unconvincing. We therefore also first bin the data to increase the signal-to-noise (S/N), and then extract the \ciii\ signal with the method described in Section \ref{sec:powlaw}.

  We perform the binning in three ways. To provide accurate line fluxes for the star clusters we perform aperture photometry with apertures of radius $r=0.125$ arcsec on all clusters, simultaneously detected in the F140LP, F25QTZ, F25CIII and F336W filters. The cluster positions were obtained with IRAF DAOFIND. The size of the aperture was chosen to both match that in the Haro 11 young cluster investigation in \citet{Adamo2010} and because it is the largest aperture that allows minimal overlap between neighboring clusters in knot A. Since the images have been PSF equalized, we apply a single aperture flux correction factor of $1.45$ ($-0.4$ mag) based on the PSF of the equalized images.

  To increase the S/N per resolution element in the space between clusters, we perform Voronoi tesselation on the F140LP image with a target S/N of $30$ per bin, using the {\tt\string vorbin} python module \citep{Cappellari2003}, with the weighted Voronoi tesselation modification proposed by \citet{Diehl2006}. Voronoi binning preserves the maximum spatial resolution of an image, given a constraint on the minimum S/N ratio inside each bin. The minimum and maximum bin sizes have been forced to $4$ and $625$ pixels, respectively, in order to avoid having single-pixel or too large bins. This means that the largest bins in the outskirts of the image will have S/N lower than $30$. The F140LP-based Voronoi bin map is applied to F25CIII, F25QTZ and all of the auxiliary images, before signal extraction. We choose F140LP as the base for the Voronoi bin map because the \ew\ is sensitive to the continuum level and so we ensure that most bins have reliable S/N in the continuum. We note that we have examined a F25CIII-based Voronoi bin map but found the bin size distribution to be non-optimal.

  Finally, we also constructed a dendrogram tree from the F140LP image using the {\tt\string astrodenro}\footnote{\url{http://dendrograms.org}} python module with $7$ pixels minimium leaf area, which is $\sim2\times$ the full width half maximum (FWHM) of the PSF. Further details on the dendrogram construction are given in Section \ref{sec:dendro} of the Appendix. We included the dendrogram signal extraction because one cannot be certain that the F25CIII PSF has not changed in the last $20$ years, and hence if our PSF equalization is satisfactory. \ciii\ signal extraction from unbinned pixels would be affected by any residual PSF unaccuracies, but also the smallest Voronoi bins would not be immune to this effect. 
  
\begin{figure*}[ht!]
  \includegraphics[width=0.4\textwidth]{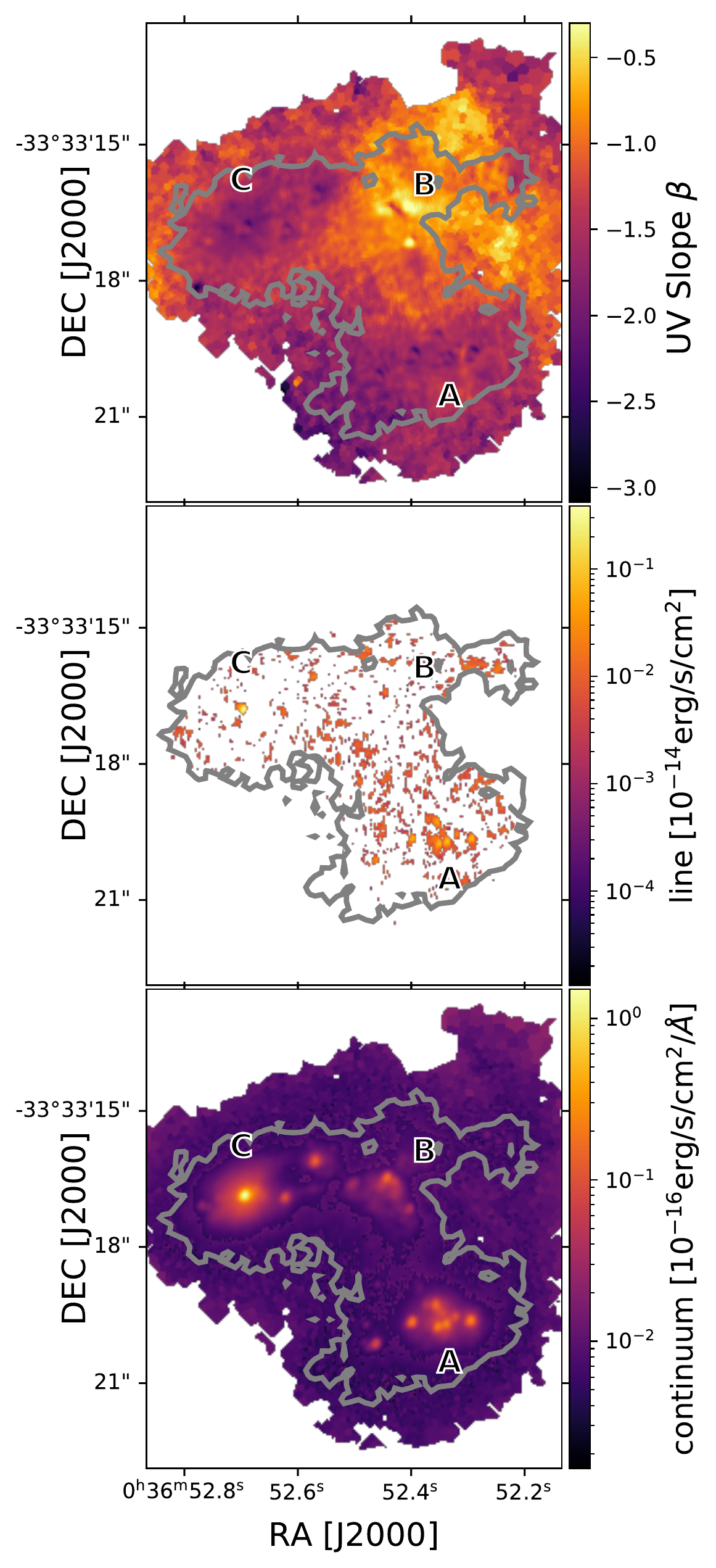}\includegraphics[width=0.4\textwidth]{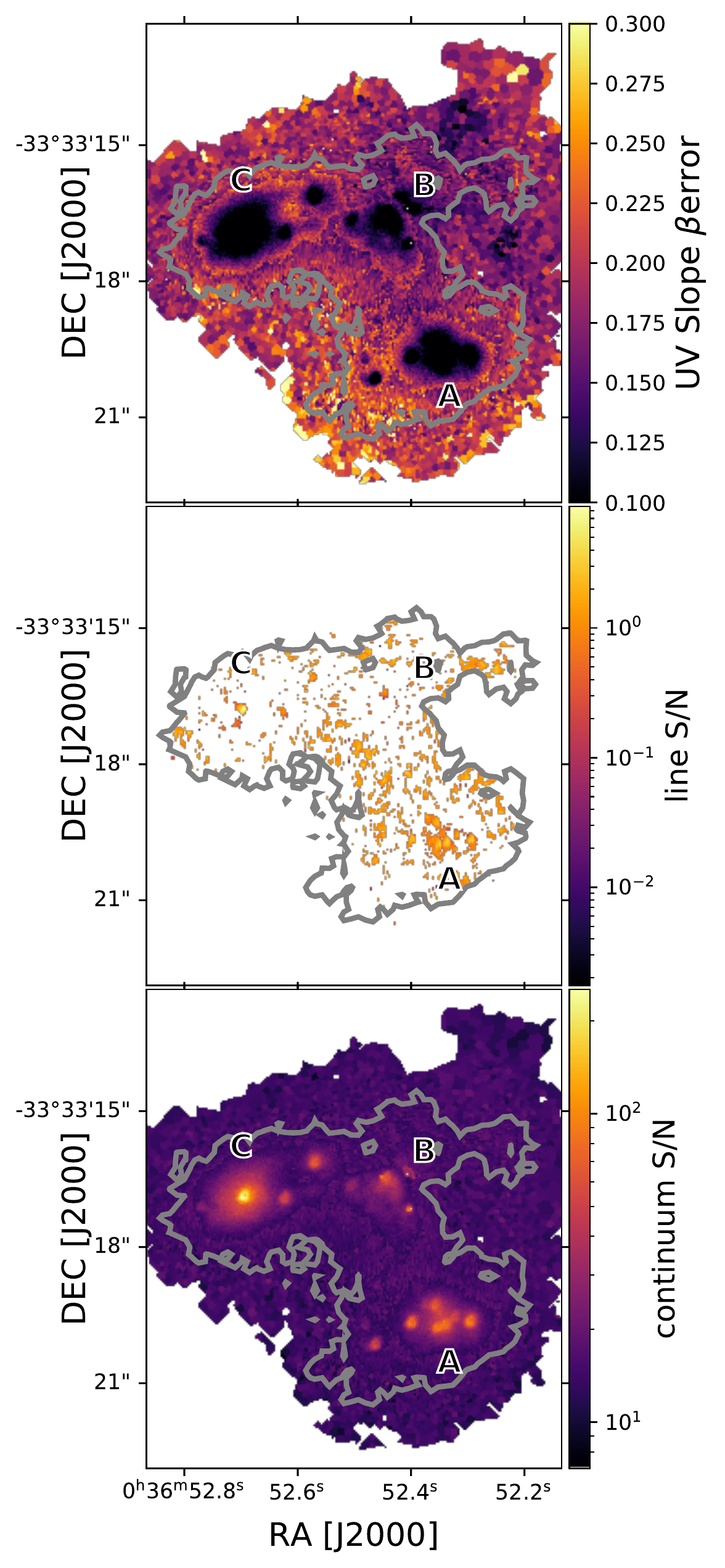}
  \caption{Voronoi bin maps of the UV slope beta (top), \ciii\ line flux (middle), and continuum at the position of the \ciii\ line (bottom), obtained via fitting a powerlaw to F140LP, F25CIII, F25QTZ, and F336W, as described in Section \ref{sec:powlaw}. The right column shows the S/N, estimated as the ratio of the signal and the standard deviation of $100$ MC realizations. Contours of S/N$\ge1$ from the unbinned, not continuum subtracted F25CIII image are overplotted in gray solid lines. Missing bins (in white) are omitted either due to negative flux, {\tt\string lmfit} uncertainty of $\ge100$ \%, or non-detection with a relative MC uncertainty $\ge100$\%. Only $13.9\%$ of the Voronoi bins inside of the gray contours in the middle panel have $S/N>1$, i.e., all of these bins may be spurious detections. \label{fig:result_vbins}  }
\end{figure*}
\begin{figure*}[ht!]
  \includegraphics[width=0.4\textwidth]{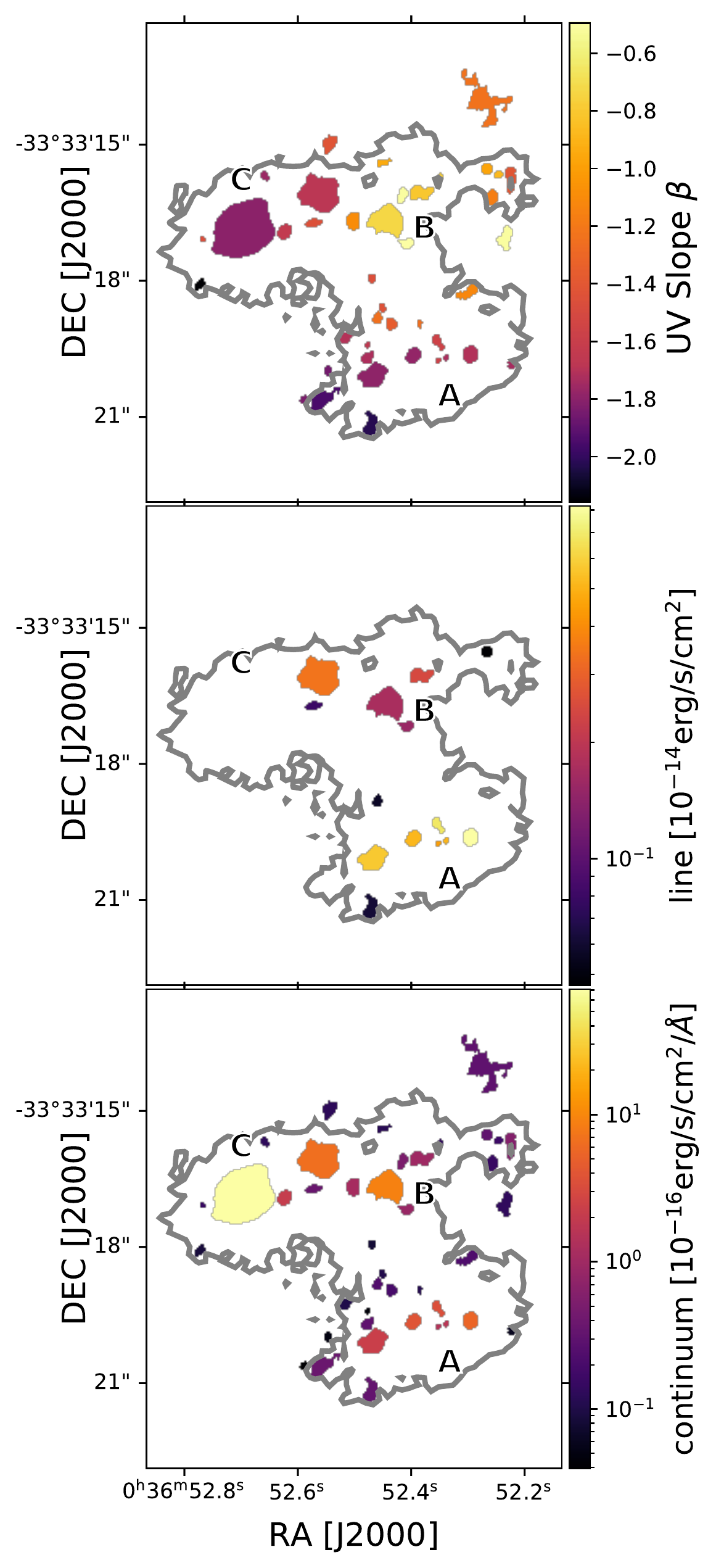}\includegraphics[width=0.4\textwidth]{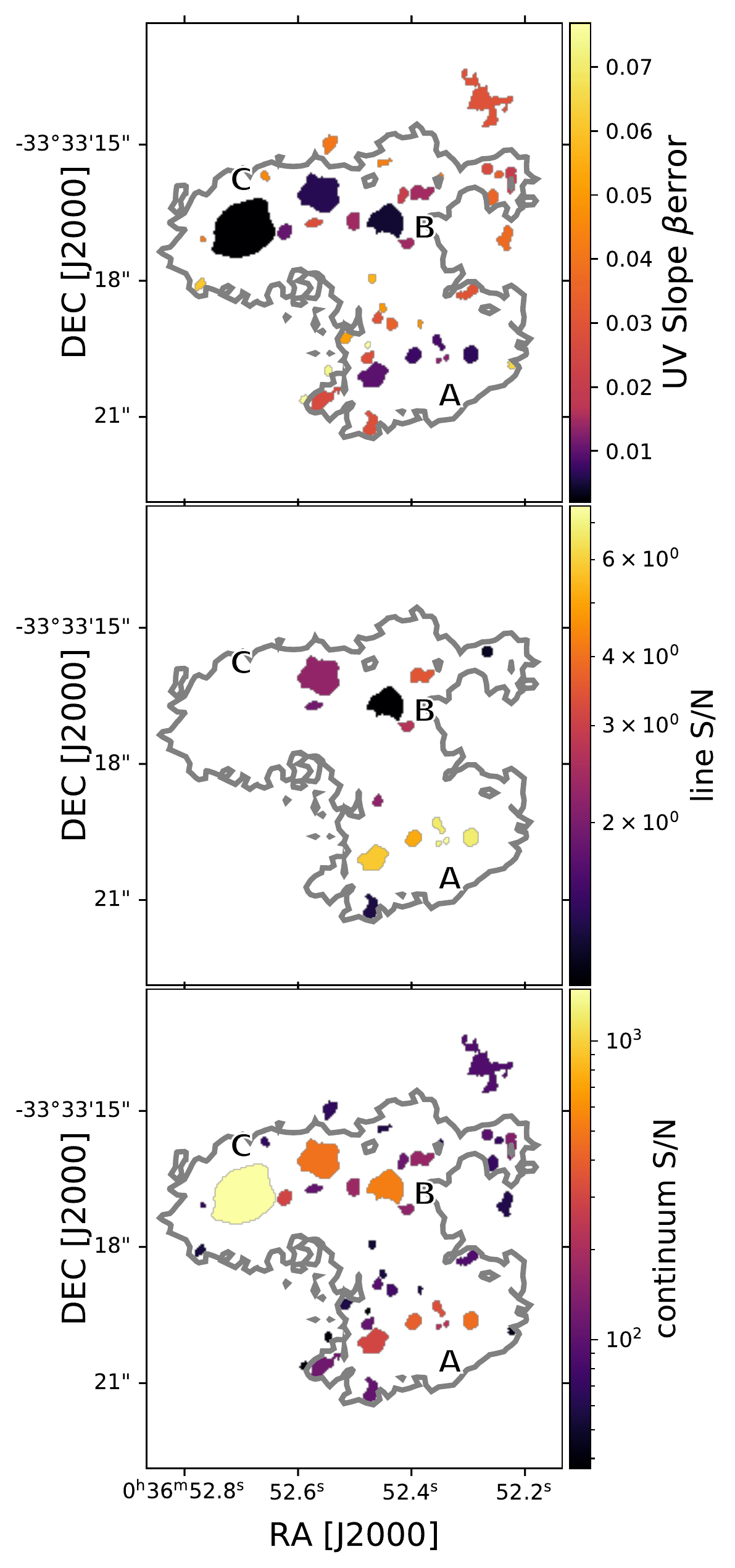}
  \caption{Same as Figure \ref{fig:result_vbins} but for dendrograms, with $N=1000$ MC realizations. \label{fig:result_dendro}  }
\end{figure*}

%\startlongtable
\begin{deluxetable*}{lcllcccc}
\tablecaption{Power law results from aperture photometry with $r=0.125$ arcsec, with the apertures ordered by UV continuum brightness. Dendrograms overlapping with apertures are displayed where available. Upper limits are given for non-detections. The errors are given as the $16^{th}$ and $84^{th}$ percentiles, with the standard deviation in brackets. Apertures 9, 10, and 14 do not overlap with any dendrogram leaf. Apertures 2 and 3 are larger than the overlapping dendrogram leaves.\label{tab:powlaw}}
\tablehead{
\colhead{ID.binning} & \colhead{UV slope $\beta$} & \colhead{\ciii\ line} & \colhead{\ciii\ continuum} &\colhead{EW(\ciii)}&\colhead{comment}\\
\colhead{} & \colhead{} & \colhead{$10^{-17}$erg s$^{-1}$ cm$^{-2}$} & \colhead{$10^{-17}$erg s$^{-1}$ cm$^{-2}$ \AA$^{-1}$} & \colhead{\AA} &\colhead{}
}
%\colnumbers
%\decimals
\startdata
1.Ap      & $-2.08^{+0.01}_{-0.01}(0.01)$ & $1701.08^{+327.90}_{-305.85}(322.30)$ & $345.67^{+0.35}_{-0.38}(0.37)$ & $4.9^{+1.0}_{-0.9}(0.9)$& knot C\\
1.Dendro  &$-1.80^{+0.01}_{-0.01}(0.01)$ & -                   &  $710.68^{+0.47}_{-0.43}(0.48)$ & -          &\\
2.Ap      & $-2.01^{+0.01}_{-0.01}(0.01)$ & $1592.71^{+142.97}_{-146.62}(144.47)$ & $59.75^{+0.15}_{-0.14}(0.15)$ & $26.7^{+2.4}_{-2.5}(2.4)$& knot A\\
2.Dendro  &$-1.57^{+0.01}_{-0.01}(0.01)$ & $444.90^{+64.16}_{-66.75}(63.54)$ & $14.61^{+0.06}_{-0.06}(0.07)$ & $30.5^{+4.5}_{-4.6}(4.4)$ & \\
3.Ap      & $-2.14^{+0.01}_{-0.01}(0.01)$ & $1548.45^{+144.72}_{-138.54}(145.38)$ & $56.11^{+0.16}_{-0.15}(0.15)$ & $27.6^{+2.6}_{-2.5}(2.6)$& knot A\\
3.Dendro  &$-1.73^{+0.01}_{-0.01}(0.01)$ & $508.17^{+63.24}_{-68.98}(67.76)$ & $16.70^{+0.07}_{-0.07}(0.07)$ & $30.4^{+3.9}_{-4.1}(4.1)$ & \\
4.Ap      & $-2.09^{+0.01}_{-0.01}(0.01)$ & $1253.37^{+146.92}_{-141.16}(137.68)$ & $53.95^{+0.14}_{-0.14}(0.14)$ & $23.2^{+2.7}_{-2.6}(2.6)$& knot A\\
4.Dendro  &$-1.71^{+0.01}_{-0.01}(0.01)$ & $819.28^{+115.68}_{-119.08}(121.71)$ & $54.91^{+0.12}_{-0.12}(0.12)$ & $14.9^{+2.1}_{-2.2}(2.2)$ & \\
5.Ap      & $-2.09^{+0.01}_{-0.01}(0.01)$ & $<916.11$          &  $40.81\pm0.12$ & $<22.4$     & knot A\\
5.Dendro  &$-1.78^{+0.01}_{-0.01}(0.01)$ & $516.97^{+101.65}_{-98.30}(100.54)$ & $38.72^{+0.09}_{-0.10}(0.10)$ & $13.3^{+2.6}_{-2.5}(2.6)$ & \\
6.Ap      & $-1.95^{+0.01}_{-0.01}(0.01)$ & $903.64^{+131.26}_{-119.80}(125.86)$ & $37.17^{+0.12}_{-0.13}(0.13)$ & $24.3^{+3.6}_{-3.2}(3.4)$ & knot A\\
6.Dendro  &$-1.57^{+0.01}_{-0.01}(0.01)$ & $679.84^{+96.26}_{-86.01}(101.25)$ & $34.02^{+0.10}_{-0.10}(0.10)$ & $20.0^{+2.8}_{-2.6}(3.0)$ & \\
7.Ap      & $-2.09^{+0.01}_{-0.01}(0.01)$ & $343.04^{+119.79}_{-122.44}(119.23)$ & $32.04^{+0.11}_{-0.12}(0.12)$ & $10.7^{+3.7}_{-3.8}(3.7)$ &\\
7.Dendro  &$-1.69^{+0.01}_{-0.01}(0.01)$ & $340.19^{+145.48}_{-153.63}(151.53)$ & $66.06^{+0.13}_{-0.15}(0.14)$ & $5.1^{+2.2}_{-2.3}(2.3)$ & \\
8.Ap      & $-0.88^{+0.01}_{-0.01}(0.01)$ & $192.73^{+124.58}_{-128.94}(124.55)$ & $36.88^{+0.15}_{-0.15}(0.14)$ & $5.2^{+3.4}_{-3.5}(3.4)$ & knot B\\
8.Dendro  &$-0.73^{+0.00}_{-0.00}(0.00)$ & $<171.53$ & $91.54^{+0.18}_{-0.18}(0.17)$ & $<1.9$ & \\
9.Ap      & $-1.93^{+0.01}_{-0.01}(0.01)$ & $<362.45$         &  $25.85\pm0.11$ & $<14.0$      & knot A\\
10.Ap     & $-1.85^{+0.01}_{-0.01}(0.01)$ & $<411.37$        &  $22.32\pm0.10$ & $<18.4$       & knot A\\
11.Ap     & $-1.99^{+0.01}_{-0.01}(0.01)$ & $184.24^{+96.80}_{-101.03}(97.76)$ & $19.71^{+0.09}_{-0.10}(0.09)$ & $9.3^{+4.9}_{-5.2}(5.0)$ & \\
11.Dendro &$-1.63^{+0.01}_{-0.01}(0.01)$ & $<45.19$ & $20.82^{+0.08}_{-0.07}(0.07)$ & $<2.2$   & \\
12.Ap     & $-2.12^{+0.02}_{-0.02}(0.02)$ & $374.47^{+85.05}_{-76.81}(80.49)$ & $14.37^{+0.07}_{-0.07}(0.07)$ & $26.1^{+6.0}_{-5.4}(5.7)$ & knot A\\
12.Dendro &$-1.79^{+0.01}_{-0.01}(0.01)$ & $572.08^{+102.84}_{-94.98}(98.59)$ & $22.86^{+0.08}_{-0.08}(0.08)$ & $25.0^{+4.5}_{-4.2}(4.4)$ & \\
13.Ap     & $-1.17^{+0.01}_{-0.01}(0.01)$ & -                &  $17.08\pm0.09$ & -             & knot B\\
13.Dendro &$-0.73^{+0.00}_{-0.00}(0.00)$ & $171.53^{+172.83}_{-176.67}(169.08)$ & $91.54^{+0.18}_{-0.18}(0.17)$ & $1.9^{+1.9}_{-1.9}(1.8)$ & \\
14.Ap     & $-1.76^{+0.02}_{-0.02}(0.02)$ & $302.08^{+87.29}_{-83.88}(84.05)$ & $14.16^{+0.08}_{-0.07}(0.08)$ & $21.3^{+6.2}_{-6.0}(6.0)$    & knot A\\
15.Ap     & $-1.47^{+0.02}_{-0.02}(0.02)$ & $<39.90$         &  $9.33\pm0.07$  & $<4.3$        &\\
15.Dendro &$-1.10^{+0.01}_{-0.01}(0.01)$ & $<63.21$ & $11.24^{+0.06}_{-0.06}(0.06)$ & $<5.6$  &\\
16.Ap     & $-0.77^{+0.02}_{-0.02}(0.02)$ & $<228.417$       &  $10.47\pm0.08$ & $<21.8$       &\\
16.Dendro &$-0.49^{+0.01}_{-0.01}(0.01)$ & $153.45^{+59.30}_{-57.43}(57.80)$ & $8.79^{+0.06}_{-0.06}(0.06)$ & $17.5^{+6.8}_{-6.6}(6.6)$ &\\
17.Ap     & $-1.89^{+0.03}_{-0.02}(0.03)$ & $<76.62$         &  $5.90\pm0.05$  & $<13.0$       &\\
17.Dendro &$-1.47^{+0.04}_{-0.04}(0.04)$ &- & $1.51^{+0.02}_{-0.02}(0.02)$ & -  &\\
18.Ap     & $-2.06^{+0.03}_{-0.03}(0.03)$ & $<37.80$         &  $3.62\pm0.04$  & $<10.4$       &knot A\\
18.Dendro &$-1.72^{+0.03}_{-0.03}(0.03)$ & $<17.00$ & $3.06^{+0.03}_{-0.03}(0.03)$ & $<5.6$        &\\
\enddata
%\tablenotetext{$\dagger$ Dendrogram is smaller than aperture}
\end{deluxetable*}

\section{Estimating line flux}\protect\label{sec:powlaw}

The F25CIII filter contains \ciii\ line and continuum emission. The latter must be estimated and subtracted from the F25CIII flux. A first try of removing the continuum with the \citet{Keenan2017} method, which derives a spatially constant continuum scaling factor by identifying a rapid change in the mode of the continuum-subtracted image flux distribution, showed that the \ciii\ levels are low and that the result is highly sensitive to the assumed continuum level. We therefore determine the continuum locally in each resolution element, thereby accounting for any color gradients due to the presence of stellar populations of varying ages, metallicity and reddening. 

In each resolution element (pixel, aperture, Voronoi bin, or dendrogram leaf), we use the python package {\tt\string lmfit} to fit a powerlaw model to the observed flux in the F140LP, F25CIII, F25QTZ, and F336W filters by minimizing the residuals between model and observations. This is illustrated for all apertures with \ciii\ detection in Section \ref{sec:powlaw2} in the Appendix. The powerlaw has the form
\begin{equation}
  f_\lambda = f_{cont}\left(\frac{\lambda}{\lambda_{obs}}\right)^\beta + \frac{F_{line}}{\sqrt{2\pi}\sigma_g} e^{\frac{-(\lambda-\lambda_{obs})^2}{2\sigma^2_g}}
\end{equation}
where $f_{cont}$ is the continuum flux density at the position of the \ciii\ line, $F_{line}$ is the flux in the line, $\lambda_{obs}=1907.7085$\AA\ $\times(1 + z)$ is the redshifted wavelength of \ciii, $\beta$ is the UV slope, and $\sigma_g$ is the standard deviation of the Gaussian, assumed to approximate the emission line. This creates a synthetic spectrum, which is then convolved with the filter transmission of the four filters using {\tt\string pysynphot}, to produce the model flux density in each filter. The residual of these values and the observations is minimized with {\tt\string lmfit}. The obtained parameters are $\beta$, $F_{line}$ and $f_{cont}$. The equivalent width is then obtained as EW(\ciii)$=F_{line}/f_{cont}$.
   
\begin{figure*}[ht!]
  \gridline{\fig{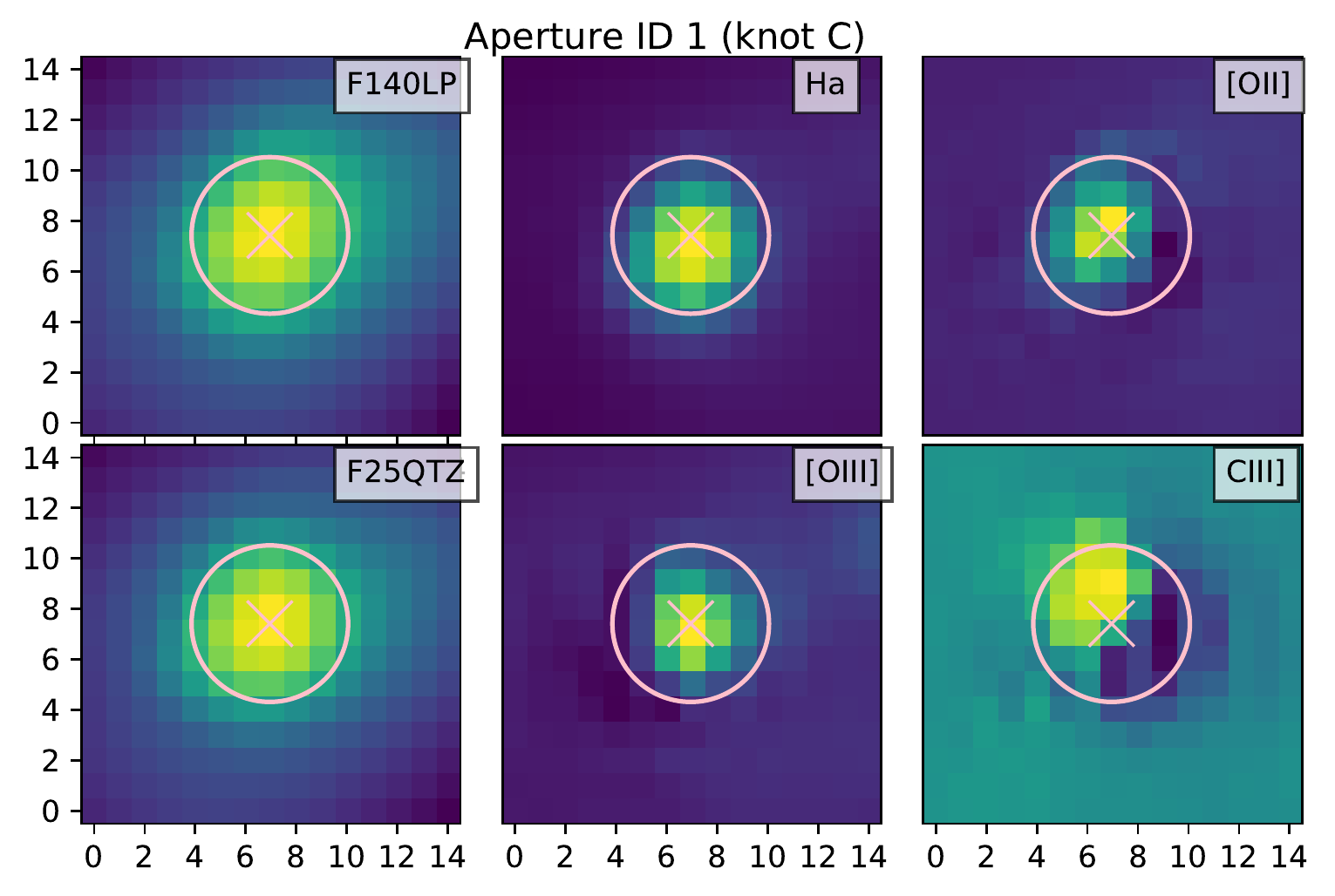}{0.4\textwidth}{}
    \fig{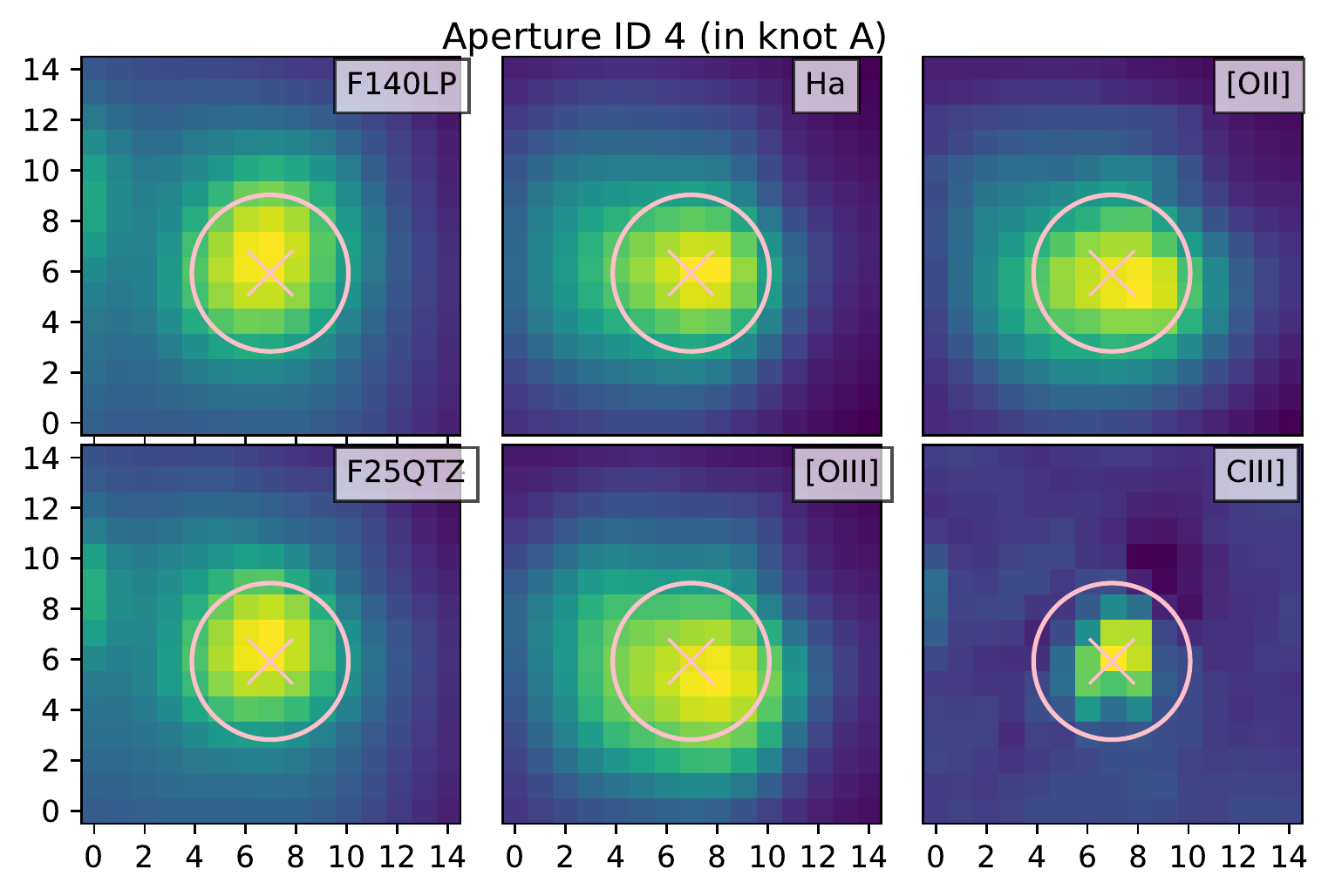}{0.4\textwidth}{}
  }
  \gridline{
    \fig{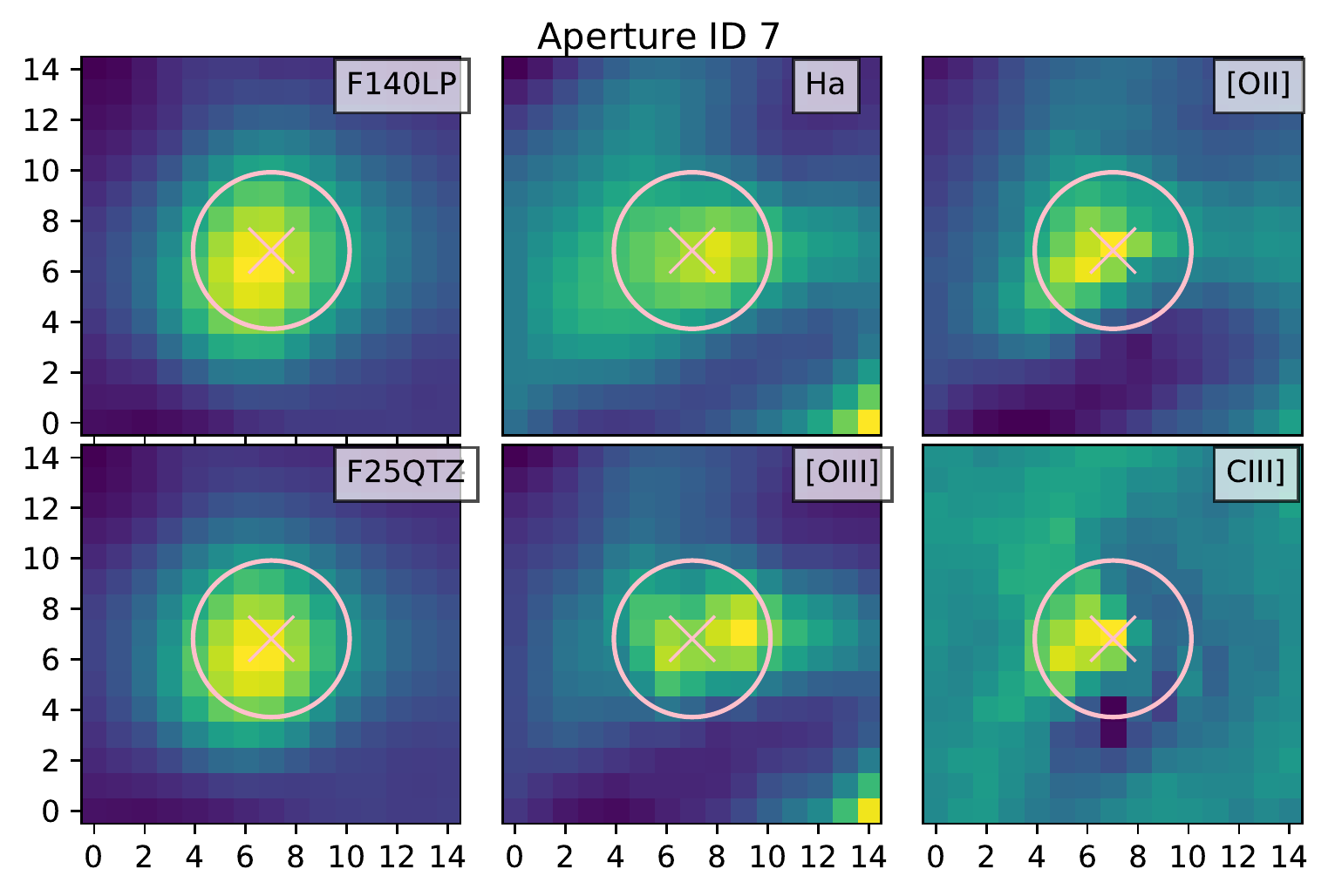}{0.4\textwidth}{}
    \fig{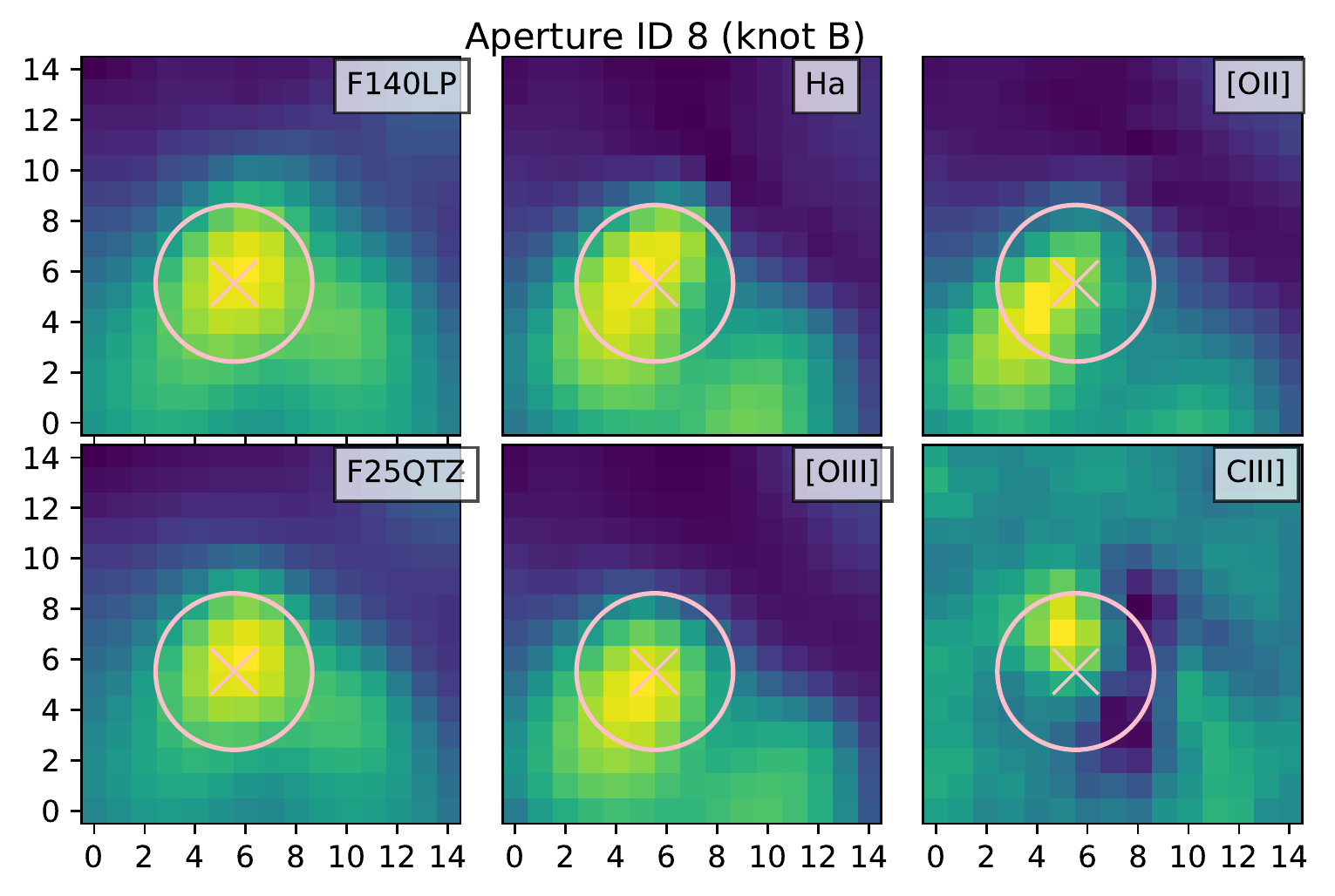}{0.4\textwidth}{}
  }
  \gridline{\fig{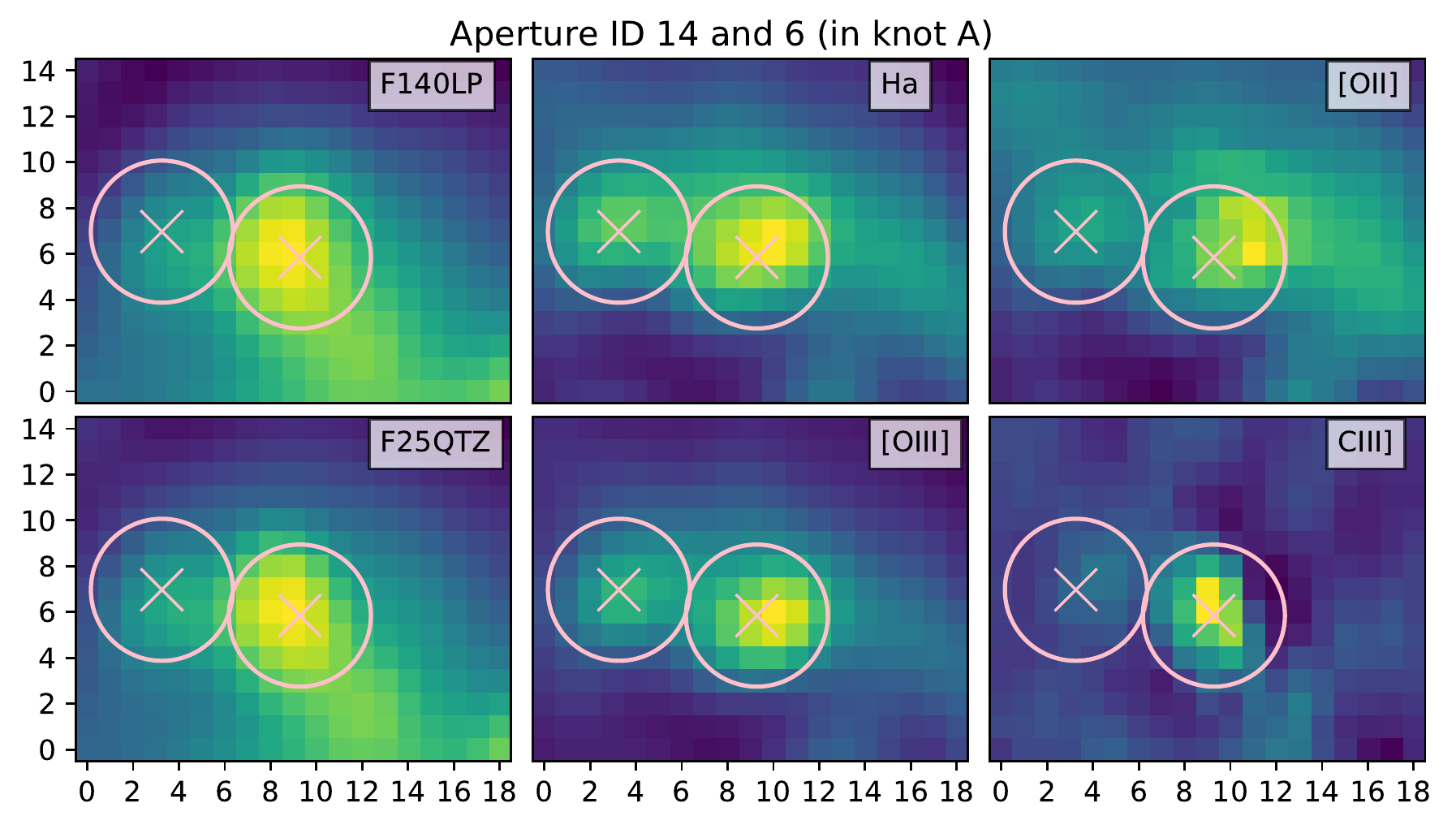}{0.4\textwidth}{}
        \fig{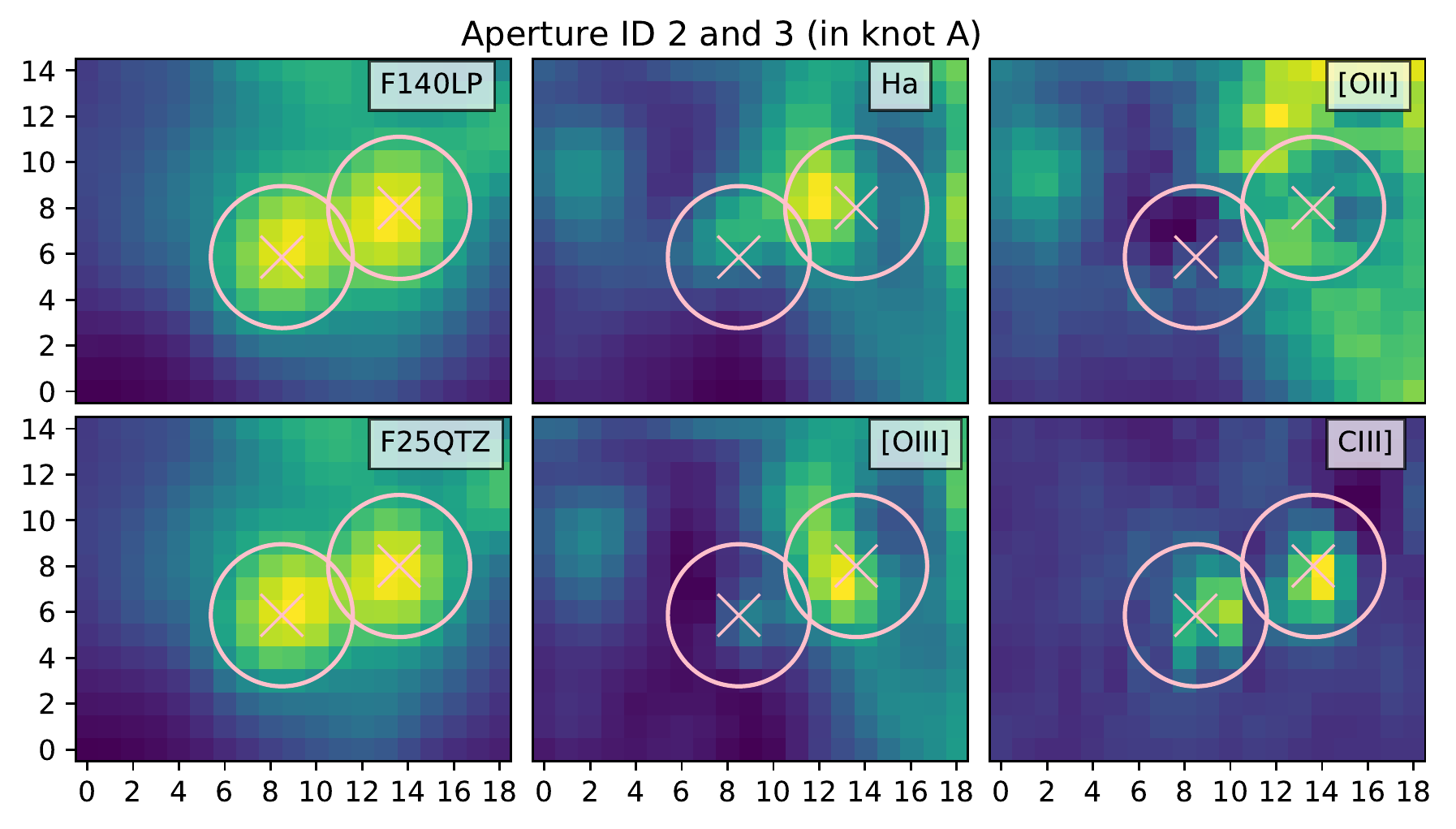}{0.4\textwidth}{}
  }
  \gridline{\fig{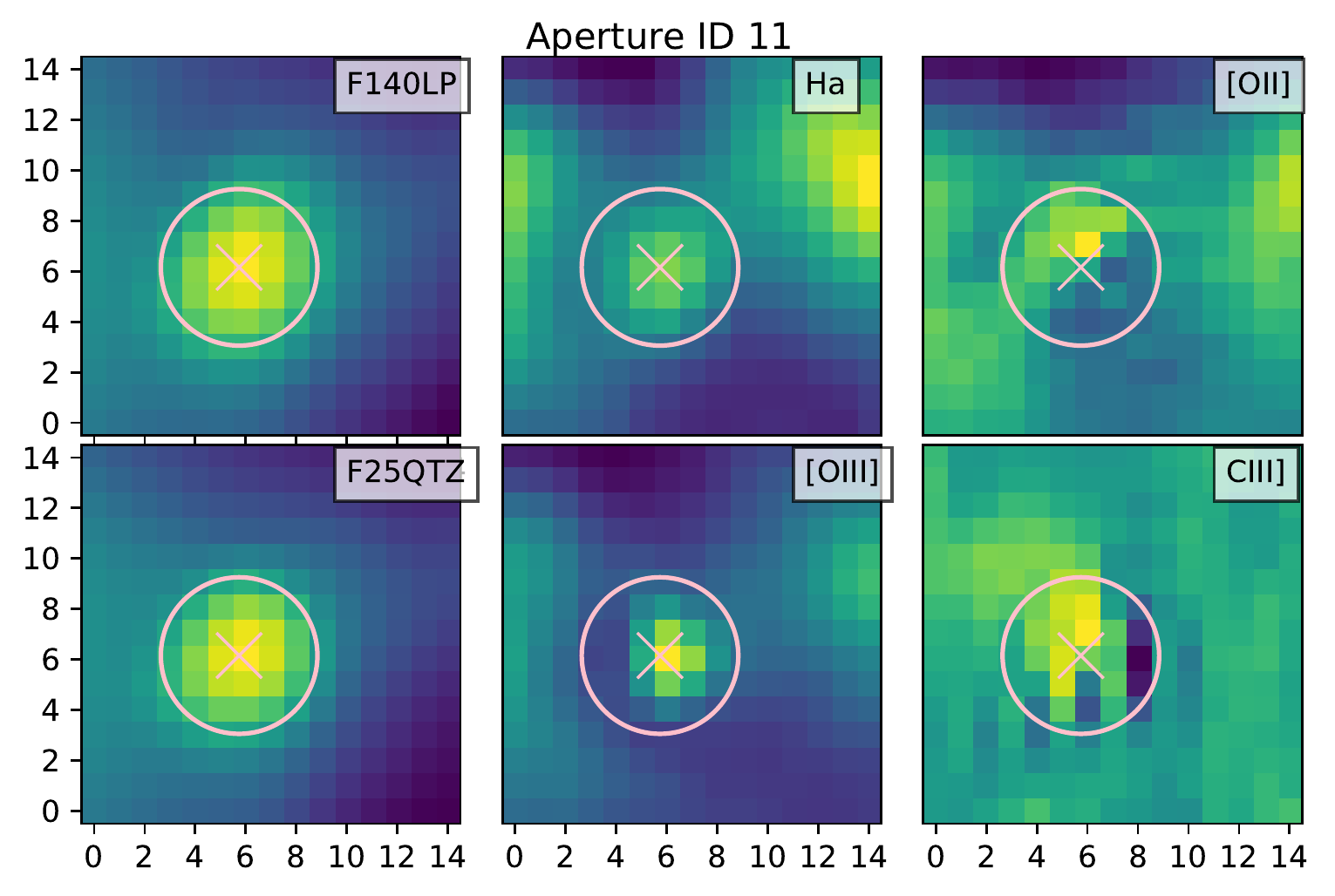}{0.4\textwidth}{}
    \fig{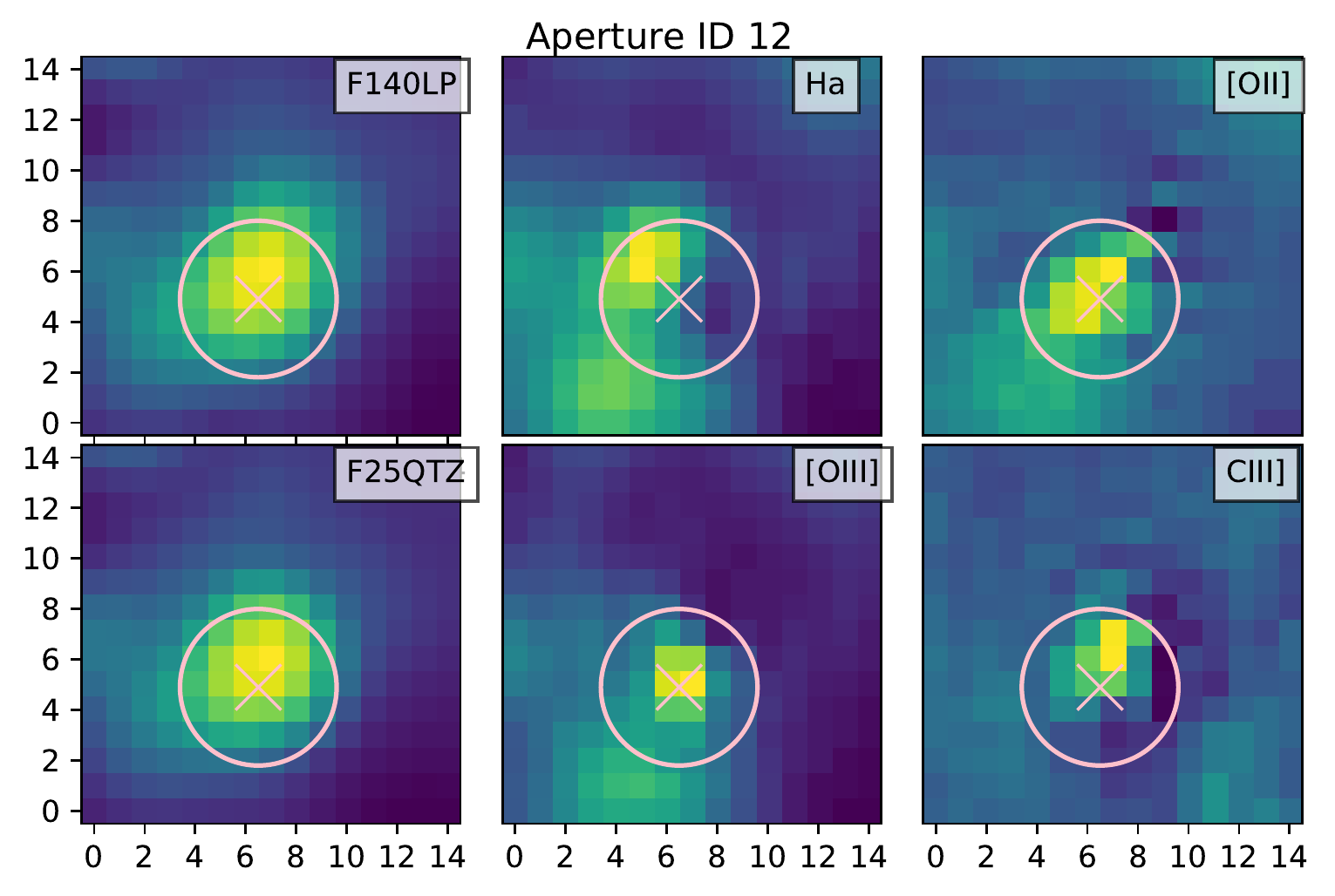}{0.4\textwidth}{}
  }
\caption{Cutouts around apertures (detections only), showing the morphology of the gas in continuum-subtracted \ha, \oiii\, \oii, and \ciii. Aperture radius is $r=0.125$ arcsec. The cross marks the F140LP image centroid position, overplotted in all other images.\label{fig:cutouts}  }
\end{figure*}

\begin{figure*}[ht!]
  \includegraphics[width=0.47\textwidth]{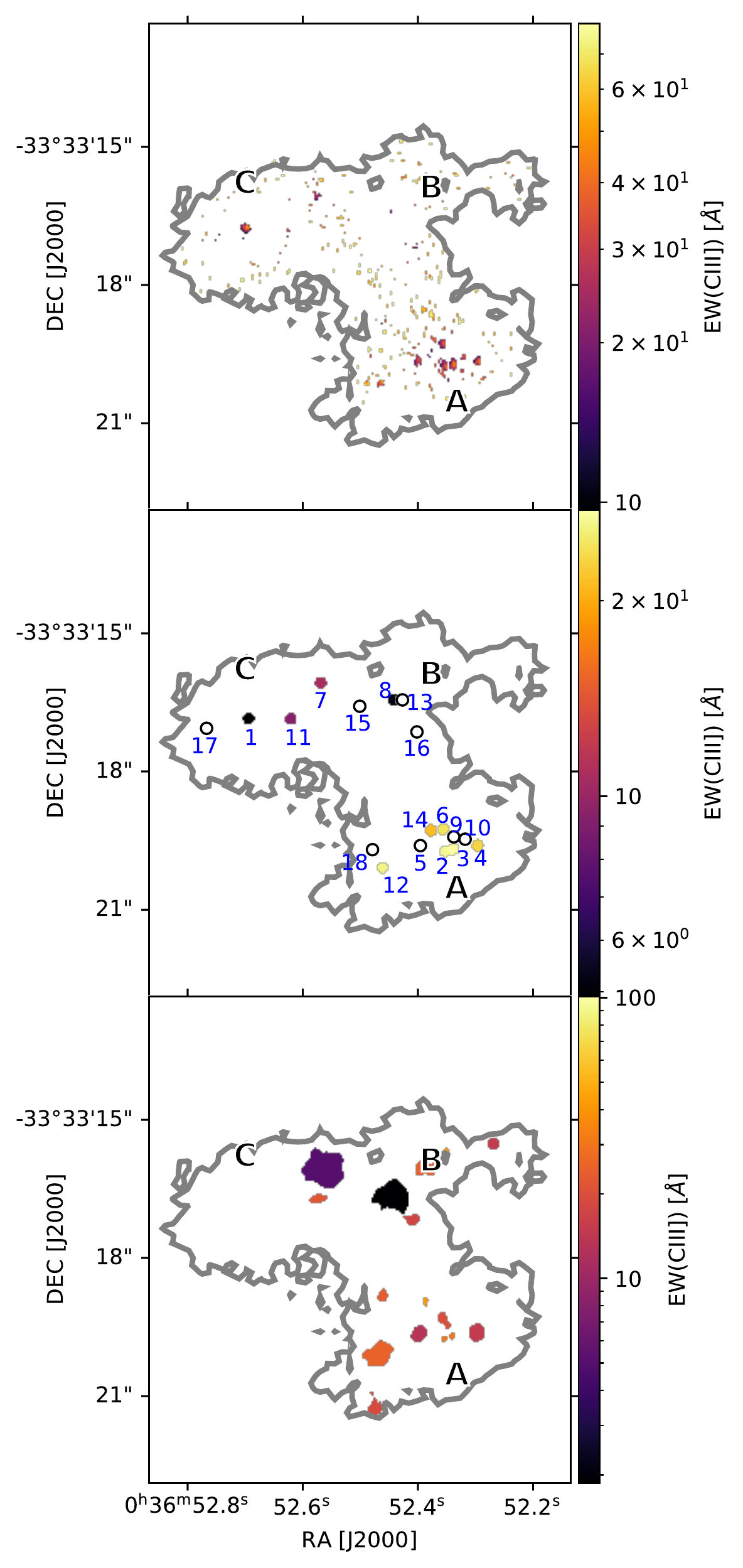}\includegraphics[width=0.47\textwidth]{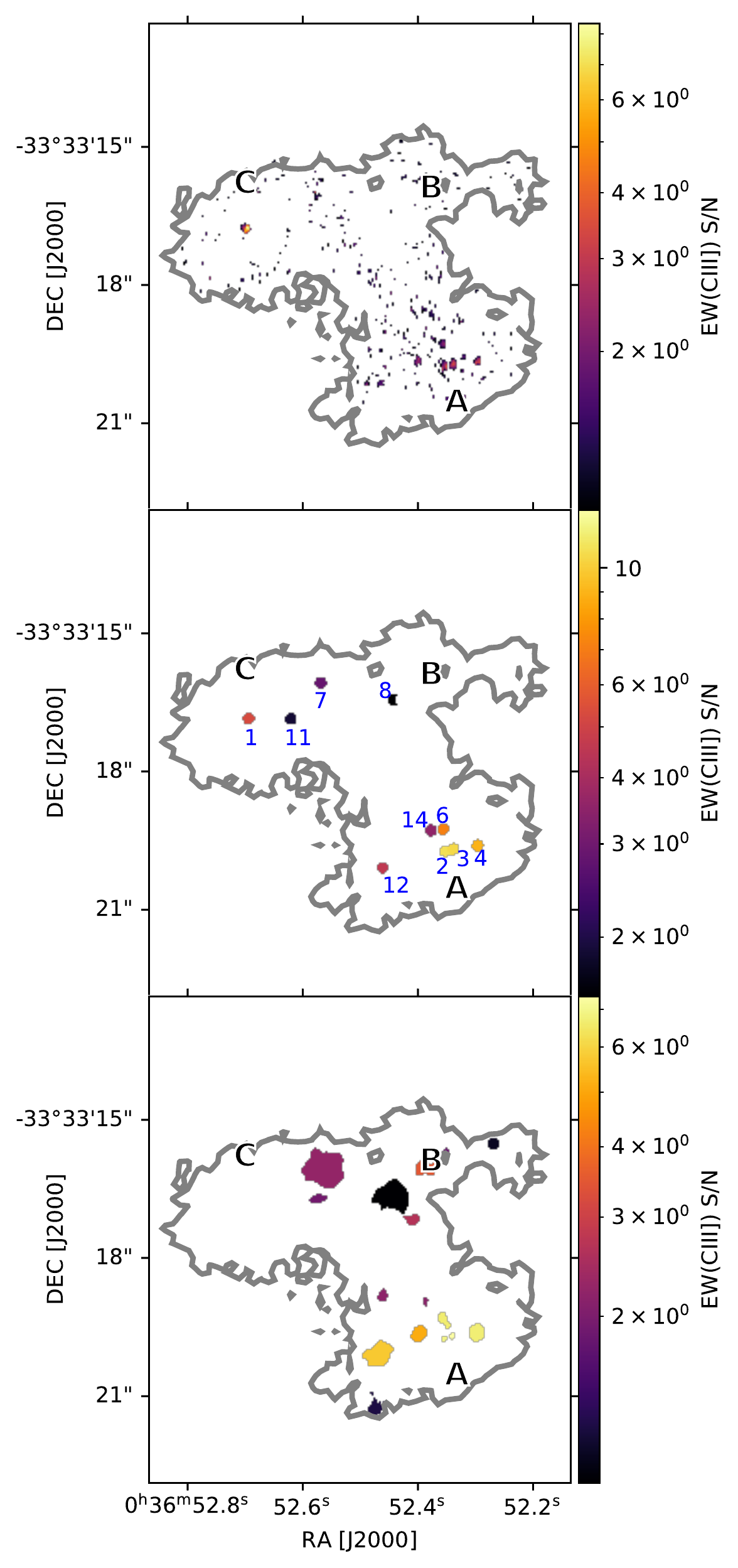}
  \caption{Maps of EW(\ciii) (left column). Top: Voronoi bin map; Middle: aperture map; Bottom: dendrogram map. The right column shows the S/N. The apertures in the middle panel are numbered according to Table \ref{tab:powlaw}, in order of decreasing UV continuum flux in F140LP. Non-detections are shown as empty circles.\label{fig:ew}  }
\end{figure*}

To estimate the uncertainties on $\beta$, $F_{line}$, $f_{cont}$, and EW(\ciii), we perform Monte Carlo (MC) simulations for each resolution element, perturbing the observed fluxes in all four filters by randomly drawing from a normal distribution with a standard deviation given by the corresponding error image (Section \ref{sec:data}). For pixels and Voronoi bins, the number of MC realizations is $N=100$, for apertures and dendrogram leaves $N=1000$. At each realization, we obtain $\beta$, $F_{line}$, $f_{cont}$, and EW(\ciii), and the final uncertainty estimate on these quantities is taken as the standard deviation of the $N$ values. In Section \ref{sec:hist} of the Appendix, we show the MC distribution of these parameters for four randomly selected apertures, demonstrating that the standard deviation is not an unreasonable measure of the uncertainty. 

\begin{figure*}[ht!]
  \includegraphics[width=0.5\textwidth]{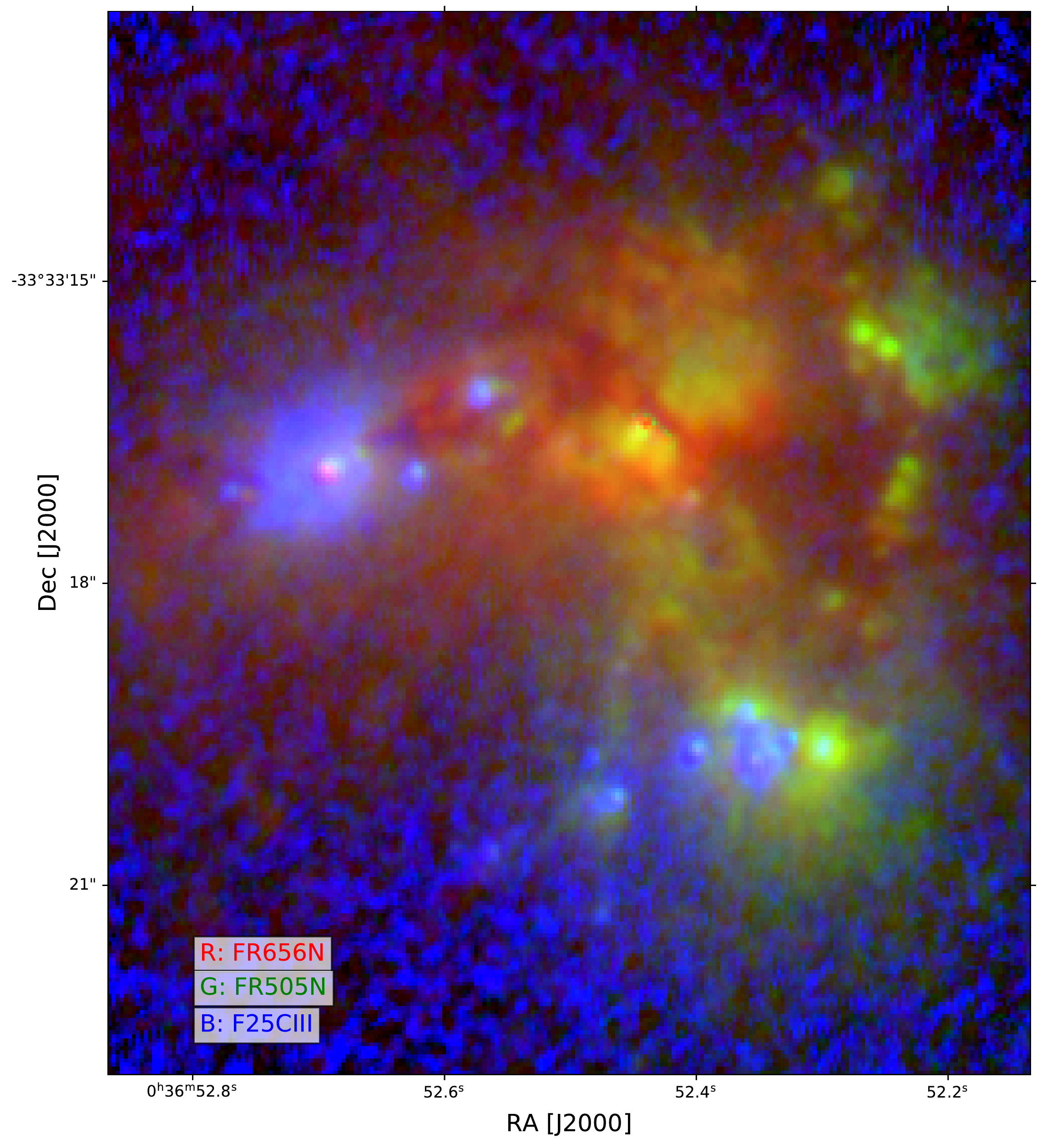}
  \includegraphics[width=0.5\textwidth]{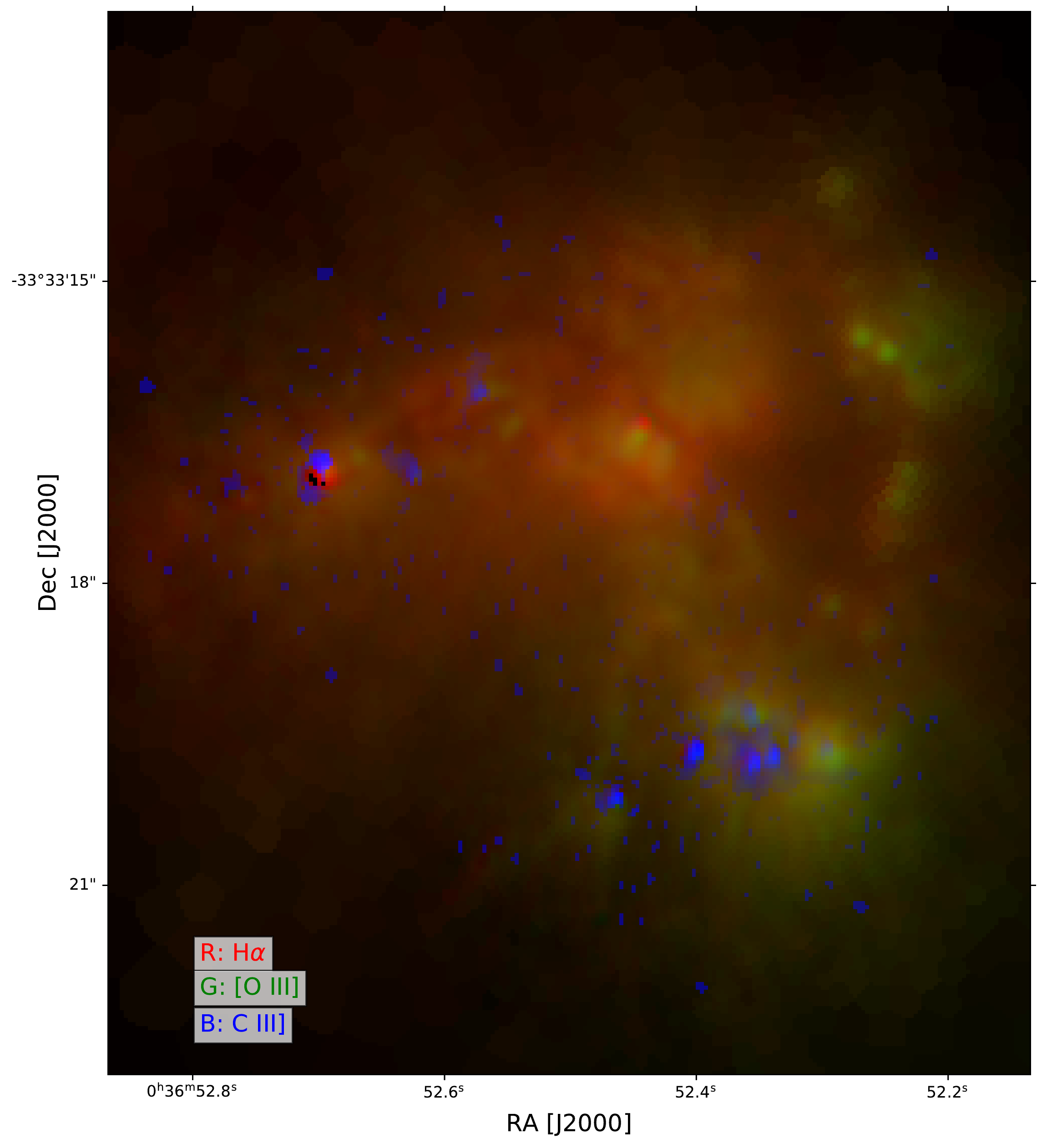}\\
  \includegraphics[width=0.5\textwidth]{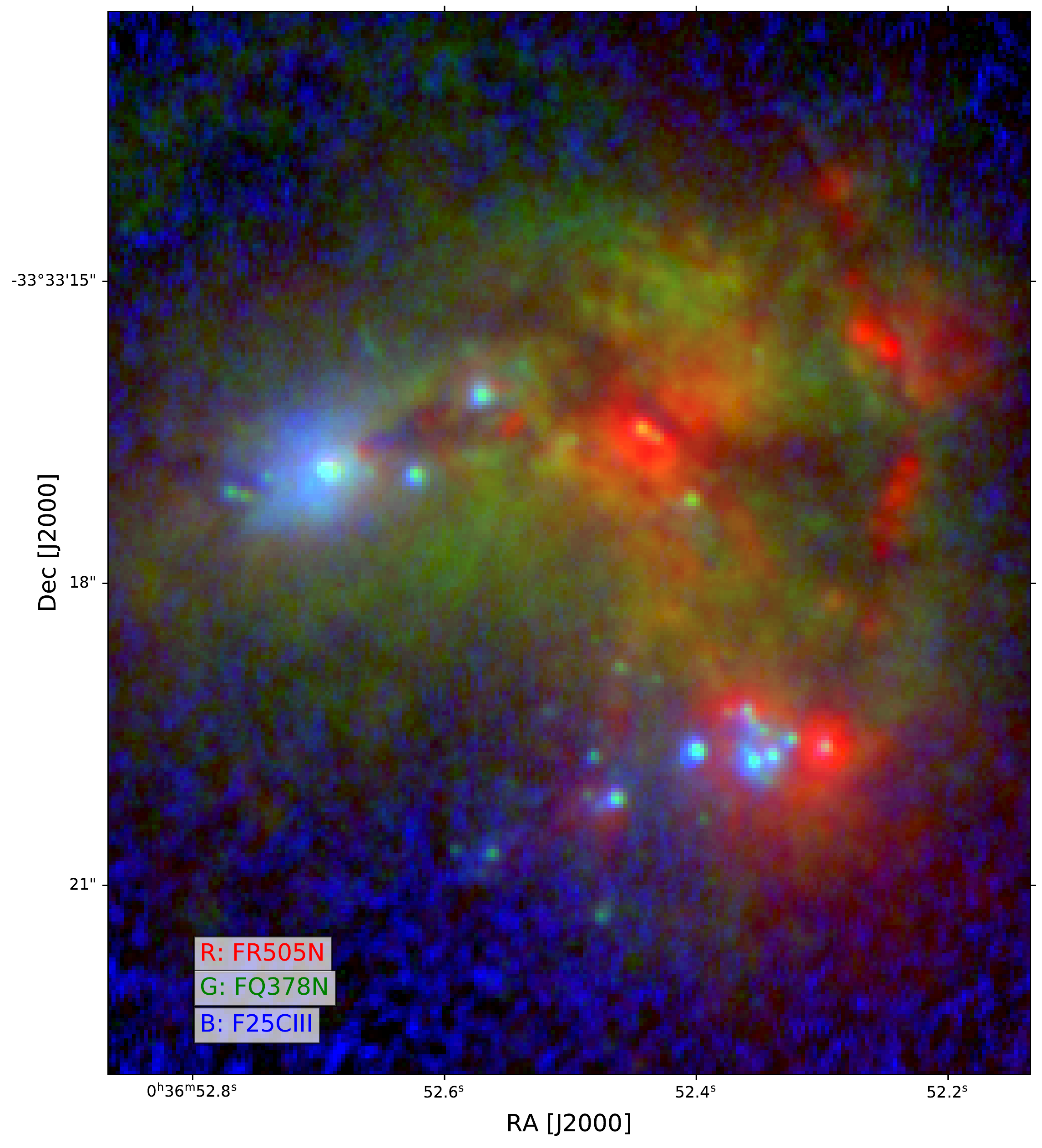}
  \includegraphics[width=0.5\textwidth]{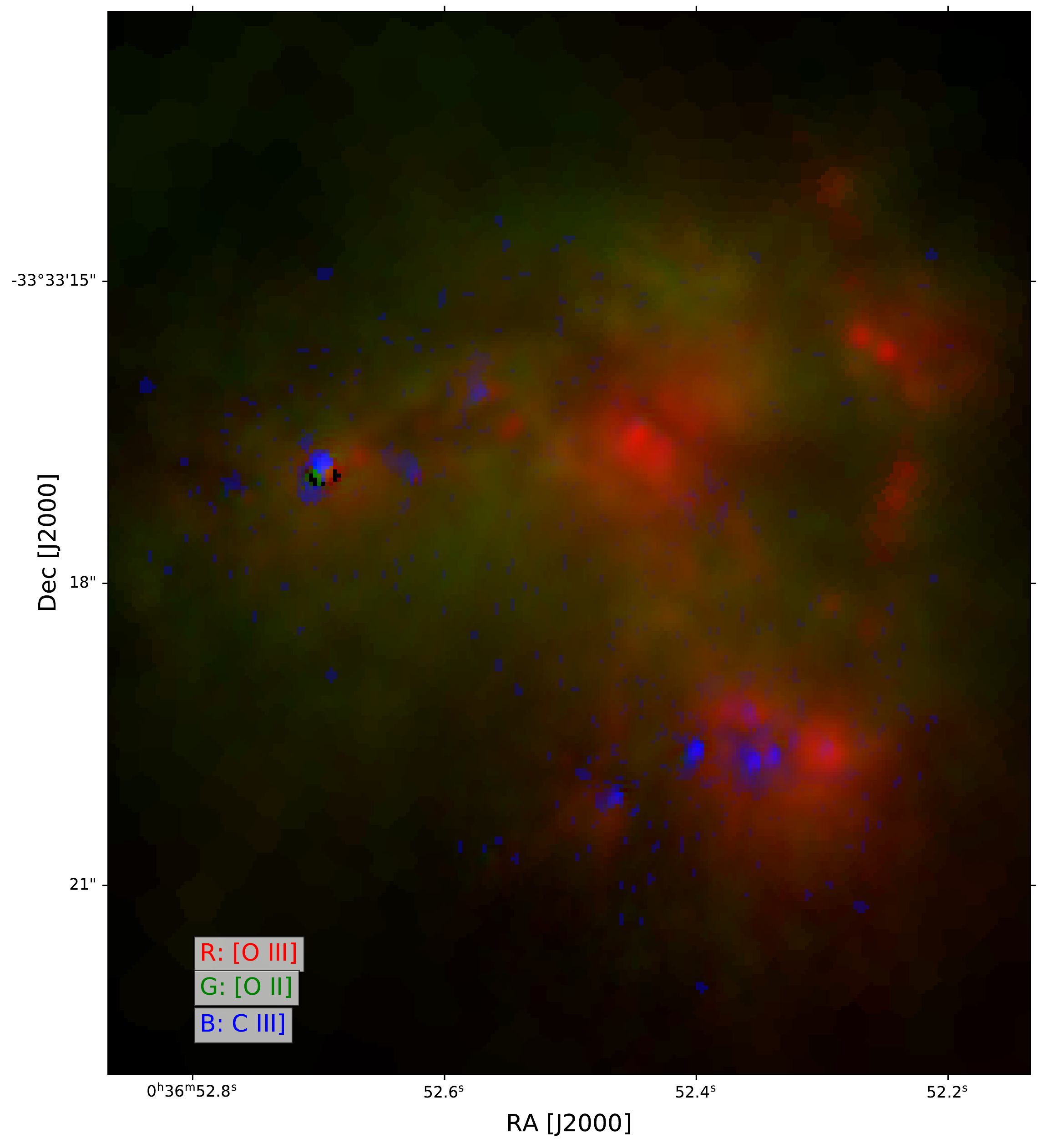}\\
  \caption{RGB composites, with individual images logarithmically scaled. The images in the left column have not been continuum subtracted. The images in the right column are pure line emission. In each image the three channels have been scaled to enhance faint features, i.e., the fluxes are not to scale. \label{fig:rgb}}
\end{figure*}

The powerlaw method is robust because it uses three broadband filters to anchor the slope of the spectrum. This minimizes any potential contamination of the \ciii\ continuum estimate by the presence of \ciii\ and possibly Mg II $\lambda\lambda2796,2803$ in the F25QTZ filter. Figures \ref{fig:result_vbins} and \ref{fig:result_dendro} show the resulting spatial maps of $\beta$, $F_{line}$, $f_{cont}$ for Voronoi bins and dendrogram leaves. Extraction from unbinned pixels gives results similar to those of the Voronoi binning, only more noisy, and is therefore not explicitly presented here. As an indicator of the reliability of these measurements, the right column of the figures shows error or S/N maps, obtained as the ratio of the parameters and their estimated MC uncertainties. The S/N in the \ciii\ line flux is in the range $S/N\in[1.0,8.7]$, with an average of $\left<S/N\right>=1.5$ for Voronoi bins, and $S/N\in[1.0,7.5]$, with an average of $\left<S/N\right>=3.0$ for dendrogram leaves. We consider the dendrogram leaves more reliable than the Voronoi bins, because only $13.9\%$ of the Voronoi bins inside the gray contour in the middle panel of Figure \ref{fig:result_vbins} have $S/N>1$, i.e, are at least a $1\sigma$ detection. This percentage is achievable by simply drawing numbers at random from a normally distributed noise image. Therefore the middle panel of Figure \ref{fig:result_vbins} should not be taken at face value to represent a spatial distribution of \ciii\ flux, and only the dendrogram and the aperture signal extractions can be considered real detections. 

The UV slope $\beta$ traces the extinction on the stellar continuum and the star formation history. The figures imply that knot B has less negative $\beta$ values, which is consistent with the higher nebular reddening obtained for knot B via mapping of the Balmer decrement $H\alpha/H\beta$ in MUSE data (Menacho et al., in prep.).

For signal extraction on apertures, in addition to applying a relative error limit of $<100\%$, i.e., at least a $1\sigma$ detection, we also visually compared the morphology of the aperture regions in \ciii\ and discarded apertures as non-detections if the morphology was not similar to what is observed in \ha\ and \oiii. The assumption that \ciii\ should be co-spatial with \oiii\ is motivated at length in the discussion (Section \ref{sec:theory}). Figure \ref{fig:cutouts} shows cutouts of the apertures, which are likely to be real detections of \ciii. As the figure shows there is a bright ``clump'' of emission inside the apertures in \ciii, co-spatial with \ha, and \oiii. Table \ref{tab:powlaw} lists the powerlaw fit results for the apertures, sorted by decreasing UV brightness, and for the overlapping dendrograms. We note that the equivalent width of \ciii\ from the apertures does not correlate with increasing UV continuum brightness.

The continuum subtraction for the H$\alpha$ (FR656N), \oiii$\lambda5007$ (FR505N), and \oii$\lambda3727$ (FQ378N) maps is done with the Lyman $\alpha$ eXtraction Software \citep[LaXs,][]{Hayes2009,Ostlin2014} which performs pixel SED fitting to find the underlying continuum for each line. The SED fitting uses the Starburst99 stellar population library with two stellar populations and four free parameters (stellar masses for the two populations, E(B-V) and stellar age for the young population). LaXs was run to extract the H$\alpha$, \oiii, and \oii\ lines on Voronoi binned images and on apertures.

\section{\ciii\ Equivalent width}\protect\label{sec:ew}

Figure \ref{fig:ew} shows maps of the equivalent width of \ciii, EW(\ciii). Only bins with a $<100\%$ fractional error \ew\ are plotted, i.e. each bin has at least a $1\sigma$ detection. Since the observations are quite shallow, the figure suggests that we detect predominantly emission from point sources, with little to no contribution from diffuse gas. We reiterate that the Voronoi binned map (top row) likely suffers from spurious detections (see Section \ref{sec:powlaw}), a suspicion which is further enhanced by the numerous measured \ew$\gtrsim50$\AA\ values. These are unphysically high and with very low S/N (right panel). The dendrogram and aperture maps on the other hand likely represent true detections because PSF effects have been mitigated by the larger spatial bins. The dendrogram map (bottom row) does show some detections of regions seemingly not overlapping with any of the star clusters in the middle panel, and may therefore be due to diffuse \ciii\ emission. The \ciii\ point sources spatially coincide with the location of the star clusters, predominantly in the resolved knot A, and the unresolved knots C and B. In Figure \ref{fig:rgb} we show RGB false-color, PSF-equalized images of Haro 11. The left column shows non-continuum subtracted images, the right shows the pure line emission. The figure indicates that the detected \ciii\ flux appears localized to the clusters and their immediate vicinity but extended, low surface brightness \ciii\ likely exists below the detection threshold.

Before proceeding, one should consider any possible sources of uncertainty or bias, which may affect the \ciii\ maps. It is not likely that we have overestimated the continuum in the F25CIII filter, because we tailor the continuum to each resolution element. If the S/N is high enough, the powerlaw fitting method should produce reliable continua \citep{Hayes2009}. Further, we consistently get signal at the position of the star clusters with four different extraction schemes (no binning, Voronoi binning, dendrogram leaves, aperture photometry). For the bins where the \ciii\ line emission was too low to be measured, we compared the $f_{cont}$ obtained by fitting a powerlaw to all four filters as described in Section \ref{sec:powlaw}, to that obtained from using just the broadband filters F140LP, F25QTZ, and F336W, and without modeling the emission line. The continuum in both cases was very simular, suggesting that the F25CIII filter indeed does not contain much line flux in these bins and therefore does not significantly influence the resulting powerlaw fit. For the apertures we have also compared the measured \ciii\ flux to the median of the corresponding MC distribution, obtaining very similar values. 

Nevertheless, we stress that the F25CIII filter has very low transmission ($\lesssim0.02\%$) and the resulting image is quite noisy despite its exposure time of $10000$ seconds. The $3\sigma$ limiting flux in the \ciii\ line is as bright as $4.8\times10^{-17}$ erg s$^{-1}$ cm$^{-2}$, obtained from $r=0.125$ arcsec apertures on the unbinned \ciii\ line flux image. In addition, the bandwidth of the F25CIII narrowband filter is $W_{F25CIII}=165.7$\AA. With this filter, an equivalent width of $2$\AA, $10$\AA, or $20$\AA, will mean that the line flux density is $1\%$, $6\%$, and $12\%$ of the continuum flux density, respectively. The S/N in the continuum, required for a detection, is then at least $100$, $16.7$, and $13.8$, respectively. Figure \ref{fig:result_vbins} shows that S/N$\gtrsim100$ in the continuum is achieved only at the location of the bright star clusters. Similarly, to detect \ciii\ emission lines with an equivalent width of $2$\AA, $10$\AA, or $20$\AA, the calibration constant PHOTFLAM would have to be known to better than $0.9\%$, $4\%$, and $8\%$, respectively. Therefore our observations cannot detect faint line emission and it is very likely that diffuse \ciii\ is simply hidden in the noise of the F25CIII image.

\begin{figure*}[ht!]
  \gridline{
    \fig{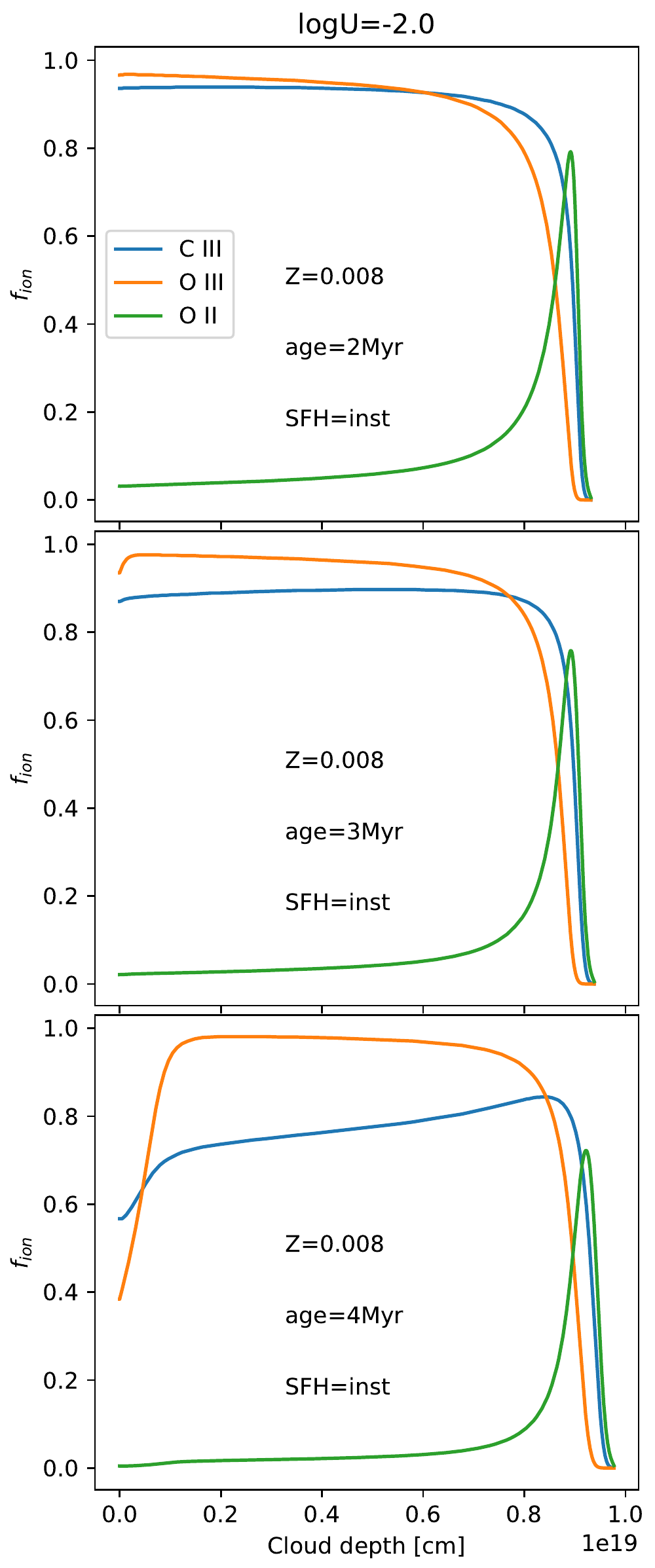}{0.3\textwidth}{a}
    \fig{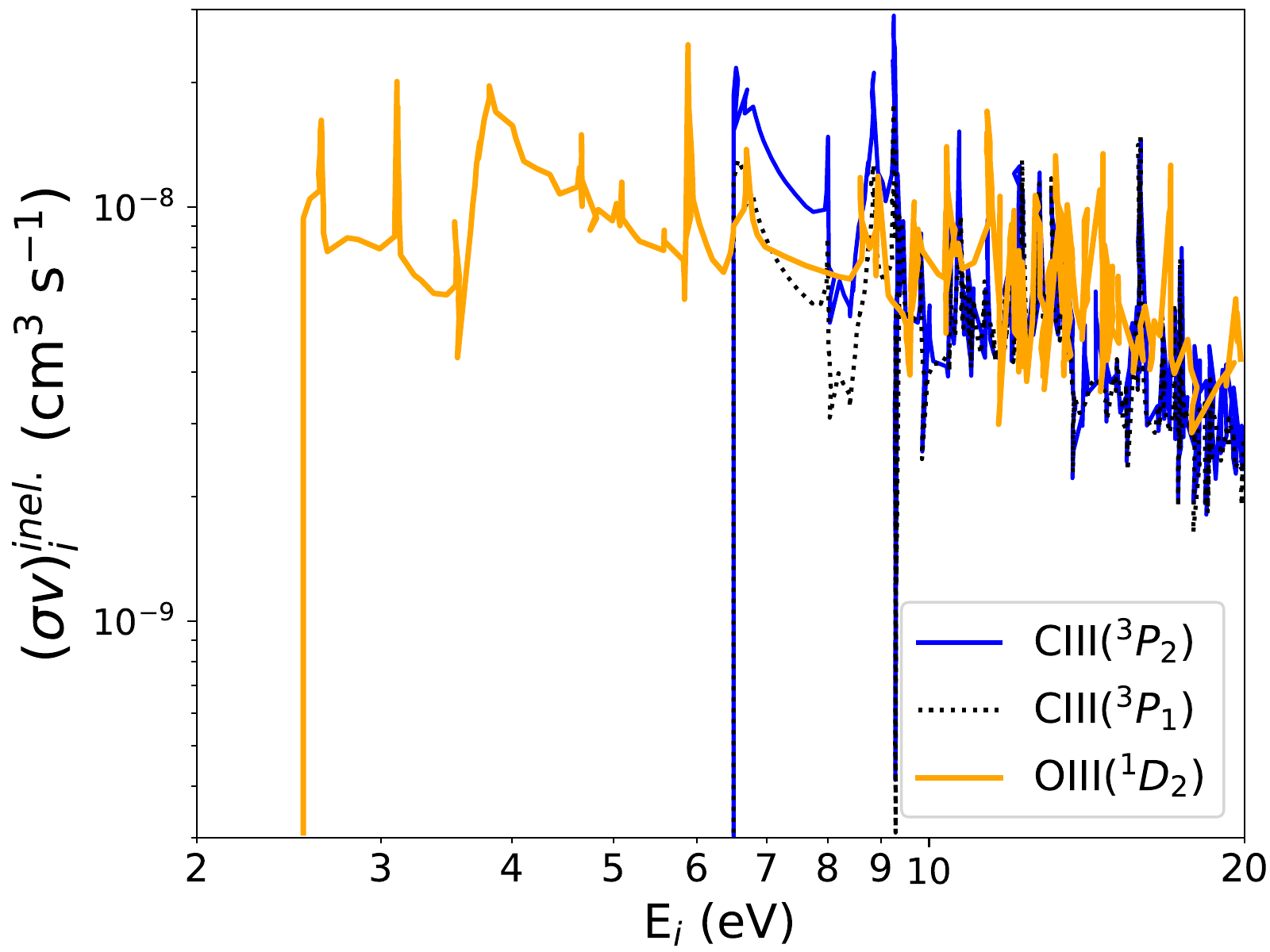}{0.3\textwidth}{b}
    \fig{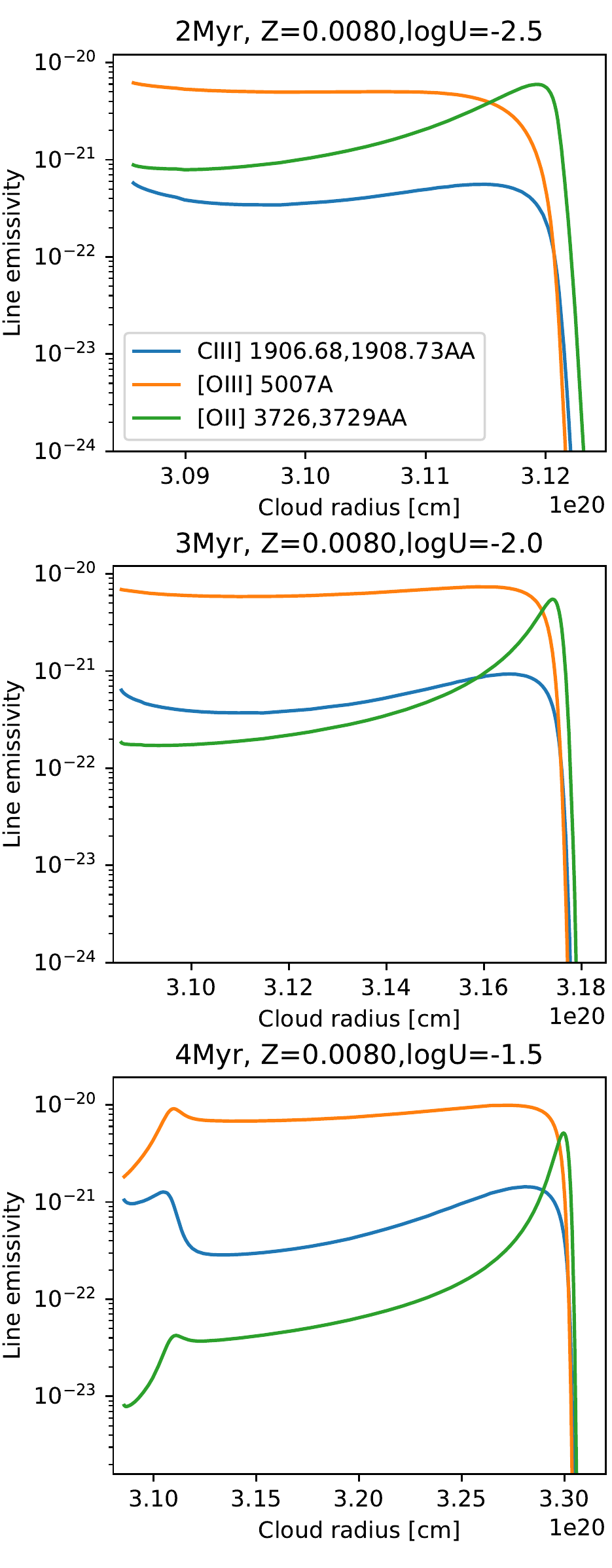}{0.3\textwidth}{c}
  }  
  \caption{(a) Ionization structure of an \hii\ region with metallicity $Z=0.008$, ionization parameter $\log{U}=-2.0$, instantaneous star formation history (SFH), and age as indicated in the inset text, for CIII, OIII, and OII ions. (b) Rate coefficients for collisional excitation of OIII ($\lambda=5007$\AA) and CIII ($\lambda\lambda=1907,1909$\AA). (c) Emissivity as a function of cloud radius for the same clouds but varying $\log{U}$. The cloud radius is equal to the distance to the illuminated face fo the cloud plus the cloud depth in (a). \label{fig:theory}  }
\end{figure*}

\section{Discussion\label{sec:discuss}}
\subsection{Should \ciii\ behave like \oiii\ or \oii?}\protect\label{sec:theory}
To the best of our knowledge, other than the data in our project, there are no spatially resolved observations of \ciii\ for any galaxy. To facilitate planning and analysis of future observations it is important to know if \ciii\ is more likely to be co-spatial with high or with low ionization lines. We therefore create Cloudy \citep[version c17.03,][]{Ferland2017} simulations to investigate if \ciii\ behaves more similar to \oiii\ or to \oii, using these lines as tracers for the high and low ionization gas, respectively. The details of the Cloudy input are given in Section \ref{sec:cloudy} of the Appendix. First we consider that the second ionization potential of carbon is $24.4$ eV, and of oxygen $35.1$ eV. This difference does not mean that \ciii\ can be considered a low-ionization line, comparable in behavior to, e.g., \oii. Figure \ref{fig:theory}a shows the ionization structure of example \hii\ regions with $Z=0.008$, $N_e=100$ cm$^{-3}$, standard C/O ratio, and $\log{U}=-2.0$. The choice of metallicity $Z=0.008$ and ages $3\pm1$ Myr is motivated in Section \ref{sec:cigale} in the Appendix. For these typical \hii\ region properties, CIII and OIII ions co-exist along most of the cloud depth, and in particular in the central regions. A harder ionizing spectrum of Wolf-Rayet (WR) stars at age $\ge4$ Myr or a spectrum with $\log{U}=-1$ will, however, ionize CIII to CIV in the center and hence most CIII ions will be found in the outer layers of the cloud. We note that very similar models are obtained for $Z=0.004$ and both instantaneous and continuous SFHs. The purpose of the models in Figure \ref{fig:theory} is simply to check how \ciii\ behaves compared to \oiii\ in a standard theoretical cloud, and we do not attempt here to fit the cluster observations.

The ionization structure by itself does not guarantee that \ciii\ emission will come from regions where CIII ions are located. The collisional excitation potential must also be considered. Given its blue wavelength of $\lambda=1909$\AA, the \ciii\ excitation potential of $6.5$ eV is higher than for \oiii\ (2.5 eV) and \oii\ (3.3 eV), which converts to temperatures of $T_{e}\sim50$, $\sim19$ and $\sim25.5$ kK, respectively. However, the Maxwellian distribution of electron velocities has a substantial tail towards increasing velocities. As an example, we consider the number of electrons with energies $E_{CIII}\ge6.5$ eV for velocity distributions of $T_{e}=10$ kK and $20$ kK. Relative to a $T_{e}=50$ kK distribution, for these temperatures the number of electrons with $\ge E_{CIII}$ is $3\%$ and $30\%$, respectively, and hence gas at such temperatures still has electrons with velocities suitable for collisional CIII excitation. One must also consider the rate coefficients for collisional excitation. Figure \ref{fig:theory}b shows that at energies slightly above $6.5$ eV the same electrons can continue to collisionally excite O$^{++}$ instead of C$^{++}$, and preferentially do so at higher energies, because the collisional excitation rate for OIII is initially comparable to, and then on average higher than for CIII. The resulting line emissivity $\epsilon$ will change with varying cloud conditions, as exemplified for a handful of clouds in Figure \ref{fig:theory}c. $\epsilon$ gets a boost close to the center of the cloud, but also near the ionization front, as the electron temperature increases with optical depth. We note that the behavior of the emissivity is driven primarily by $\log{U}$, and the age is effect is very small for these young and similar ages. We conclude that CIII ions can be collisionally excited in all layers of the \hii\ region, and \ciii\ can be assumed to be co-spatial with \oiii. 

\subsection{Observed strength of \ew}\protect\label{sec:strength}
For aperture photometry, the average of all clusters is $\left<\ew\right> =17.9$\AA, with the lowest (non-zero) and highest values of $4.9$\AA\ and $27.6$\AA, respectively. Similarly for dendrograms the numbers are $18.8$\AA, $1.9$\AA, and $30.4$\AA\ for the average, lowest and highest, respectively. A population of single stars can account for \ew$\le10$\AA, and binary stars increase the limit to $\le20$ \AA\ \citep{Jaskot2016,Nakajima2018}. In other words, the observed strength of \ew\ in our observations, listed in Table \ref{tab:powlaw}, is generally consistent with the theoretical stellar maximum. Apertures 2, 3, 4, 6, and 12 are all in knot A and above $20$\AA. A top-heavy initial mass function (IMF), a higher C/O ratio, and the presence of shocks would all increase the observed \ciii\ flux \citep{Jaskot2016}. It is possible that all or a combination of these factors are boosting the \ew. On the other hand, the theoretical models are for an ``effective'' \hii\ region, i.e., integrated over the entire galaxy, and may not fully capture the physical conditions on the resolved scales of individual custers. For example, the electron density and temperature may fluctuate significantly, and local overdensities would provide more electrons for collisional excitation of C{\tt III}, the geometry of individual clouds may be influenced by the proximity of neighboring \hii\ regions (e.g., in the crowded knot A), and shocks may significantly contribute to heating and ionizing the gas \citep{Jaskot2016}.

An additional point is lingering PSF issues. The apertures with \ciii $>20$\AA\ may possibly be affected by a not perfect PSF equalization. For apertures 2 and 3 (Figure \ref{fig:ew}, middle panel), the corresponding dendrograms are of similarly small physical size (Figure \ref{fig:ew}, bottom panel), and hence may also be affected. For apertures 4 and 6, the corresponding dendrograms show values of $14.9\pm2.2$, and $20.0\pm3.0$, respectively, which are consistent with a purely stellar ionizing source. Aperture 12 is a borderline case, with the dendrogram having \ew$=25.0\pm4.4$.

However, we note that Knot A is the most highly ionized knot in Haro 11.  The observed emission line flux ratio \oiii$\lambda5007$/\oii$\lambda3727$ (hereafter\oiii/\oii) is $\gtrsim10$ in this region \citep{Keenan2017}, which generates the properties that make Haro 11 a Green Pea analog. It is somewhat surprising to find enhanced \ciii, i.e., \ew$\gtrsim20$\AA, with the highest S/N in this region, since C III exists at somewhat lower ionization energies than O III; the ionization potentials for C III and O III are $47.9$ eV and $54.9$ eV, respectively.  We show below in Section 6.4 that the observed \ciii\ emission is difficult to explain with photoionization models.

Another possibility is a harder ionization field. \citet{Nakajima2018} find that to explain \ew\ beyond $20$\AA\ a mix of stellar and AGN components is necessary. For galaxies with $10\mbox{-}20\%$ solar metallicities and $\log{U}\le-1.7$, an enhanced C/O ratio is further required. X-ray emission has indeed been detected from Haro 11. Knot B, with $L_X\sim10^{41}$ erg s$^{-1}$, is likely a black hole binary and knot C, with $L_X\sim5\times10^{40}$ erg s$^{-1}$ is an Ultra Luminous X-ray binary \citep{Prestwich2015}. Although not explicitly discussed by the authors, figure $1$ in \citet[][]{Dittenber2020} shows a possible X-ray point source co-incident with Knot A, with X-ray emission at the $\ge0.3$ keV level. If X-rays are present this would imply a harder ionizing field than a pure stellar spectrum, which would increase the expected \ew. 

%It is also conceivable that knot A is similar to the case of the N44C fossile X-ray ionized nebula \citet{Pakull1989} suggest that N44C is being ionized by a presently-extinct X-ray source, which would explain the hard ionizing spectrum necessary to produce the spectral features they observe, and it would simultaneously also account for the lack of an X-ray detection.

\begin{figure*}
  \includegraphics[width=0.5\textwidth]{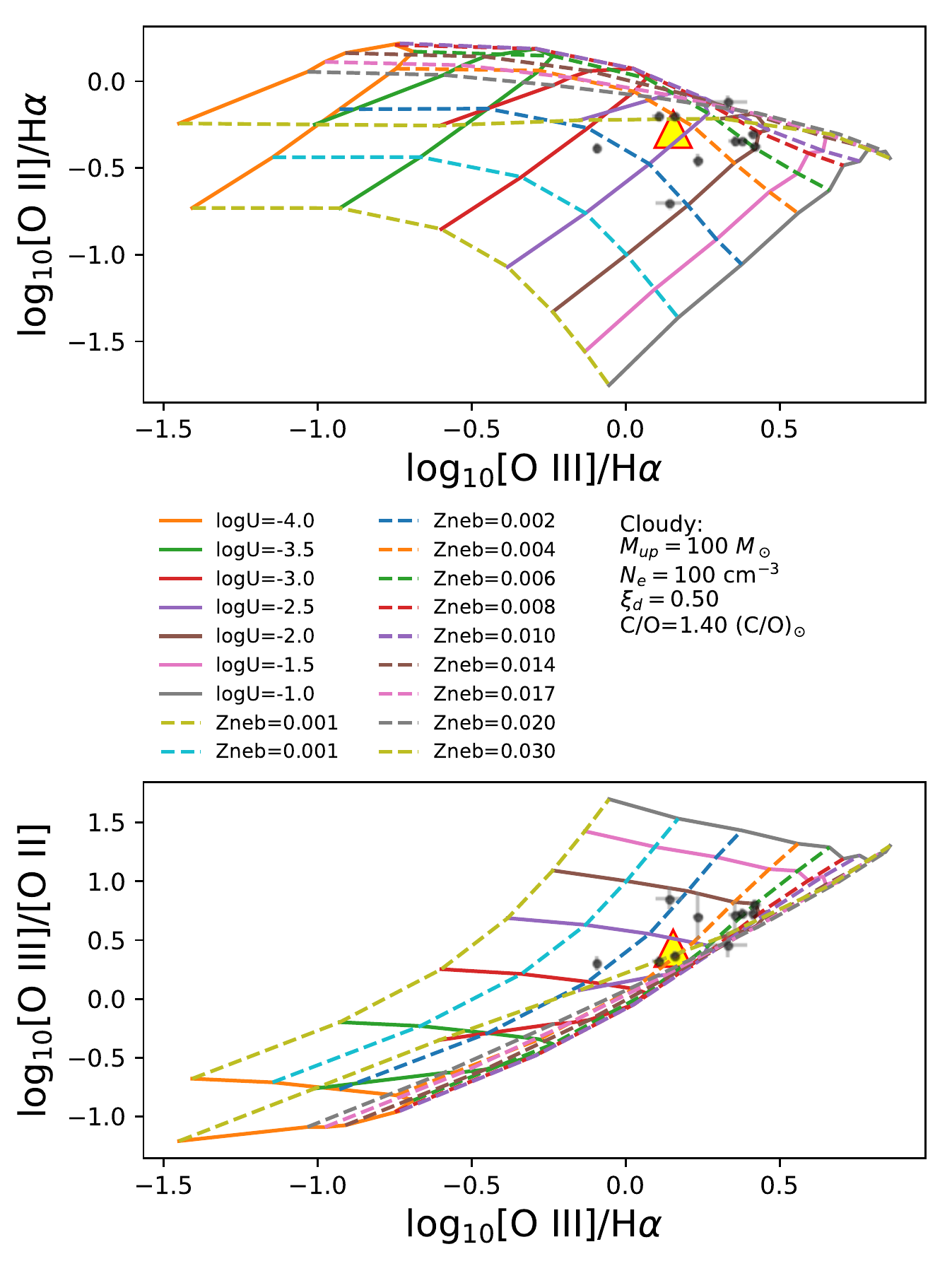}
  \includegraphics[width=0.5\textwidth]{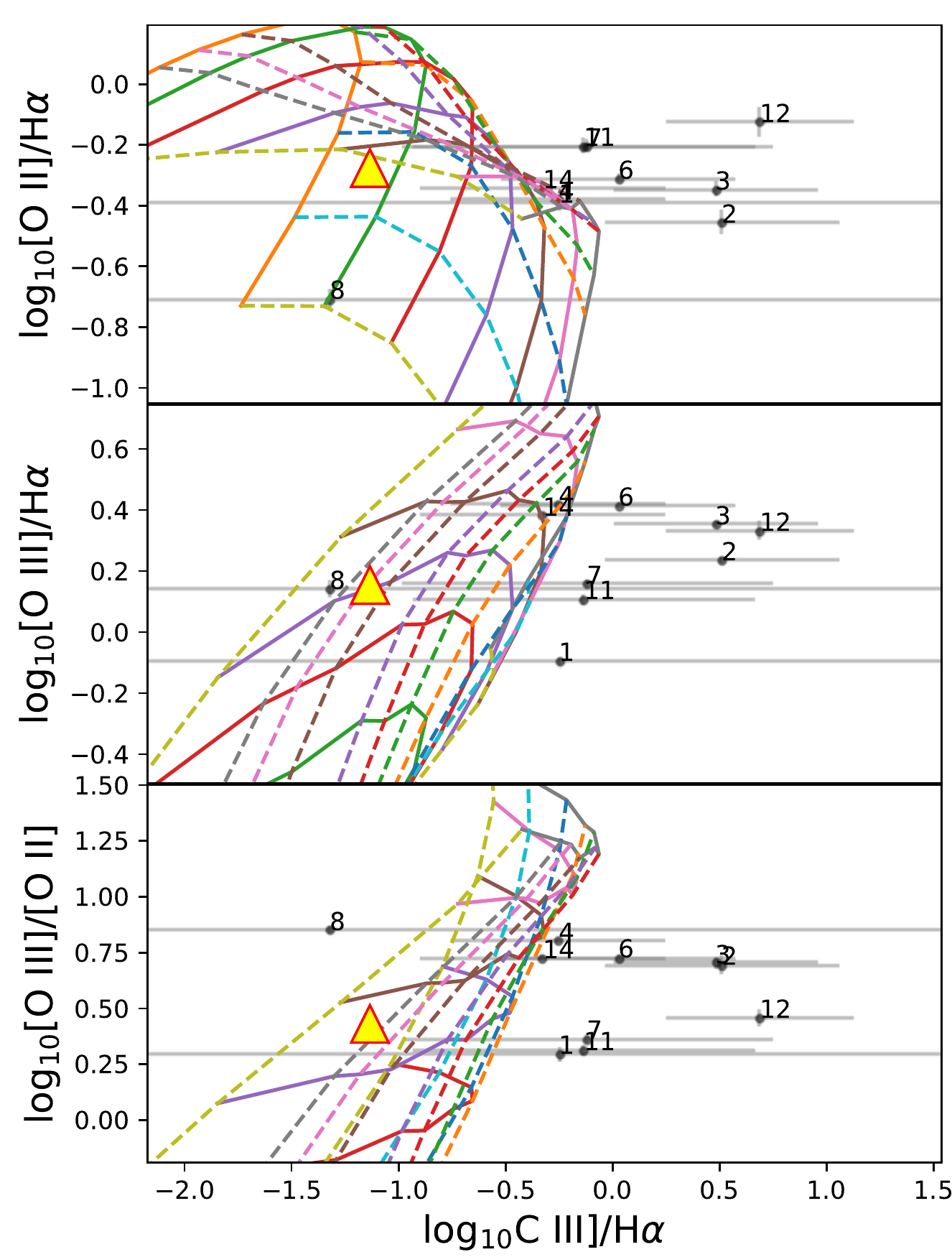}
\caption{Left: Emission line ratios of \oiii, \oii, \ha\ versus \oiii/\ha\ for $r=0.125$ arcsec apertures on the clusters (black circles). The ratios for Haro 11 as a whole are indicated by a yellow triangle. Right: the same ratios versus \ciii/\ha. The Cloudy model grids are for a purely stellar ionizing source.  \label{fig:lines}}
\end{figure*}

\begin{figure*}[ht!]
  \gridline{
    \fig{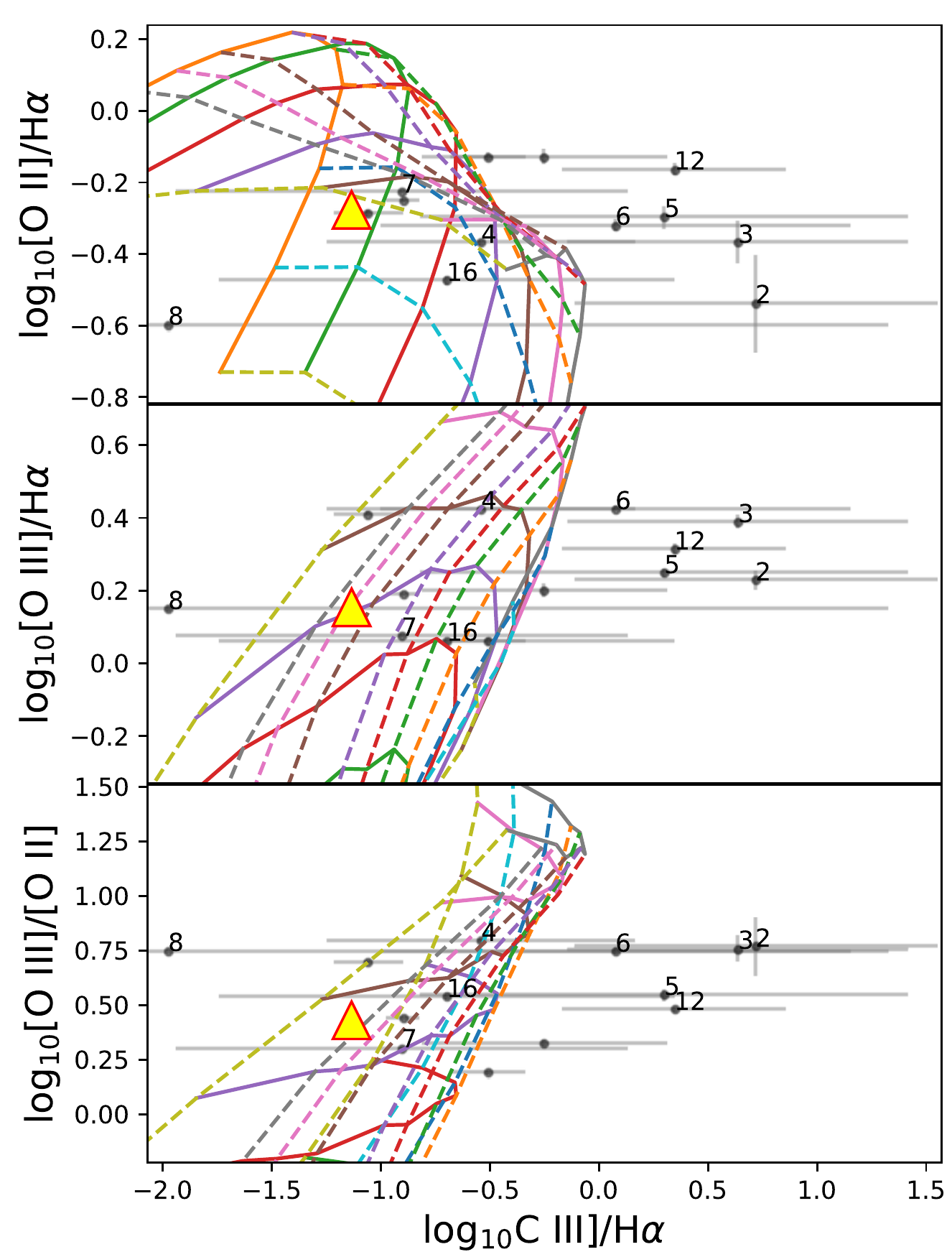}{0.3\textwidth}{a}
    \fig{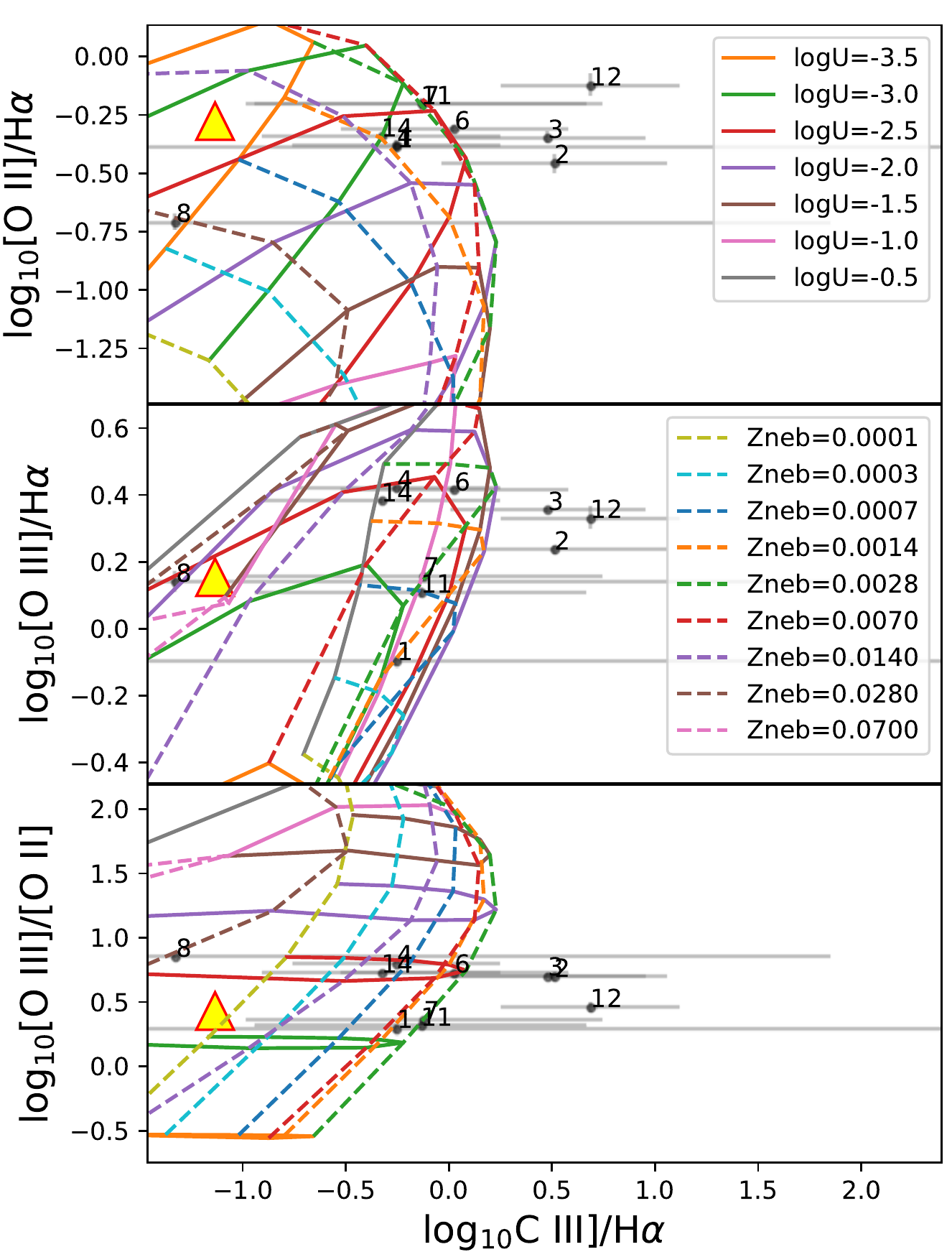}{0.3\textwidth}{b}
    \fig{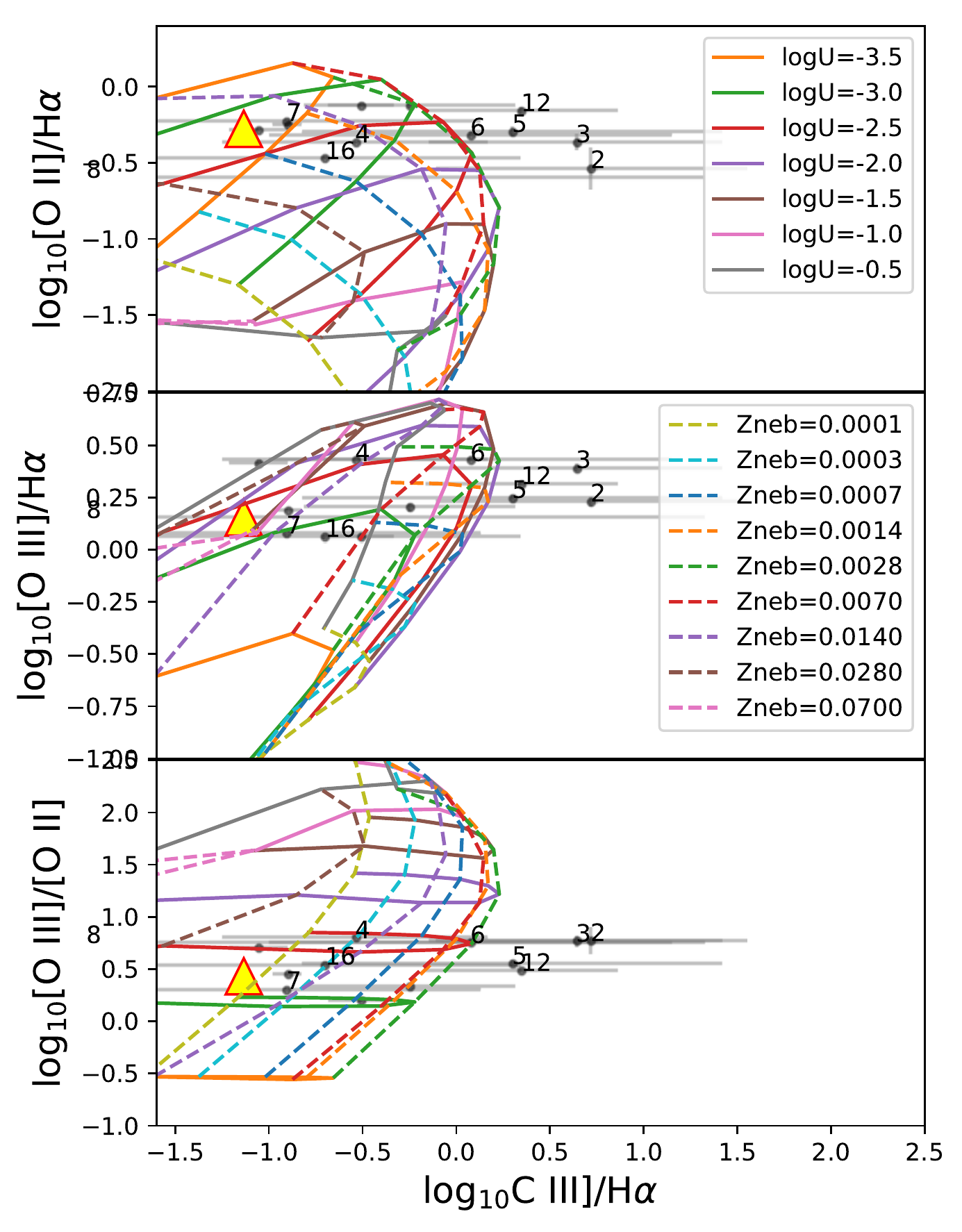}{0.3\textwidth}{c}
  }    
\caption{Same as Figure \ref{fig:lines} but (a) for dendrograms. Numbers indicate IDs of overlapping apertures. (b) for apertures and models with an AGN ionizing source. (c) for dendrograms and models with an AGN as ionizing source. \label{fig:experiment}}
\end{figure*}

\subsection{\ew\ versus cluster properties}
\citet{Adamo2010} perform SED fits to the cluster population in Haro 11. They obtain ages, stellar masses and ISM attenuation $E(B-V)$. We crossmatched our clusters with Adamo and compared the strength of \ew\ to these properties, finding no convincing correlations. The comparison is not very telling partially due to the fact that all clusters, except knot C, have been assigned identical ages. Another potential problem is that the Adamo SED fits use the F140LP filter and may therefore be affected by its wrong zeropoint.

We therefore redo the SED fitting for the clusters shown in Table \ref{tab:powlaw} using Cigale \citep{Noll2009, Serra2011} as detailed in Section \ref{sec:cigale} of the Appendix, this time with the new and correct F140LP zeropoint. An additional benefit of redoing the fits is that Cigale provides a more detailed parameter space, which includes the ionization parameter $\log{U}$, the young stellar mass, and the dust fraction internal to the \hii\ region. The comparison is shown in Figure \ref{fig:cprops} of the Appendix for five parameters. The Cigale SED fitting does not reveal any significant trends lurking in the data, we observe no correlatioins with any of the cluster parameters. This could be due to the fact that nebular physics, more than stellar physics, governs the strength of \ciii, while Cigale is a pure stellar population synthesis code, which does not model the nebular emission. 

\subsection{\ew\ versus other emission lines}
In Figure \ref{fig:lines} we show diagrams of the line ratios of \oiii/\ha\ versus \oii/\ha\ and \oiii/\oii\ for the cluster apertures shown in the middle panel of Figure \ref{fig:ew}. The line ratios have been corrected for Galactic extinction using a \citet{Fitzpatrick99} attenuation law with \citet{Schlegel1998} dust maps. Overplotted are Cloudy \citep[version c13.03,][]{Ferland2013} models of nebular emission powered by a purely stellar ionizing source, as modeled and presented in detail in \citet{Gutkin2016}\footnote{These authors provide their model grids for public download: \url{http://www.iap.fr/neogal/models.html}}. We choose to compare our observations primarily to published, vetted and publicly available Cloudy simulations. In the same figure, we also show the \ciii/\ha\ line ratio versus \oii/\ha, \oiii/\ha\, and \oiii/\oii. None of the line ratios in the figure have been corrected for internal attenuation, because the Cloudy models contain dust grains and hence attenuate the emerging spectra. As is obvious from the figure, the cluster line ratios are consistent with the Gutkin grid when comparing to \oiii/\ha\ (left column). When comparing to \ciii/\ha\ (right column), the clusters are also consistent with the grid within their errorbars for all but three apertures (2, 3, and 12, all in knot A). Aside from the comparison in the figure, we have also compared the observations to Cloudy grids with an upper mass limit $M_{up}=300$ M$_\odot$, dust-to-metal mass ratios of $\xi_d=0.1$ and $0.3$, electron density of $N_e=10^3$ and $10^4$ cm $^{-3}$, and changing the C/O ratio from enhanced to normal. The best match, i.e. the maximum number of clusters consistent with the model grid, is obtained for the Cloudy run in Figure \ref{fig:lines} with $M_{up}=100$ M$_\odot$, dust-to-metal ratio $\xi_d=0.5$, electron density $N_e=100$ cm $^{-3}$, and $C/O=1.4 (C/O)_{\odot}$. The latter is the highest $C/O$ ratio available from the Gutkin collection. For all other Gutkin grids even more cluster data points are offset from the Cloudy predictions. In the next section \ref{sec:alt} we investigate this in more detail. Plots with the \ciii/\oiii\ line ratio look similar and are not explicitly presented here.  

Taking Haro 11 as a whole, integrating all emission line signal inside the gray contour in Figure \ref{fig:ew}, and constructing total galaxy line ratios, gives a \ciii/\ha\ ratio fully consistent with Cloudy predictions, as indicated by the yellow triangle in Figure \ref{fig:lines}. The data point for Haro 11 falls between lines of constant ionization parameter $\log{U}=-2.0$ and $-2.5$, and metallicity lines between $Z=0.004$ and $0.01$, with average $Z=0.008$, which are within the realm of expectations for this galaxy \citep{Bergvall2002}.

\subsection{Reconciling \ciii\ with models}\protect\label{sec:alt}
In an attempt to reconcile all cluster observations with models, we consider if $r=0.125$ arcsec is the physically appropriate size for the apertures. Assuming $N_e=100$ cm $^{-3}$, our data would not resolve the Str\"omgren sphere for $\log{U}\le-1.5$, because the corresponding Str\"omgren radius is $\le11.8$ pc, which is less than a pixel.

For most clusters, increasing the extraction region from aperture to dendrogram significantly lowers the \ew. This is demonstrated in Table \ref{tab:powlaw} and Figure \ref{fig:experiment}a. For example, compare table entries 4.Ap of $23.2\pm2.6$ to 4.Dendro of $14.9\pm2.2$, where the latter is fully consistent with pure stellar photoionization.  The exceptions are apertures 2, 3, and 12, all in knot A and discussed below. The fact that larger bins decrease the \ew\ could also be interpreted as the \hii\ regions leaking LyC photons. A drop in the \ew is expected for a non-zero LyC escape fraction \citep{Jaskot2016}. This would be in line with results from \citep{Keenan2017} who use ionization parameter mapping to identify Knot A is a prime candidate for the escape of LyC photons. We investigate this possibility in Section \ref{sec:cloudy} in the Appendix, alas with inconclusive results. 

Recall from Section \ref{sec:strength} that X-ray emission from an ULX has been detected in knots B and C, and so investigating Cloudy models with an AGN as the ionizing source is also of interest. In Figure \ref{fig:experiment}b and \ref{fig:experiment}c we show apertures and dendrograms, respectively, with overplotted  models by \citet{Nakajima2018}, with an AGN as the ionizing source and with powerlaw index $\alpha=-1.2$, $N_e=100$ cm$^{-3}$ and $C/O=1$ ($\sim2$ times solar). Apertures 2 and 3 are now consistent with the model grid within their errorbars, and aperture 12 is $\sim1\sigma$ away. However, all of these apertures are in knot A, which is at best a weak X-ray source and is likely dominated by stellar photoionization. For apertures 2 and 3, the dendrograms are smaller than the aperture radius, and hence they may suffer from PSF equalization problems. The dendrogram overlapping aperture 12, however, is large enough to not have PSF problems and yet the data point is clearly deviating from the grids in figures \ref{fig:lines} and \ref{fig:experiment}. We mention here that we have also created and tested Cloudy models with $N_e=10,10^3,10^4$ cm$^{-3}$, as well as with a non-constant electron density $N_e$ declining with cloud depth as a powerlaw with index $-2$, alas that too did not match aperture 12. We have verified that the models in \citet{Feltre2016}, which are similar in setup to \citet{Gutkin2016} but with an AGN ionizing source and assume a solar C/O value, are not consistent with most data points.

We note that a high C/O ratio of $1.4\times (C/O)_\odot$ is needed to match other apertures in knot A, i.e. 4, 6, and 14. For blue compact galaxies \citet{Izotov1999} find an average $C/O\sim0.3$ (i.e., sub-solar) at the metallicity of Haro 11, $12+\log{O/H}=7.9$. The C/O ratio appears not to have been explicitly measured for Haro 11 and it is therefore possible that it is higher than what is expected, in the entire galaxy, or locally in knot A. Further increasing the C/O ratio substantially will have the counter effect of decreasing \oiii\ emission as the cooling via \ciii\ becomes significantly higher. We have explicitly verified that a $C/O=3$, $6$, $10$ times solar would still not reconcile a stellar ionizing source model with aperture 12. We conclude that we are unable to fit aperture 12 with any model in our arsenal. 

We have also compared the observations to the Cloudy model predictions from \citet{Nakajima2018} for a pure binary population \citep[BPASS v2.0,][]{Stanway2016} with $M_{up}=300M_\odot$, instantaneous SFH at $1$ Myr age, $N_e=100$ cm$^{-3}$, a standard and twice solar C/O ratios, for a radiation- and density-bounded nebula. In additional, we compared observations to a mixed population of BPASS and four versions of an AGN component, the latter with AGN contributions of $3$ and $10\%$ of the ionizing photons for the powerlaw slopes of $-1.2$ and $-1.8$. The results were similar to what we explicitly present in figures \ref{fig:lines} and \ref{fig:experiment}, in the sense that only the models with the high C/O ratio come close to matching all data points.

The result from the Gutkin and Nakajima Cloudy model comparison does not imply that all the clusters we analyse must have an increased C/O ratio. In fact, the uncertainties on the cluster measurements are such that we cannot reliably distinguish between different models for individual data points. We can only note that increased C/O ratios are necessary to come close to reconciling model predictions with the clusters with the highest \ew. The Cloudy models do not account for attenuation in the ISM, and therefore applying a non-zero ISM attenuation to the models would only increase the mismatch between models and data.

Since an increased C/O ratio for a subsolar metallicity galaxy is problematic, we conclude that photoionization seems unable to account for the enhanced \ciii\ emission in Knot A. \citet{Keenan2017} suggest that Knot A is the LyC emitting source in Haro 11, based on ionization parameter mapping. These conditions, with extremely young, newly formed, super star clusters, are those conducive to catastrophic cooling of adiabatic superwinds \citep[e.g.,][]{Silich2004}, and \citet{Gray2019} predict that such strongly cooling flows can produce enhanced \ciii. Thus, the observed enhanced \ciii\ may be a diagnostic of these conditions, which have been suggested in other extreme objects \citep[e.g.,][]{Oey2017,Jaskot2017}.

\subsection{\ew\ variation among galaxies}\protect\label{sec:ewvary}
As mentioned in the introduction, \ciii\ equivalent widths of widely varying strengths have been measured among galaxies with seemingly similar \ciii-enhancing properties. In light of our observations, the explanation for this is somewhat trivial. If strong \ciii\ predominantly manifests as emission from a compact source, and barring any additional sources of $C^{++}$ excitation like shock heating of the ISM outside of \hii\ regions, then the effective equivalent width of a galaxy would depend on its morphology, i.e. on how many star-forming young clusters there are and how they are distributed across the galaxy. In a galaxy with few \ciii\ point sources the \ew\ will decrease as the integrated continuum flux increases with the addition of continuum-dominated, emission-line-free galaxy regions. For the case of Haro 11, the \ew\ of individual clusters ranges between $4.9$\AA\ and $27.6$\AA\ within a $r=0.125$ arcsec aperture, with average $17.9$\AA\ (Table \ref{tab:powlaw}). When integrating over the entire galaxy area (gray contour in Figure \ref{fig:ew}), with equivalent radius of $r=3.7$ arcsec, we obtain a non-detection in \ciii\ within the uncertainties. This aperture is much smaller than the rectangular aperture of $10\times23$ arcsec$^{2}$ of the IUE observations of Haro 11, from which we measure a value of \ew$\sim3$\AA. This is another indication that diffuse emission must be hidden in the noise of the \ciii\ image.

\section{Conclusions}\label{sec:conclusions}
We present the first ever spatially resolved map of \ciii\ line emission in the starforming galaxy Haro 11. To verify the reality of the signal, we extract the \ciii\ flux in four different binning schemes. In all cases the \ciii\ emission is observed to originate from star clusters and their immediate vicinity. With the quality of the current data we are unable to detect any significant contribution of diffuse \ciii. However, the strength of the \ew\ for the dense nebular cluster cores varies between $4.9$ to $27.6$\AA, with an average of $17.9$\AA. Fitting the spectral energy distribution of the clusters shows no significant correlations between \ew\ and the ionization parameter $\log{U}$, the dust fraction $f_{dust}$, the dust attenuation $E(B-V)$, or the young mass fraction $M_{\star}^{young}$.

Most clusters have \ciii/\ha, \ciii/\oiii, and \ciii/\oii\ line ratios which can be reconciled with Cloudy models either of clouds powered by pure stellar populations or by AGN. In both cases a super-solar C/O ratio of $\ge1.4(C/O)_\odot$ is needed to match the clusters with the highest \ew. Multiple clusters in Knot A appear to show unusually enhanced \ciii\ emission that cannot be explained by massive star ionization.  The extreme nebular excitation in this region suggests conditions that may harbor catastrophic cooling of adiabatic superwinds, which may explain the enhanced \ciii\ emission.

Our observations offer an explanation for the large scatter in \ew\ observed from galaxies with otherwise similar SF properties and metallicities and at all redshifts. Since the observed \ciii\ originates only from star clusters and its distribution is point-source-like, then the effective total \ew\ of a galaxy would depend on its morphology and on the surface area of continuum-dominated versus line emission-dominated regions. 

\appendix

\section{Dendrogram tree structure}
\begin{figure}[ht!]
  \includegraphics[width=0.9\textwidth]{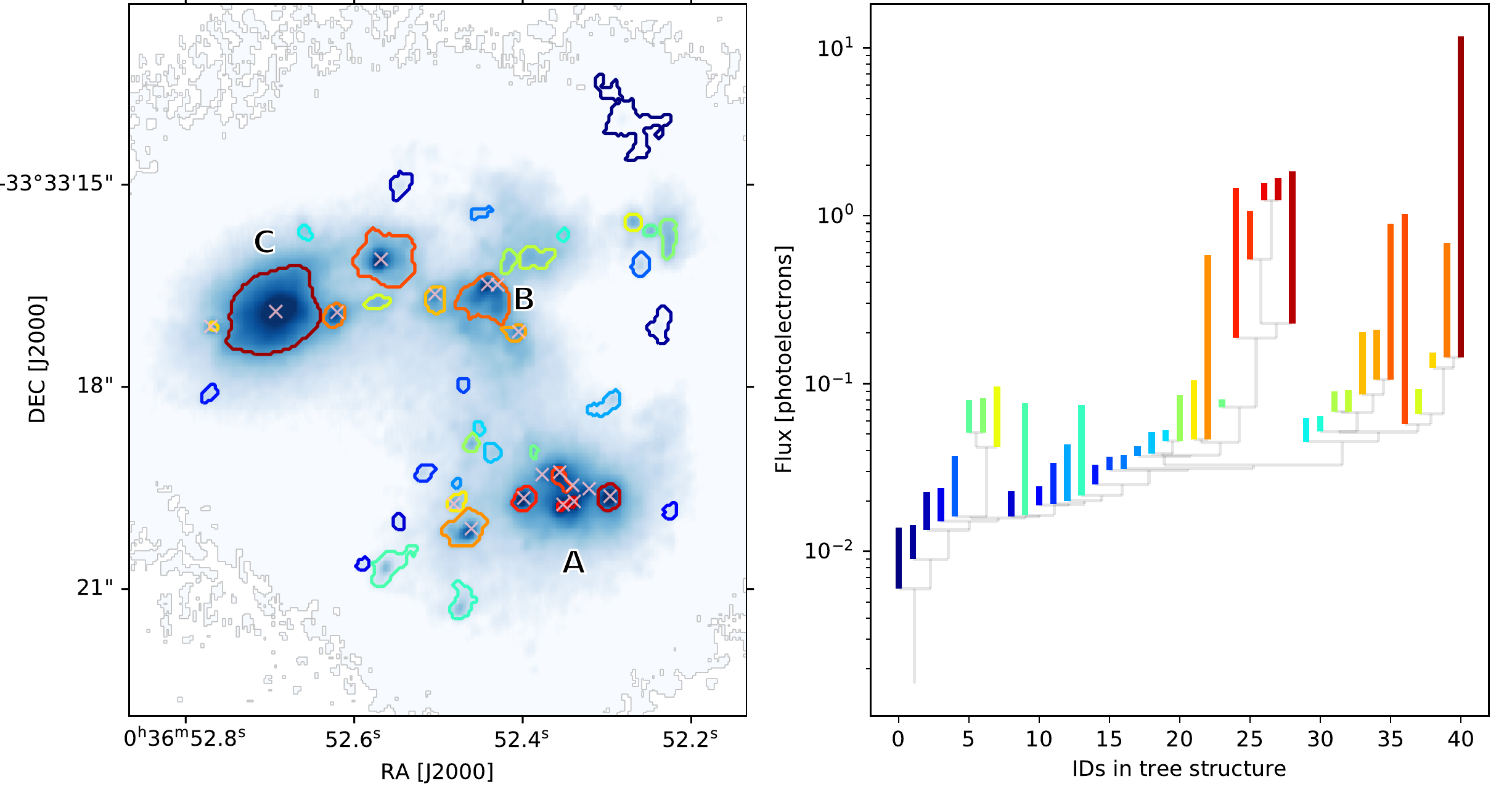}
  \caption{(Left) Dendrogram of Haro 11, with all $41$ detected leaves. Crosses indicate the location of the $18$ clusters, used for aperture photometry in Table \ref{tab:powlaw} and Figure \ref{fig:ew}. (Right) The tree structure, with the leaves color-coded to match the contours in the left panel. \label{fig:dendromap}  }
\end{figure}

\protect\label{sec:dendro}
A dendrogram is a diagram representing a tree of hierarchical clustering, obtained by a process similar to that used by source detection software. In the context of this paper, each leaf from the tree is a region of pixels that seemingly ``belong together'', based on the strength of their flux in the F140LP UV continuum filter. Dendrograms are often used to decide upon the apparent size of \hii\ regions \citep[e.g.,][]{Weilbacher2018}. Figure \ref{fig:dendromap} shows all $41$ resulting leaves of the tree, their spatial location in Haro 11, and the dendrogram tree itself. In the construction of the tree, we did not set a limiting flux, but imposed a minimum height of structures to be retained to {\tt\string min\_delta}$=0.005$, and the minimum size of a leaf was set to {\tt\string min\_npix}$=7$ pixels. Changing these parameters would increase or decrease the number of retained leaves, but would not change their shape in Figure \ref{fig:dendromap}, except to merge neighbouring leaves. This set of parameters optimizes the number of matches to the cluster apertures, shown with crosses in the figure.  

\section{Fitting photometry to synthetic powerlaw spectra}\protect\label{sec:powlaw2}

\begin{figure*}
  \gridline{\fig{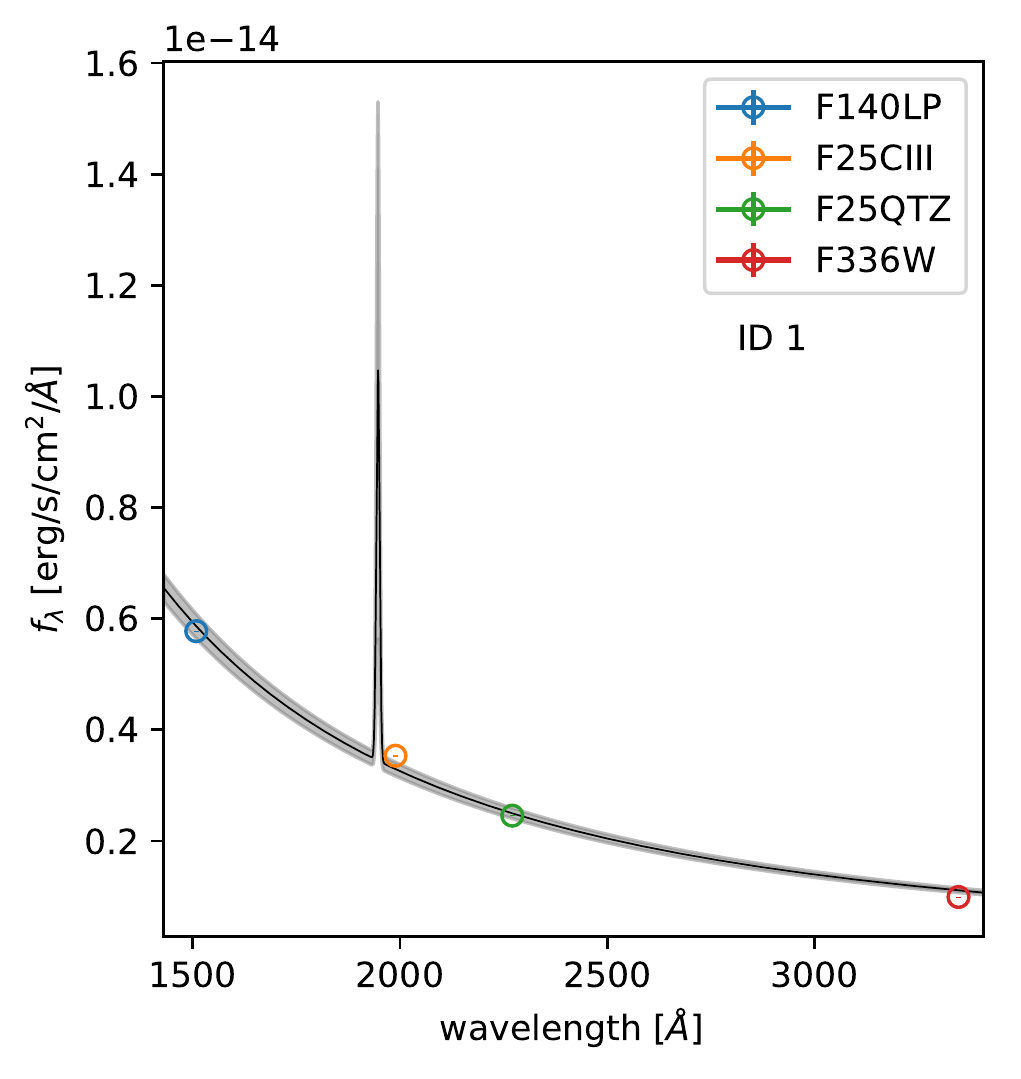}{0.25\textwidth}{}
    \fig{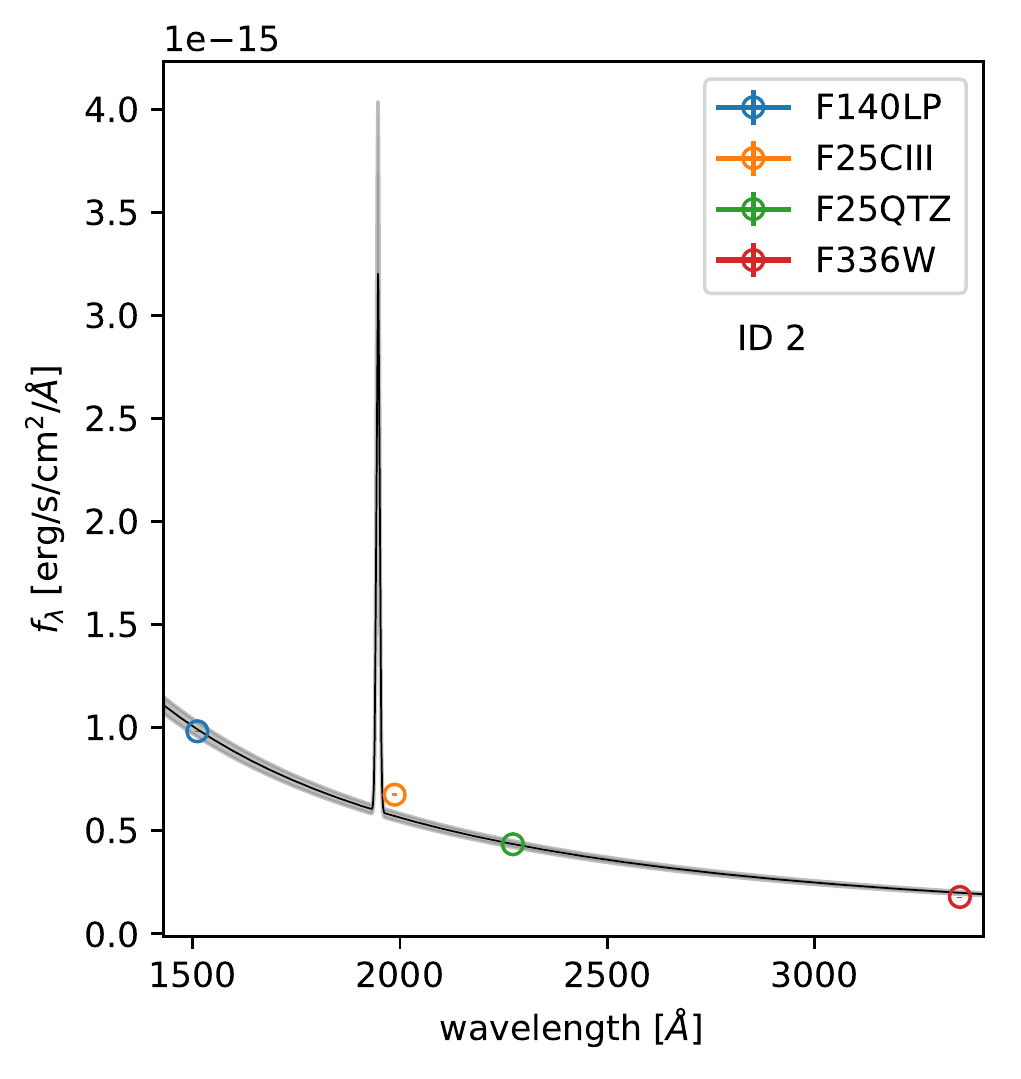}{0.25\textwidth}{}
    \fig{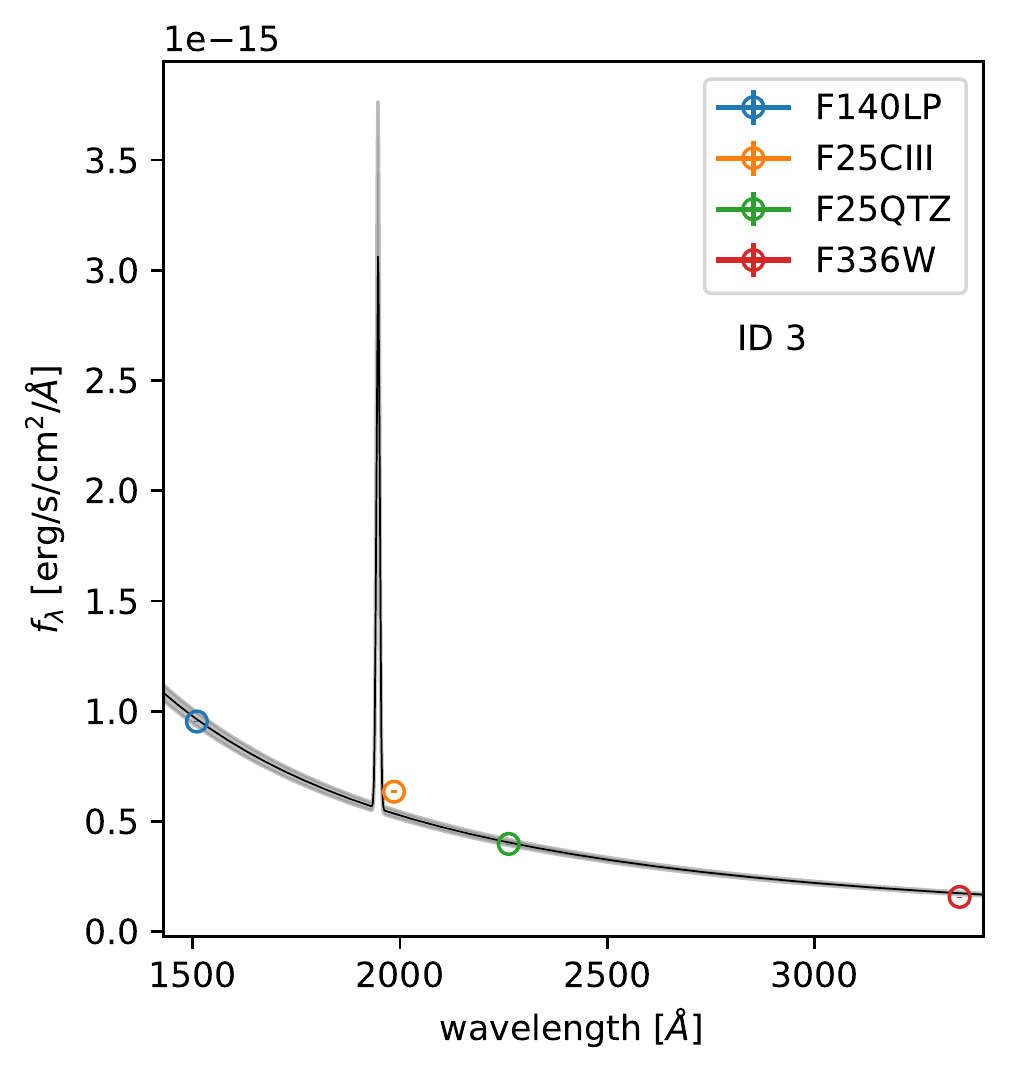}{0.25\textwidth}{}
    \fig{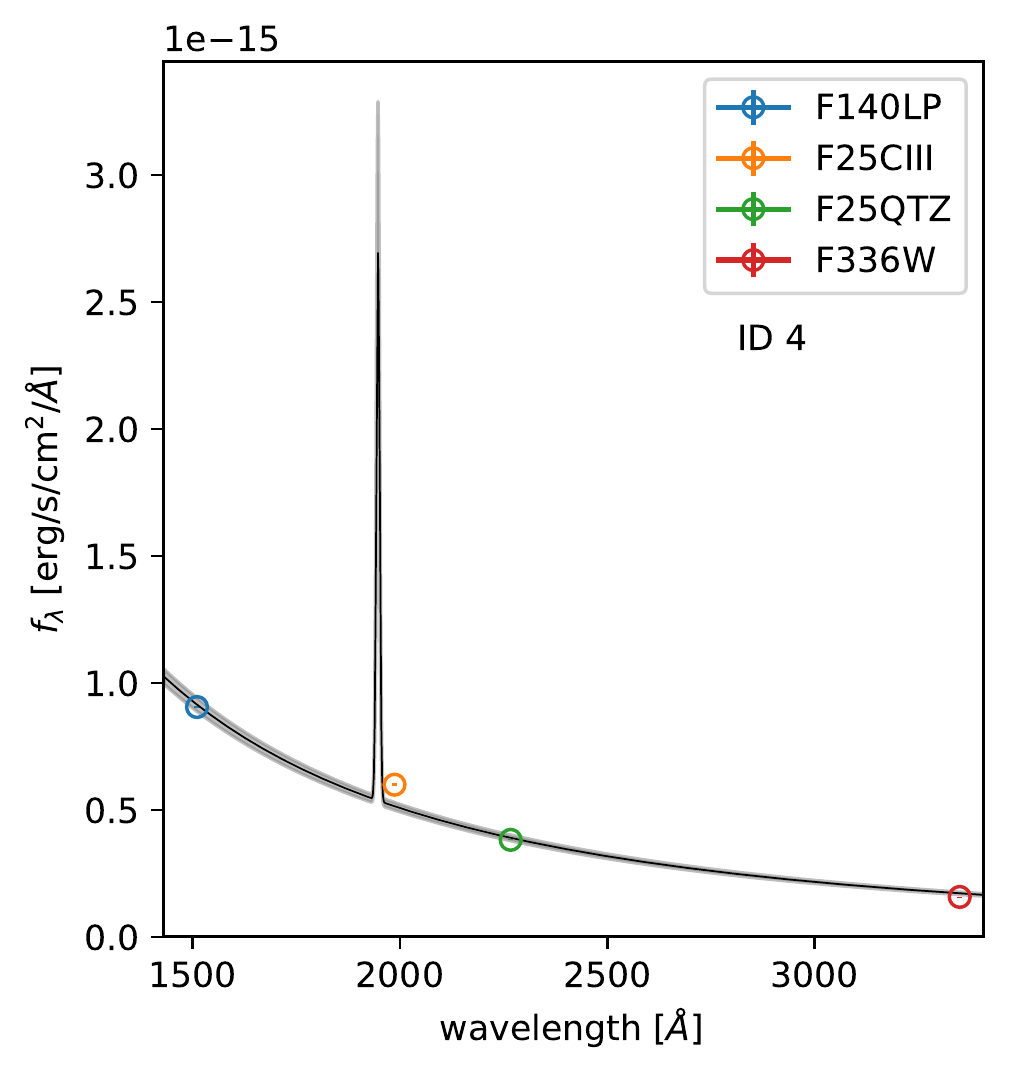}{0.25\textwidth}{}
  }
  \gridline{\fig{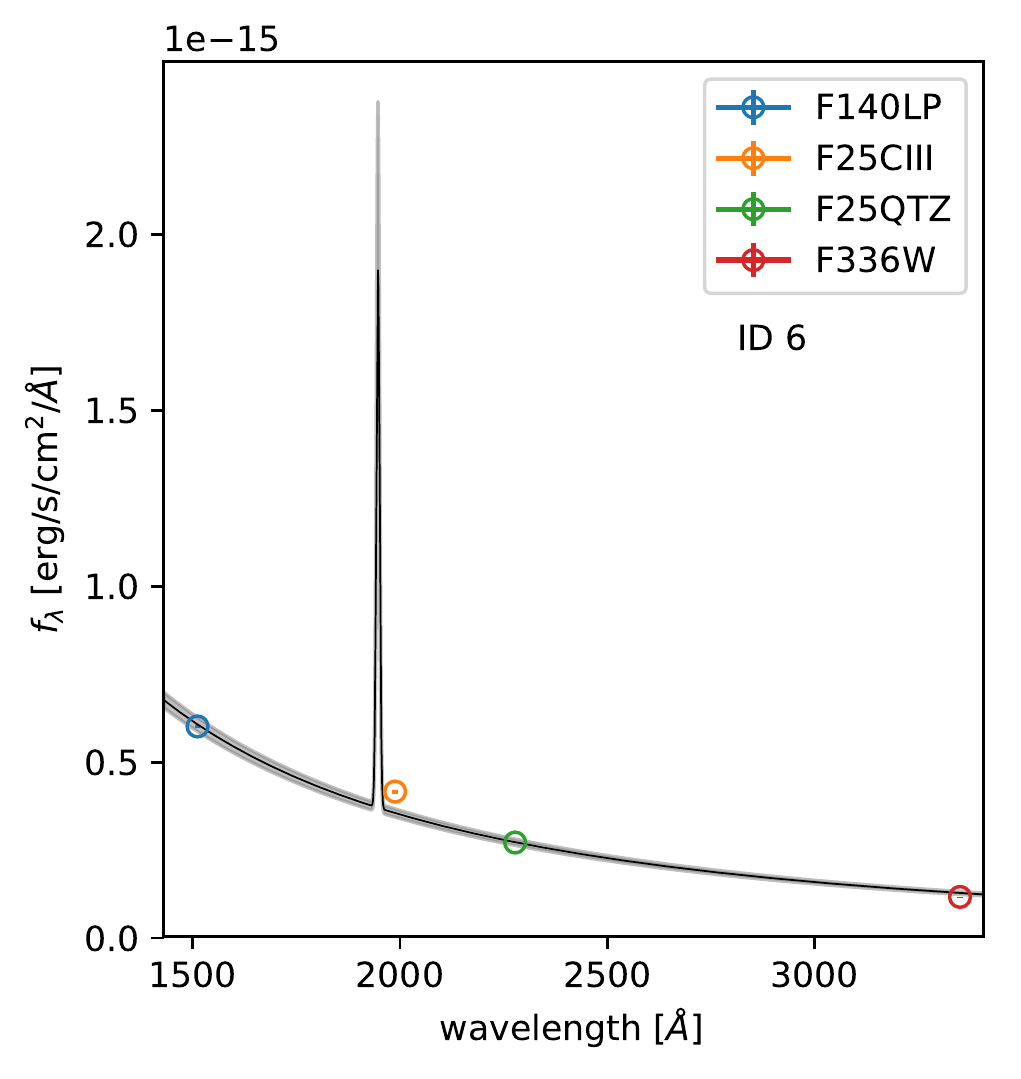}{0.25\textwidth}{}
    \fig{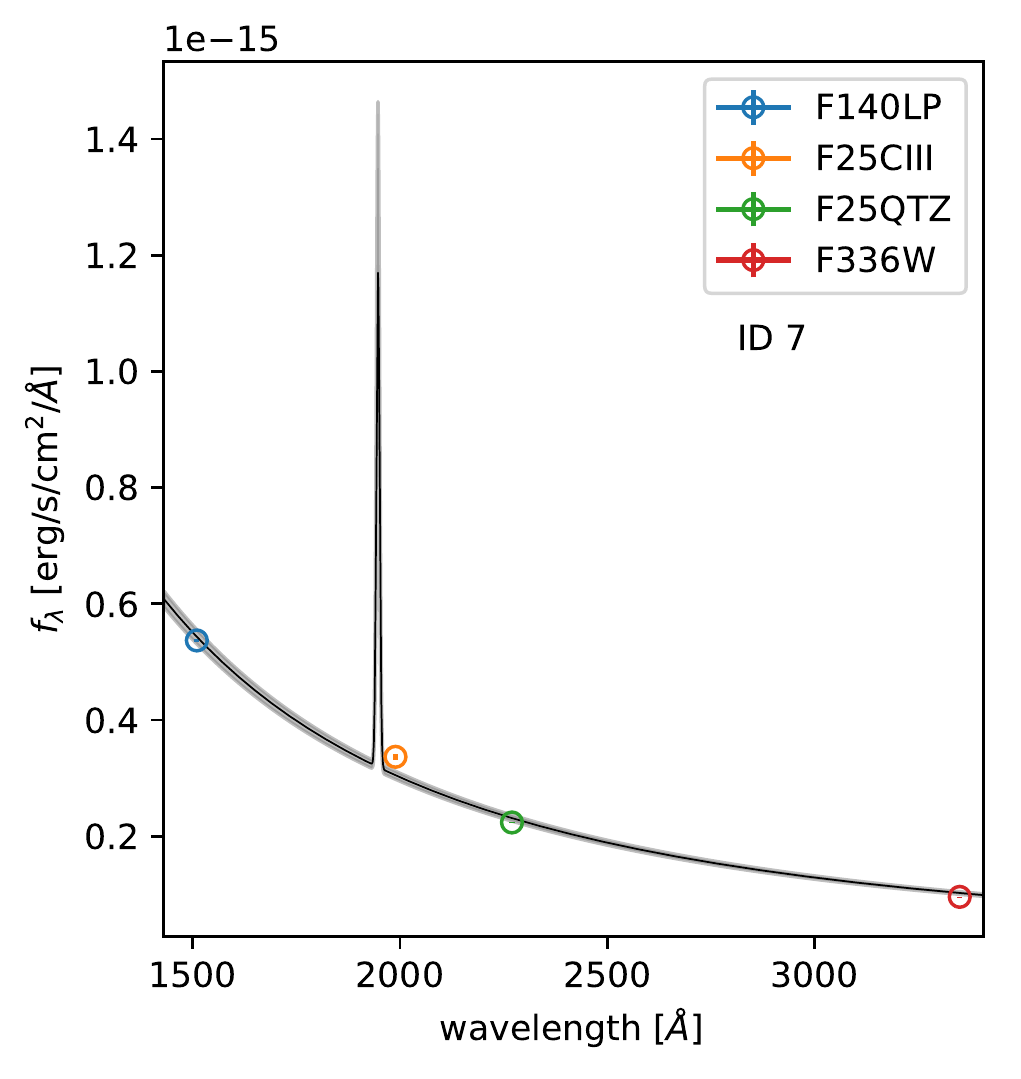}{0.25\textwidth}{}
    \fig{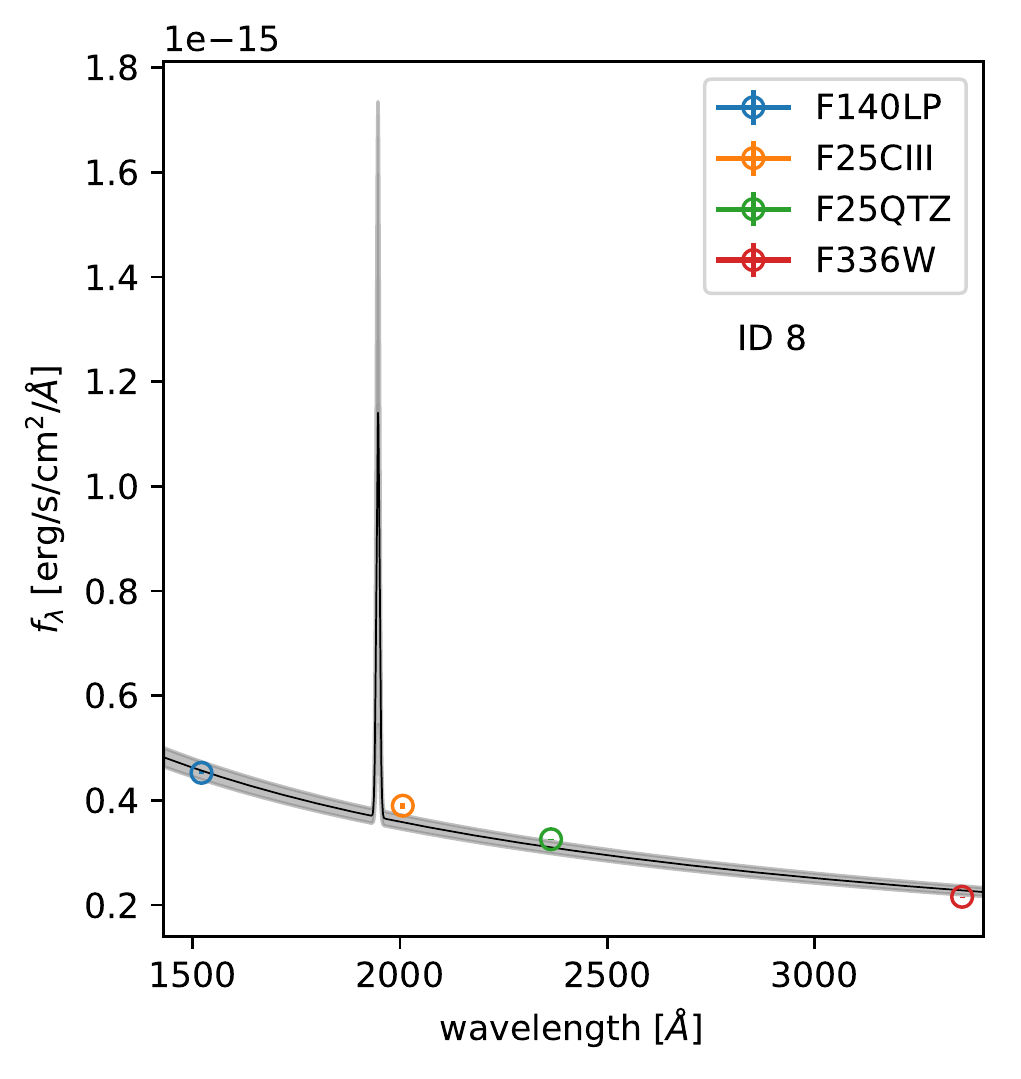}{0.25\textwidth}{}
    \fig{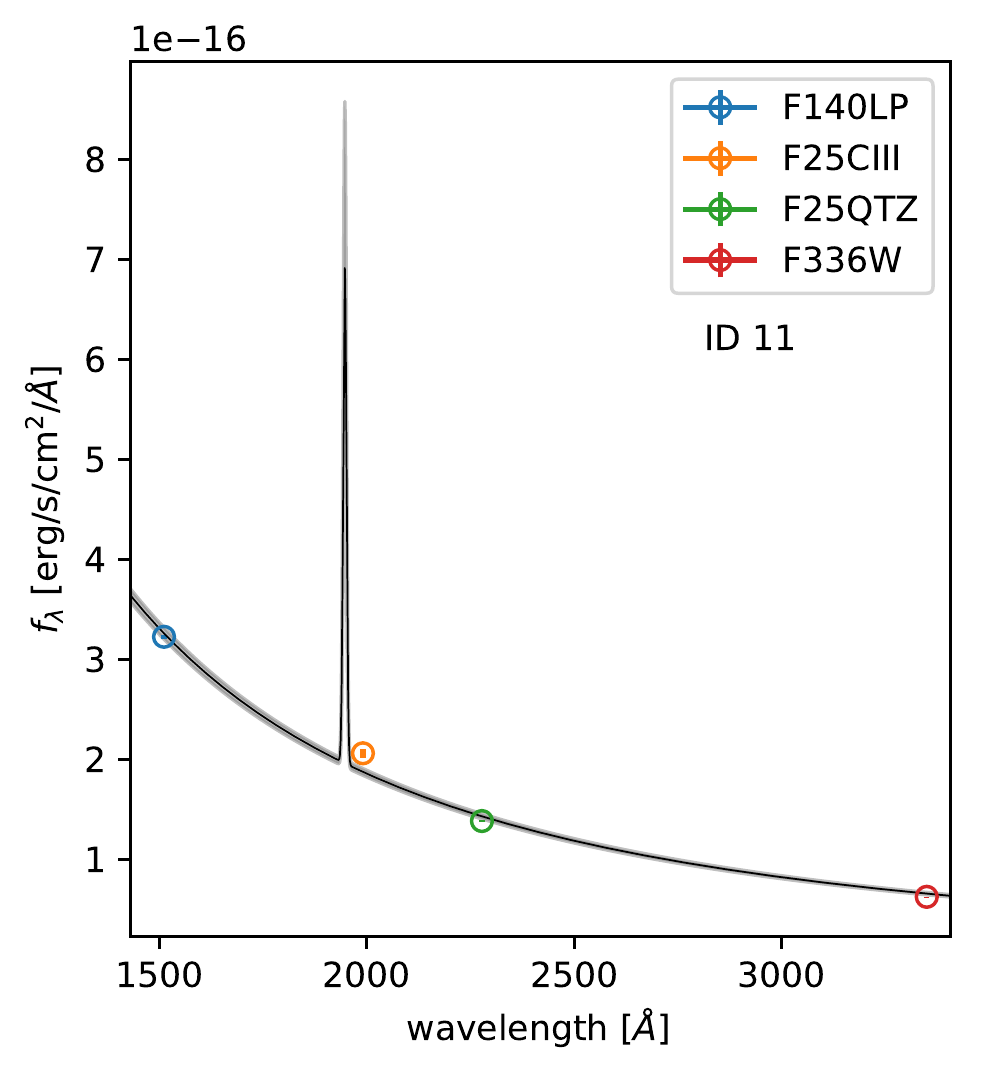}{0.25\textwidth}{}
  }
  \gridline{\fig{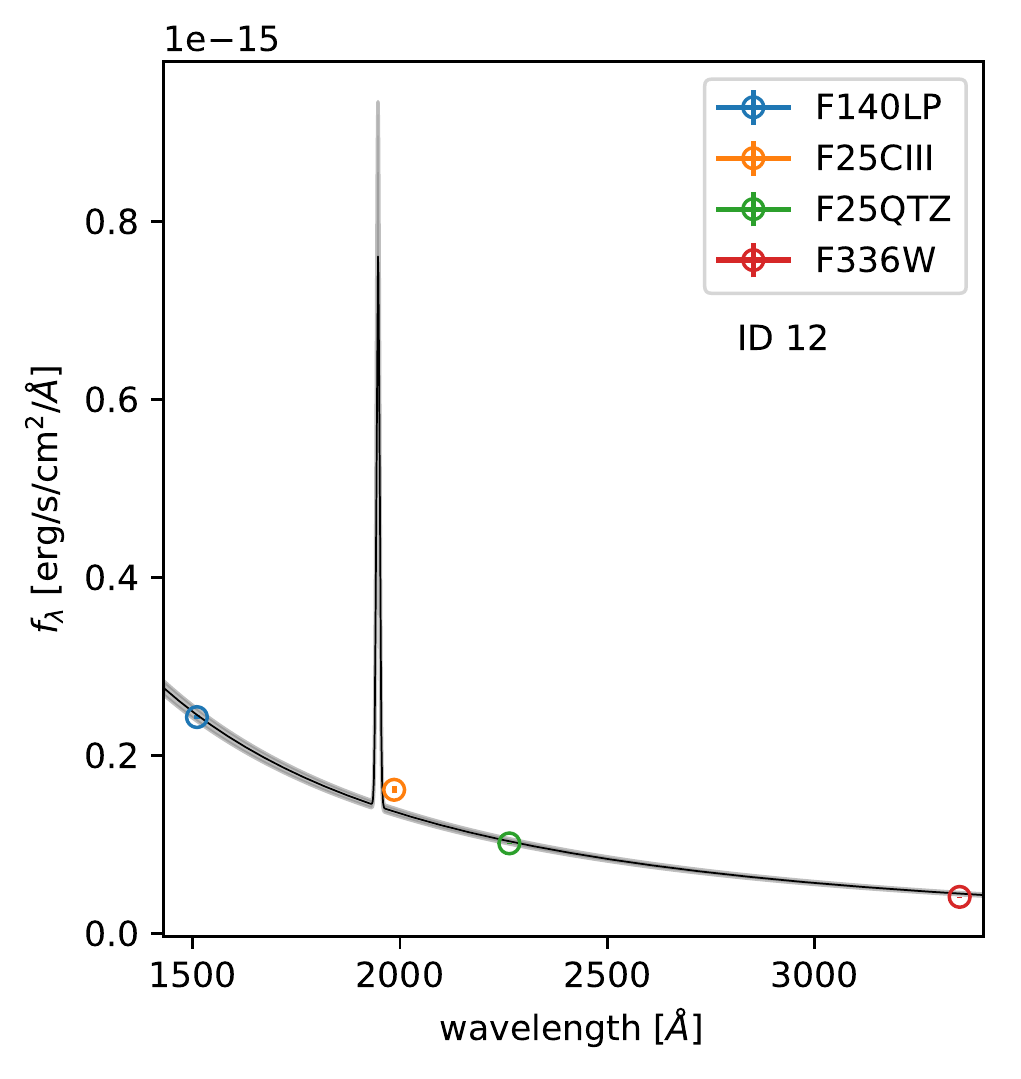}{0.25\textwidth}{}
    \fig{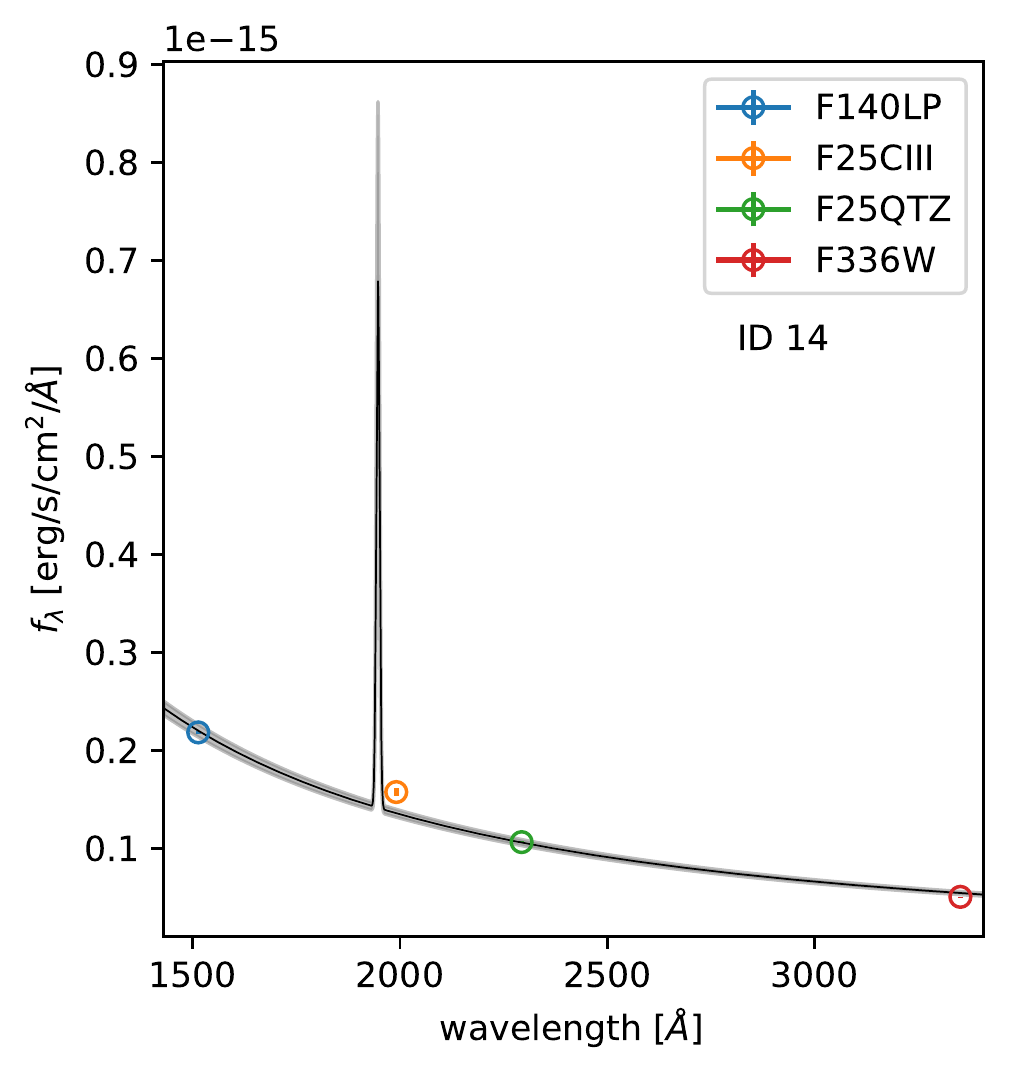}{0.25\textwidth}{}
  }
\caption{Powerlaw fits to all apertures with detections. The position of the markers changes slightly from aperture to aperture because the pivot wavelength has been recalculated to account for the shape of each spectrum. Gray-shaded area represents the uncertainty in the resulting fit parameters. \label{fig:powlaw2}  }
\end{figure*}

The pivot wavelength of a filter depends on the shape of the spectrum if the filter is broad enough and if the slope of the spectrum changes significantly in the wavelength range covered by the filter. To account for this, for every resolution element (pixel, Voronoi bin, aperture, or dendrogram) we first obtain an initial powerlaw fit to the three UV broadband filters F140LP, F25QTZ, and F336W, {\it excluding} the Gaussian line in the equation in Section \ref{sec:powlaw}. The resulting synthetic spectrum is a first-guess, and gives a better handle on the slope of the spectrum. It is used to obtain a more precise pivot wavelength for each filter and each resolution element via
\begin{equation}
  \lambda_{pivot} = \sqrt{ \frac{\int{F_\lambda \lambda d\lambda}}{\int{(F_\lambda / \lambda)d\lambda}}  }
\end{equation}
where $F_\lambda$ is the source flux distribution, i.e. the synthetic spectrum convolved with the bandpass throughput. In practical terms this amounts to using {\tt\string Observation.pivot()} instead of {\tt\string ObsBandpass.pivot()} in {\tt\string pysynphot}. Using these re-calculated pivot wavelengths, the powerlaw is then repeated for each resolution element, using all four filters, i.e. F140LP, F25CIII, F25QTZ, and F336W, and the formula given in Section \ref{sec:powlaw}. Figure \ref{fig:powlaw2} shows all (final) powerlaw fits for the apertures with detections. 

\section{Histograms of MC runs}
\begin{figure*}[ht!]
  \fbox{\includegraphics[width=0.5\textwidth]{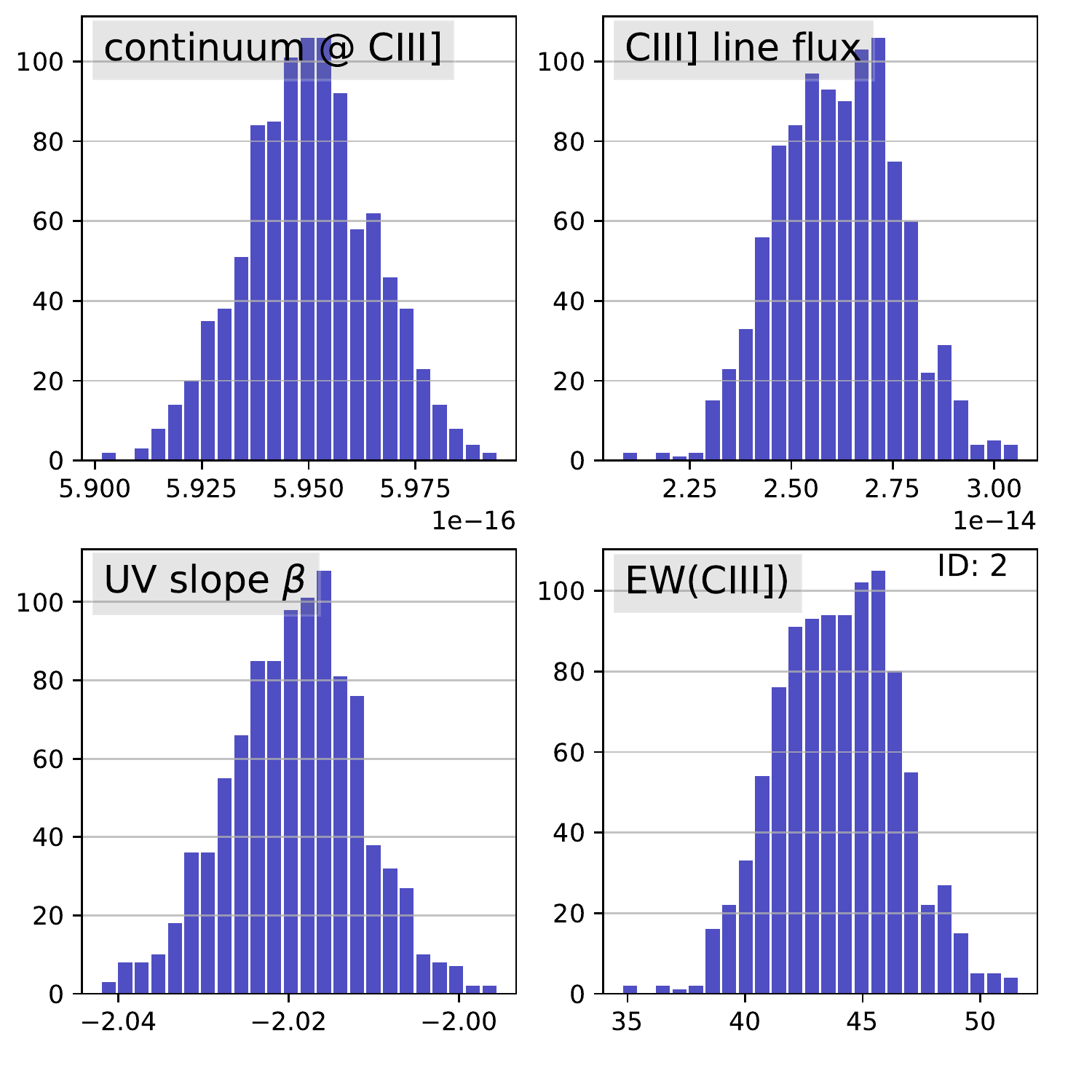}}\fbox{\includegraphics[width=0.5\textwidth]{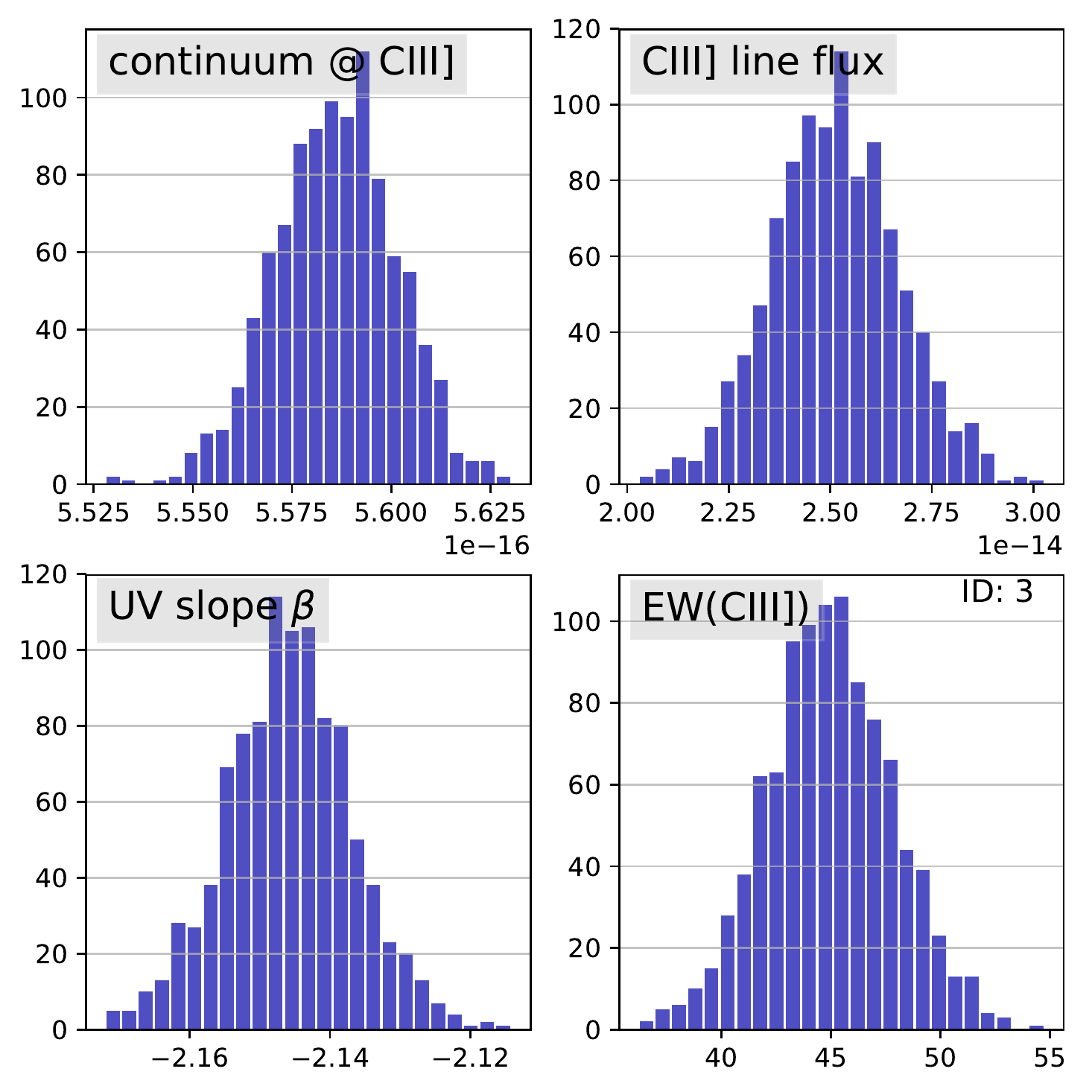}}\\
  \fbox{\includegraphics[width=0.5\textwidth]{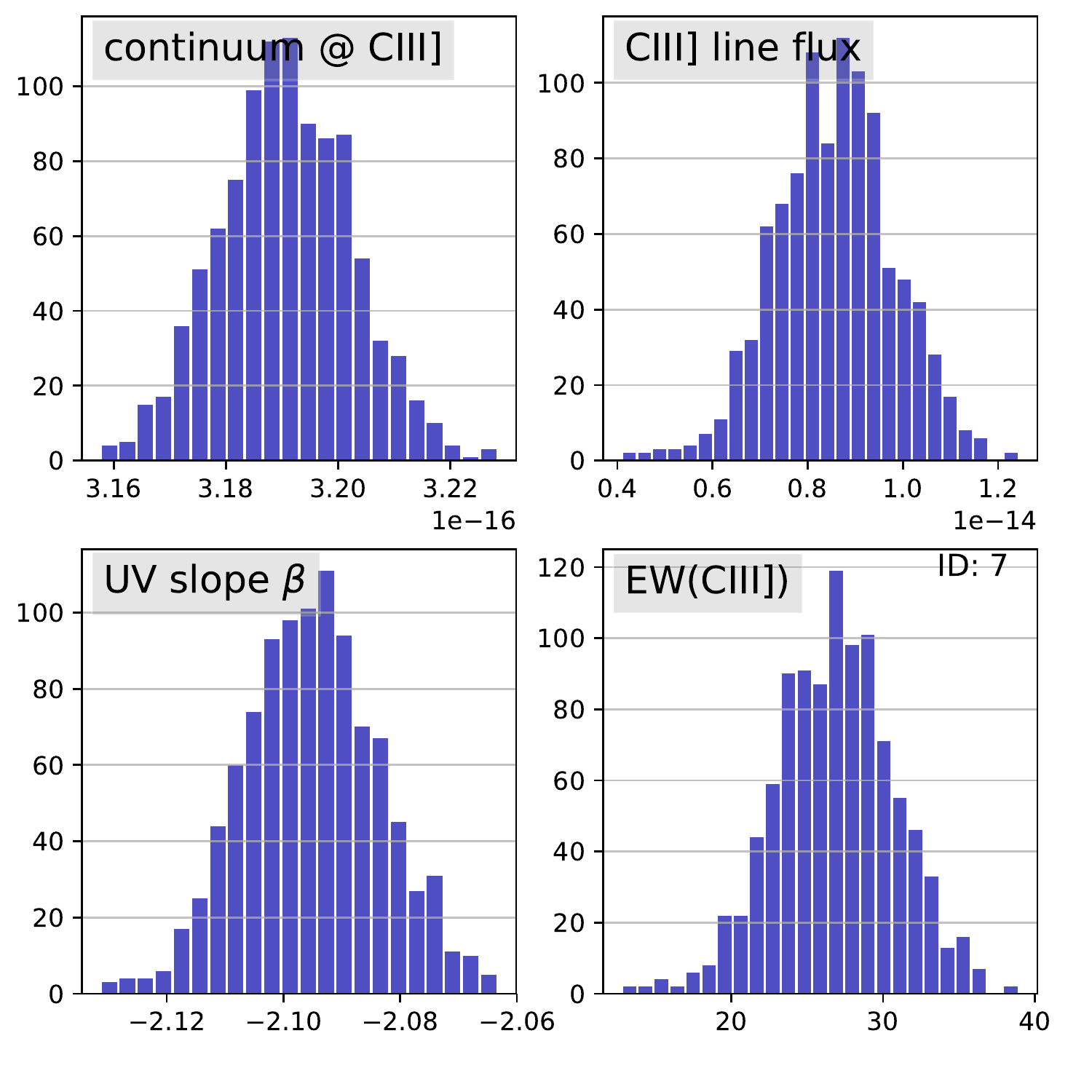}}\fbox{\includegraphics[width=0.5\textwidth]{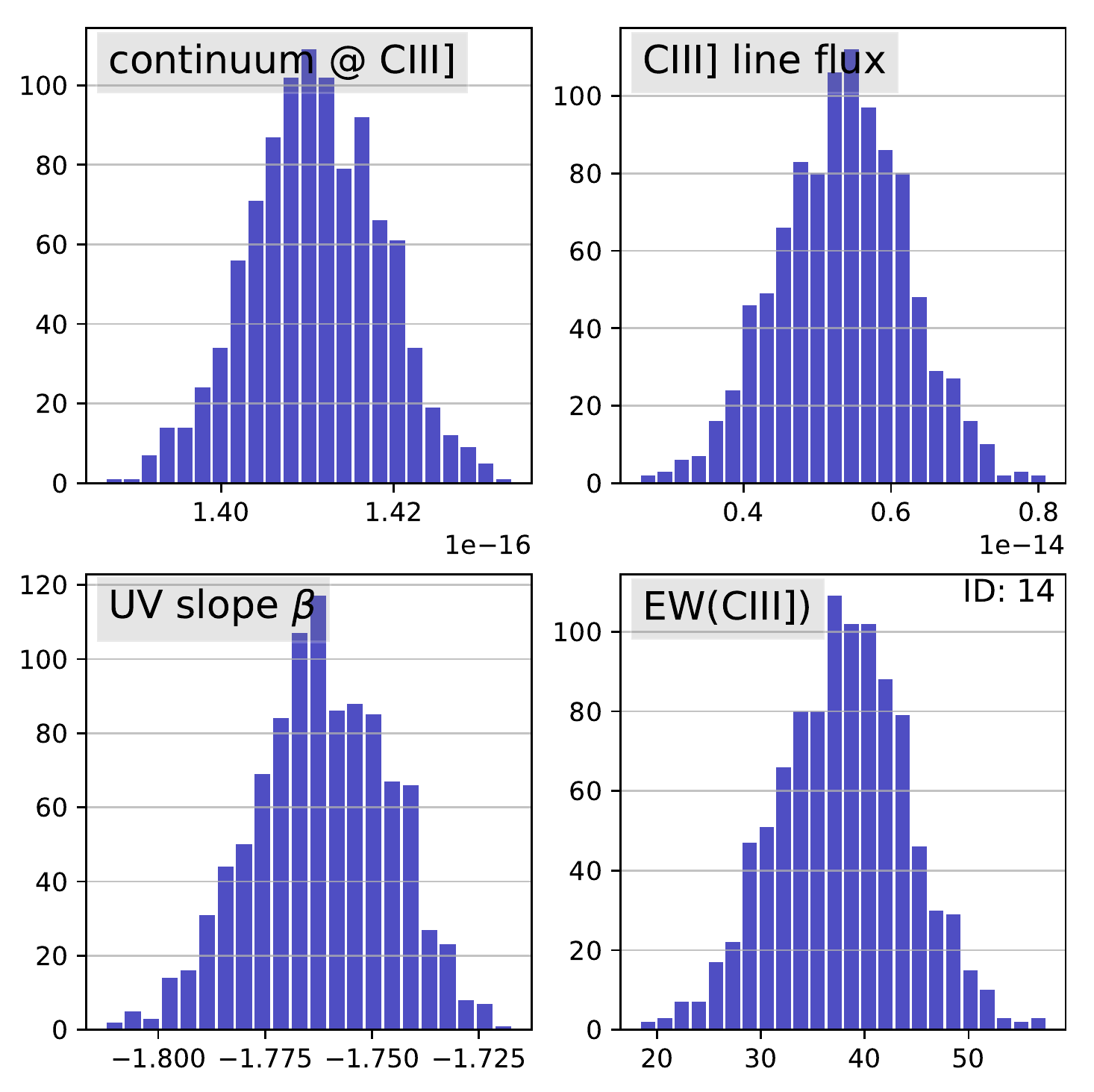}}\\  
  \caption{Histograms of MC runs with $N=1000$ for the continuum, line flux, UV slope $\beta$, and EW(\ciii) for four random apertures, with ID numbers given in the EW(\ciii) subplots. Histograms belonging to the same aperture are framed together. \label{fig:hist}  }
\end{figure*}

\protect\label{sec:hist}
The uncertainties on the $f_{cont}$, $F_{line}$, $\beta$, and, most importantly, on EW(\ciii) are estimated as the standard deviation of $N=100$ [$1000$] MC realizations. EW(\ciii) is obtained as the ratio of the distributions of two correlated parameters ($F_{line}$ and $f_{cont}$), and hence one can neither estimate its errors via error propagation, nor can one blindly assume that the standard deviation (stddev) of its distribution is a valid approximation of the errors. The stddev is only a meaningful statistic if the distribution is normal. In Figure \ref{fig:hist} we show the distribution of these parameters with $N=1000$ MC runs, for signal extraction via aperture photometry for four out of $18$ apertures, selected at random. The figure shows that assuming a normal distribution is not unreasonable, and hence using the standard deviation is an acceptable estimate of the uncertainties.

\section{Cloudy simulations}\protect\label{sec:cloudy}
We used the photoionization code Cloudy \citep[v17.01][]{Ferland2017} to model \hii\ regions in order to examine their ionization structure and line emissivity as a function of cloud depth. The models were used to obtain figure \ref{fig:theory}. The input Starburst99 SEDs were for instantaneous SFH and metallicity $Z=0.008$. The metallicity of the stars and the gas was equal. To obtain a volume-averaged ionization parameter of  given $\log{U}$ value, we first calculated the corresponding production rate of ionizing photons following \citet{Stasinska2015}, then obtained the Str\"omgren radius $R_S$, and finally the radius $r_{0}$ to the inner face of the cloud by selecting $r_0\sim R_S$. This ensures plane parallel geometry. We note that assuming spherical geometry does not change the ionization structure or the general pattern of line emissivity variations with cloud depth. The hydrogen density was set to $n_H = 100$ cm$^{-3}$. Following \citet{Gutkin2016}, we adopted the solar abundance distribution (their table 1), accounted for secondary nitrogen production, and adjusted the abundance of each element to compensate for the increase in nitrogen abundance. Standard dust grains were loaded, with dust sublimation and depletion on metals turned on. Both a cosmic ray and a cosmic microwave background were also added via the default commands. The covering and the filling factors were constant and unity. The stopping criterion was reaching an electron density of $1$ cm$^{-3}$. We obtained models for $\log{U}=-1, -1.5,-2, -2.5$, ages $1\mbox{-}6$ Myr in steps of $1$ Myr. PyCloudy \citep{Morisset2013} was used create the input files, run the models, and plot the output.

To add a density-bounded cloud to the \citet{Gutkin2016} set of models, we adopted their setup (plane parallel geometry, continuous SFH, age $100$Myr, $N_e=100$ cm $^{-3}$, $C/O=1.4 (C/O)_\odot$, secondary Nitrogen production), but stopped the cloudy calculation when a neutral column density of $N_{HI}=10^{17.2}$ cm$^{-2}$ was achieved. The metallicity varies as $Z=0.0004, 0.004, 0.008, 0.020$, and the ionization parameter covers the range $\log{U}=-1.0$ to $-3.5$ in steps of $-0.5$. 

\begin{figure*}[ht!]
  \includegraphics[width=0.5\textwidth]{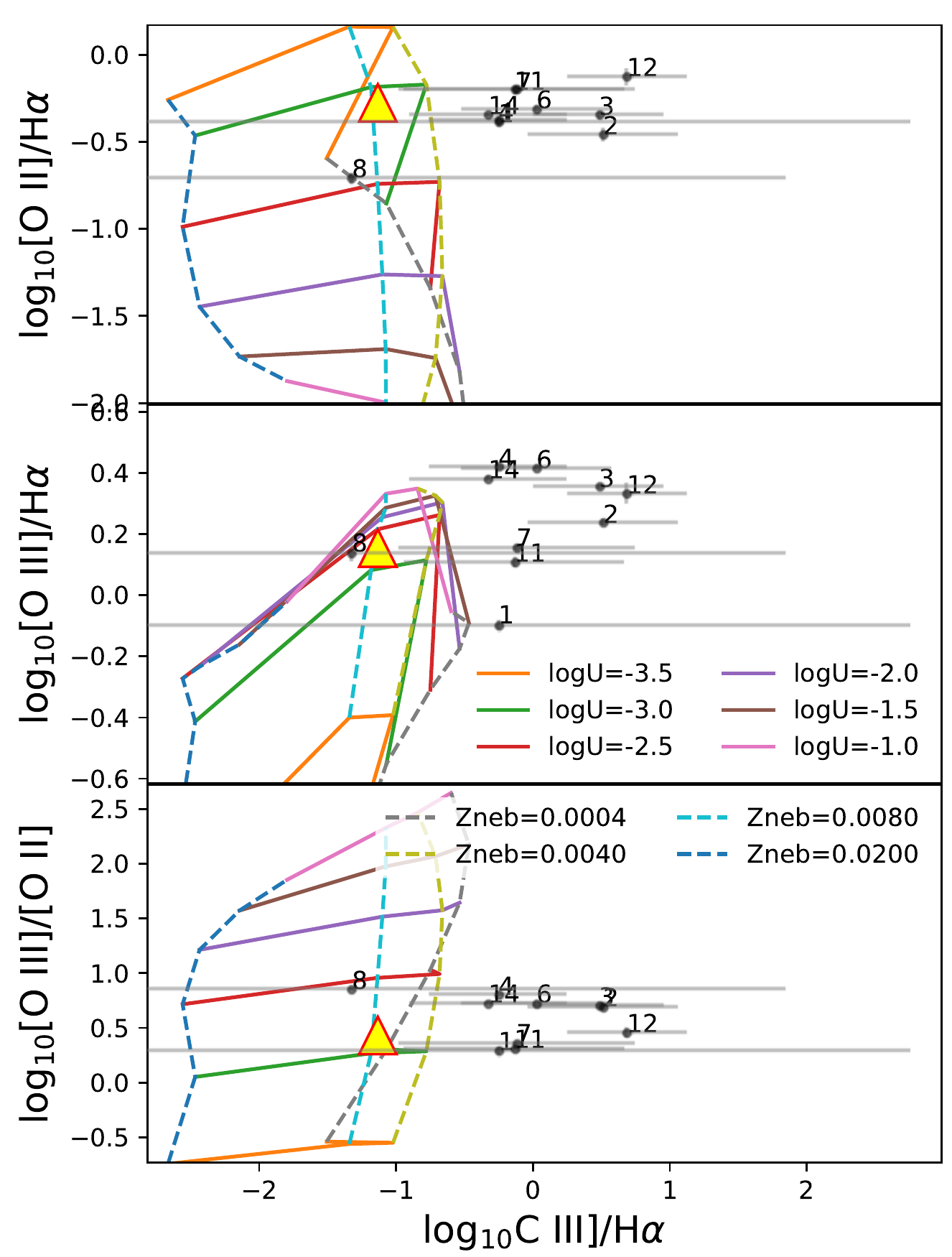}
  \includegraphics[width=0.5\textwidth]{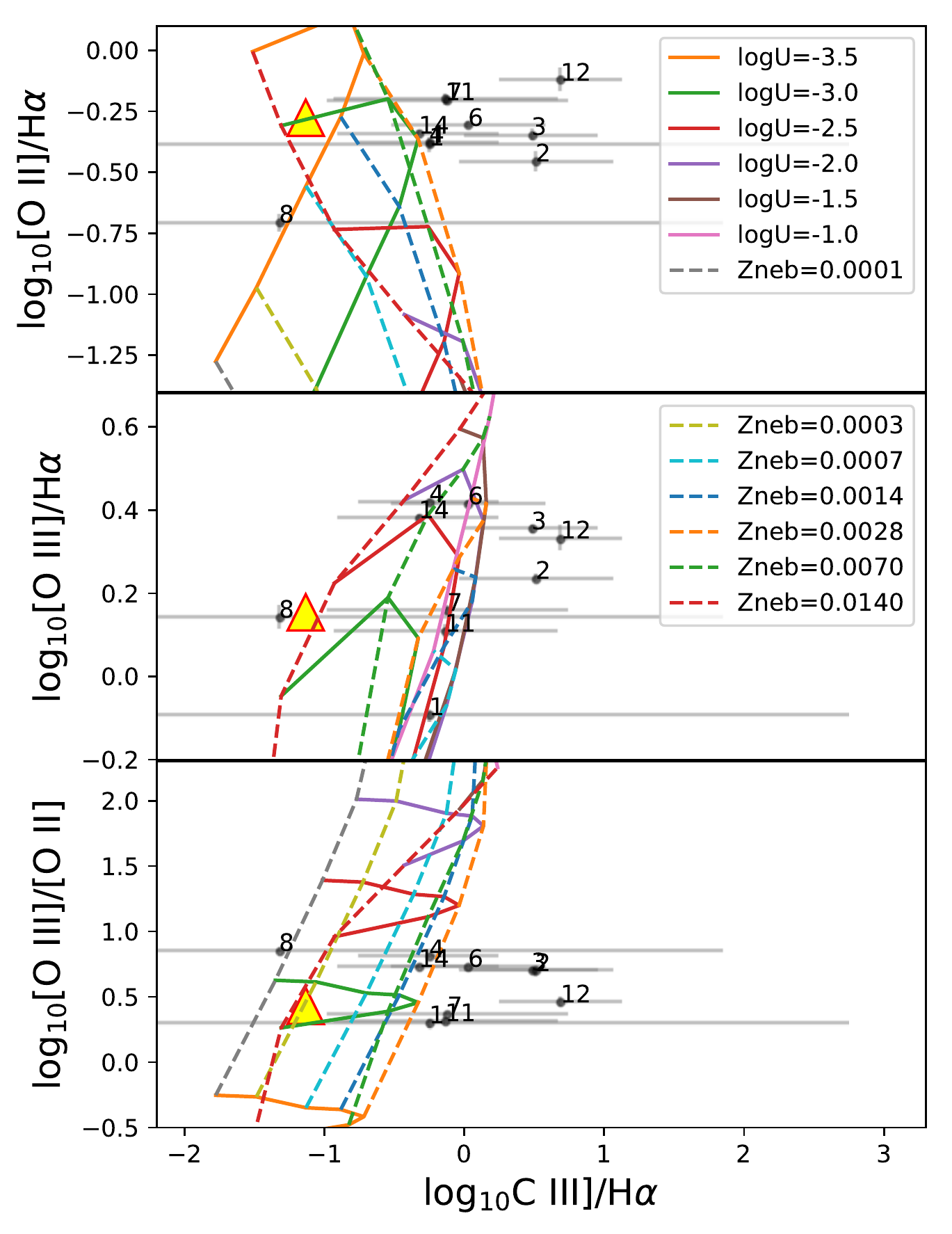}\\
  \caption{Left: Same as Figure \ref{fig:lines} but with an optically thin cloud, ionized by single stars, (C/O)$=1.4(C/O)_\odot$. Right: for an optically thin cloud, ionized by a mixture of binary stars and AGN, (C/O)$=2(C/O)_\odot$.\label{fig:leak}  }
\end{figure*}

In the left panel of Figure \ref{fig:leak} we show the resulting Cloudy grid for density-bounded \hii\ regions. The grid is very coarse due to the limited metallicity range available in Starburst99, but it appears inconsistent with apertures 2, 3, and 12. We also note in passing that the errorbar on knot C (aperture ID 1) is too large to determine if it is better matched with a leaking cloud or with the optically thick grid in Figure \ref{fig:lines}, and hence we cannot help determine if knot A or C is the origin of the LyC leakage in Haro 11, a question posed by \citet{Keenan2017}. 

In the right panel of Figure \ref{fig:leak}, we show a density-bounded Cloudy model grid, based on the \citet{Nakajima2018} setup. The ionizing source is a mixture of binary stars (BPASS) and a $10\%$ contribution of AGN ionizing photons, with an AGN powerlaw slope of $-1.2$, an upper mass limit of $M_{up}=300M_\odot$ and a super solar C/O$=1$ (about twice solar). Similar to the Starburst99 grid in the left panel, the calculation was stopped at $N_{HI}=10^{17.2}$ cm$^{-2}$. Apertures 2, 3, and 12 are again inconsistent with the models. A density-bounded cloud therefore cannot explain their enhanced \ciii\ emission.

\section{SED fitting with Cigale}\protect\label{sec:cigale}
For the SED fit of the clusters in Table \ref{tab:powlaw} we used six broadband filters. In addition to our F25QTZ filter, from HST archival data we added F140LP from proposal ID (PID) $9470$, F336W and F763M from PID $13702$, and F435W and F550M from PID $10575$. The star formation history was assumed to be a double exponential, with an e-folding time of $\tau_{old}=6$ Gyr for the main stellar population, and $\tau_{burst}=0.01$ Myr for the young population. The young mass fraction was allowed to vary from $f_{burst}=0.01$ to $0.9$ in steps of $0.05$. The burst age spanned $2$ to $12$ Myr in steps of $1$ Myr. The gas and stellar metallicities were assumed to be the same and allowed to take on values of $Z=0.004$ or $Z=0.008$, which covers the metallicity estimates for Haro 11 as a whole \citep[e.g.,][$12+\log{O/H}=7.9$]{Bergvall2002} and the individual knots A, B, and C \citep[][$8.09\pm0.2$, $8.25\pm0.15$, and $7.8\pm0.13$, respectively]{James2013}. The ionization parameter was allowed to vary from $\log{U}=-1.1$ to $-2.7$ in steps of $-0.1$. The escape fraction was set to zero, while the dust fraction was allowed to vary from $f_{dust}=0.1$ to $0.95$ in steps of $0.05$. Finally, the ISM attenuation was varried from $E(B-V)=0.1$ to $0.7$ in steps of $0.02$. 

\begin{figure*}[ht!]
  \includegraphics[width=0.9\textwidth]{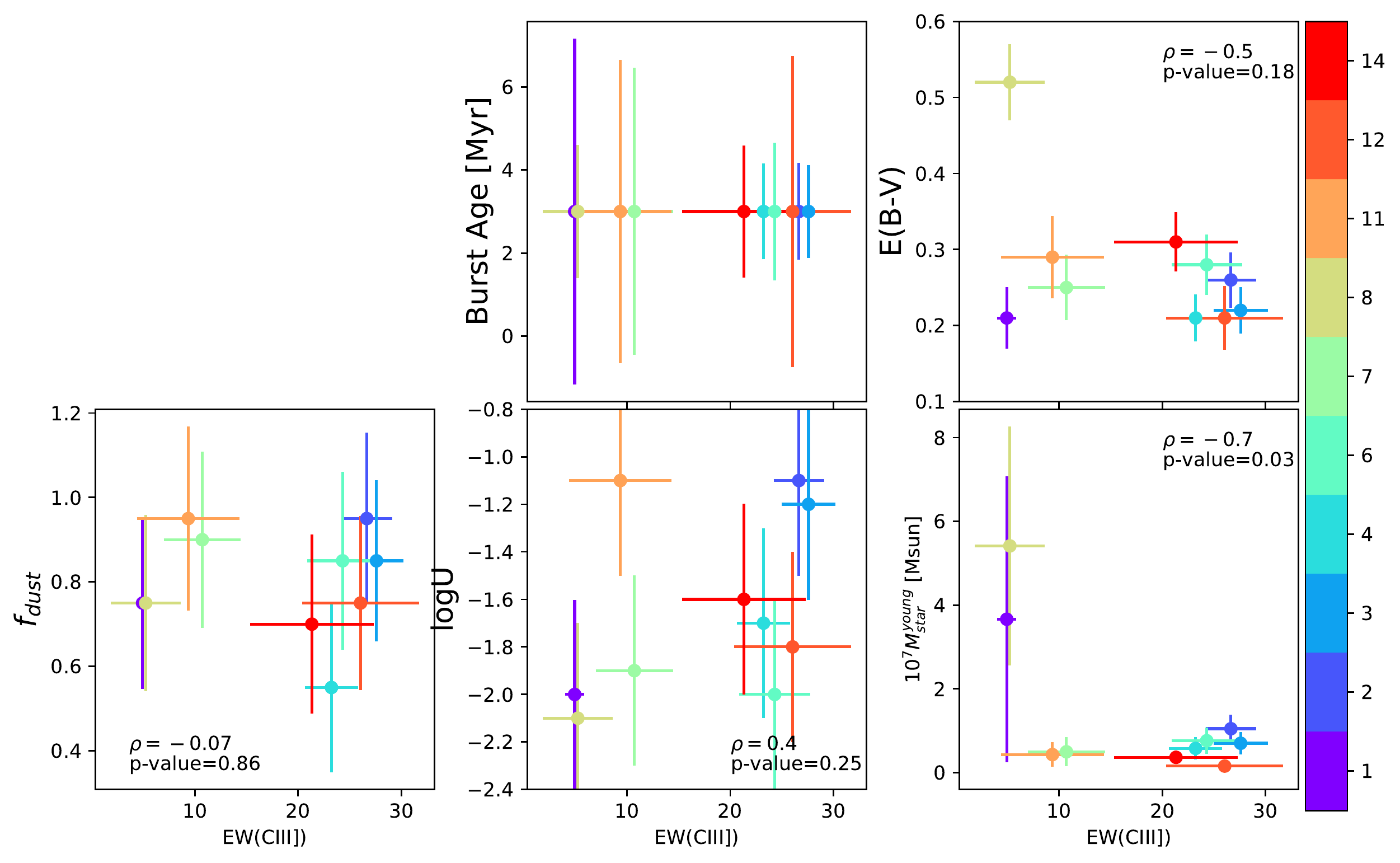}
\caption{\ew\ versus cluster properties from the Cigale SED fit. Pearson's $\rho$ and p-values are indicated in each panel. Markers are color-coded by the aperture IDs in Table \ref{tab:powlaw}. The location of the clusters inside of Haro 11 can be seen in the middle panel of Figure \ref{fig:ew}. The best fit for each cluster was with a metallicity of $Z=0.008$, and hence the metallicity is identical for all data points. \label{fig:cprops}}
\end{figure*}

Our Cigale fits compare well to the \citet{Adamo2010} fits. The results of the major parameters are listed in Table \ref{tab:cigale}. For clusters 2, 3, 4, 7, 12, and 14, Adamo have an age of $3.5$ Myr, while we have ages of 3 Myr. Cigale only has steps of $\pm1$ Myr, and hence this is perfectly consistent with the Adamo results. Exceptions are clusters 1 (knot C) and 5, for which Adamo find much older ages of $\ge10$ Myr.  

It is certainly possible to produce better SED fits by modeling the individual nebulosity for each cluster with Cloudy. The purpose of our SED fitting is only to check if the wrong F140LP zeropoint has affected the SED fitting in \citet{Adamo2010} to the point of destroying any correlation with \ew. Since we also find no correlations using Cigale with the correct F140LP zerorpoint, we judge that the F140LP filter has not affected the Adamo fit significantly and we do not further explore the SEDs of the clusters. 

\begin{deluxetable*}{lCCCCC}
\tablecaption{Parameters from the SED fit with Cigale. \label{tab:cigale}}
\tablehead{
\colhead{ID} & \colhead{Age} & \colhead{E(B-V)} & \colhead{$f_{dust}$} &\colhead{log U}&\colhead{$M_\star^{young}$}\\
\colhead{} & \colhead{Myr} & \colhead{mag} & \colhead{} & \colhead{} &\colhead{$10^7M_\odot$}
}
\startdata
1  & 3\pm 4.2 & 0.21\pm 0.04  & 0.75\pm 0.20 & -2.0\pm 0.4 & 3.66\pm 3.42\\
2  & 3\pm 1.2 & 0.26\pm 0.04  & 0.95\pm 0.20 & -1.1\pm 0.4 & 1.05\pm 0.34\\
3  & 3\pm 1.1 & 0.22\pm 0.03  & 0.85\pm 0.19 & -1.2\pm 0.4 & 0.70\pm 0.26\\
4  & 3\pm 1.2 & 0.21\pm 0.03  & 0.55\pm 0.20 & -1.7\pm 0.4 & 0.58\pm 0.27\\
6  & 3\pm 1.7 & 0.28\pm 0.04  & 0.85\pm 0.21 & -2.0\pm 0.4 & 0.76\pm 0.31\\
7  & 3\pm 3.5 & 0.25\pm 0.04  & 0.90\pm 0.21 & -1.9\pm 0.4 & 0.50\pm 0.35\\
8  & 3\pm 1.6 & 0.52\pm 0.05  & 0.75\pm 0.21 & -2.1\pm 0.4 & 5.41\pm 2.85\\
11 & 3\pm 3.7 & 0.29\pm 0.05  & 0.95\pm 0.22 & -1.1\pm 0.4 & 0.43\pm 0.29\\
12 & 3\pm 3.7 & 0.21\pm 0.04  & 0.75\pm 0.21 & -1.8\pm 0.4 & 0.16\pm 0.15\\
14 & 3\pm 1.6 & 0.31\pm 0.04  & 0.70\pm 0.21 & -1.6\pm 0.4 & 0.36\pm 0.16\\
\enddata
\end{deluxetable*}

\acknowledgments
\noindent AKI is supported by JSPS KAKENHI Grant Number JP17H01114.\\
%The authors kindly thank Kimihiko Nakajima for providing us with his Cloudy grid for AGN.\\
The authors thank the anonymous referee for the very thorough and careful reading of the paper.
The authors thank Norberto Castro Rodriguez, Philipp Girihidis, Charles Cowley, B.T. Draine, Jakob Walcher, and Matthias Steffen for useful and enlightening discussions.\\
This research made use of astrodendro, a Python package to compute dendrograms of Astronomical data (http://www.dendrograms.org/)\\
This research made use of Astropy,\footnote{http://www.astropy.org} a community-developed core Python package for Astronomy \citep{astropy:2013, astropy:2018}.\\

\bibliographystyle{aasjournal.bst}
\bibliography{ciii_haro11}

\end{document}